\documentclass[twocolumn,twocolappendix,showkeys]{openjournal}

\usepackage[utf8]{inputenc}
\usepackage[english]{babel}
\usepackage{apjfonts}
\usepackage{booktabs}
\usepackage{amsmath}
\usepackage{placeins}
\usepackage[colorlinks=true,citecolor=blue,linkcolor=blue,urlcolor=blue]{hyperref}
\newcommand{\orcidlink}[1]{\href{https://orcid.org/#1}{%
  \raisebox{-0.2ex}{\includegraphics[height=1.45ex]{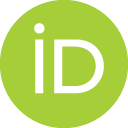}}}}
\DeclareGraphicsExtensions{.pdf,.png,.jpg}

\newcommand{\ha}{\ensuremath{{\rm H}\alpha}}
\newcommand{\hb}{\ensuremath{{\rm H}\beta}}

\newcommand{\oiii}{[O\,\textsc{iii}]}
\newcommand{\oii}{[O\,\textsc{ii}]}
\newcommand{\nii}{[N\,\textsc{ii}]}
\newcommand{\lya}{\ensuremath{{\rm Ly}\alpha}}

\providecommand{\arcsec}{\ensuremath{^{\prime\prime}}}
\newcommand{\authorname}[2][]{\textsc{#2}\if\relax\detokenize{#1}\relax\else\,\orcidlink{#1}\fi}
\makeatletter
\renewcommand{\abstractname}{\MakeUppercase{Abstract}}
\def\apjsecfont{\normalsize}
\def\frontmatter@authorformat{\small\scshape\centering}
\def\frontmatter@title@format{\normalsize\bfseries\boldmath\centering}
\makeatother
\newcommand{\zlc}{z}
\graphicspath{{../figures/}}

\begin{document}
\renewcommand{\abstractname}{ABSTRACT}

\title{MARTA: An overmassive black hole hidden in a young, clumpy, and metal-poor galaxy at $\zlc=3.8$}

\author{\authorname[0000-0002-2678-2560]{Mirko Curti}\altaffilmark{1},
\authorname[0009-0004-7069-1326]{Bianca Moreschini}\altaffilmark{2,3,4},
\authorname[0009-0005-5187-4001]{Elisa Cataldi}\altaffilmark{2,3},
\authorname[0000-0002-2545-5752]{Francesco Belfiore}\altaffilmark{4},
\authorname[0000-0001-8706-2252]{Ivan Delvecchio}\altaffilmark{1},
\authorname[0000-0002-1528-5091]{Sophie Koudmani}\altaffilmark{5,6},
\authorname[0000-0003-1282-7454]{Anthony J. Taylor}\altaffilmark{7,8},
\authorname[0000-0002-9889-4238]{Alessandro Marconi}\altaffilmark{2,3},
\authorname[0000-0002-4803-2381]{Filippo Mannucci}\altaffilmark{3},
\authorname[0000-0001-5487-2830]{Elena Bertola}\altaffilmark{3},
\authorname[0000-0002-5281-1417]{Giovanni Cresci}\altaffilmark{3},
\authorname[0000-0001-6865-2871]{Anna Feltre}\altaffilmark{3},
\authorname[0000-0002-4985-3819]{Roberto Maiolino}\altaffilmark{6,9},
\authorname[0000-0002-7412-647X]{Francesca Pozzi}\altaffilmark{10,1},
\authorname[0009-0002-6167-6728]{Letizia Scaloni}\altaffilmark{1,10},
\authorname[0000-0003-1006-924X]{Alberto Traina}\altaffilmark{11},
\authorname[0009-0009-0580-2604]{Caterina Bracci}\altaffilmark{2,3,4},
\authorname[0000-0002-6719-380X]{Stefano Carniani}\altaffilmark{12},
\authorname[0000-0002-5836-4056]{Carlotta Gruppioni}\altaffilmark{1},
\authorname[0000-0001-8349-3055]{Giacomo Venturi}\altaffilmark{3},
\authorname[0000-0001-7214-3009]{Magda Arnaboldi}\altaffilmark{4},
\authorname[0000-0002-6444-8547]{Meriem Behiri}\altaffilmark{1},
\authorname[0000-0002-9122-1700]{Michele Ginolfi}\altaffilmark{2,3},
\authorname[0000-0003-1041-7865]{Mahsa Kohandel}\altaffilmark{1},
\authorname[0000-0002-3258-3672]{Livia Vallini}\altaffilmark{1}
}
\email{mirko.curti@inaf.it}
\affiliation{%
\altaffilmark{1} INAF -- Osservatorio di Astrofisica e Scienza dello Spazio di Bologna, Via Piero Gobetti 93/3, I-40129 Bologna, Italy\\
\altaffilmark{2} Universit\`a di Firenze, Dipartimento di Fisica e Astronomia, Via G. Sansone 1, I-50019 Sesto Fiorentino, Florence, Italy\\
\altaffilmark{3} INAF -- Osservatorio Astrofisico di Arcetri, Largo E. Fermi 5, I-50125 Florence, Italy\\
\altaffilmark{4} European Southern Observatory, Karl-Schwarzschild-Strasse 2, D-85748 Garching bei M\"unchen, Germany\\
\altaffilmark{5} Centre for Astrophysics Research, Department of Physics, Astronomy and Mathematics, University of Hertfordshire, College Lane, Hatfield AL10 9AB, UK\\
\altaffilmark{6} Kavli Institute for Cosmology, University of Cambridge, Madingley Road, Cambridge CB3 0HA, UK\\
\altaffilmark{7} Department of Astronomy, The University of Texas at Austin, Austin, TX 78712, USA\\
\altaffilmark{8} Cosmic Frontier Center, The University of Texas at Austin, Austin, TX 78712, USA\\
\altaffilmark{9} Cavendish Laboratory, University of Cambridge, 19 JJ Thomson Avenue, Cambridge CB3 0HE, UK\\
\altaffilmark{10} Dipartimento di Fisica e Astronomia ``Augusto Righi'', Alma Mater Studiorum -- Universit\`a di Bologna, Via Piero Gobetti 93/2, I-40129 Bologna, Italy\\
\altaffilmark{11} MIT Kavli Institute for Astrophysics and Space Research, Massachusetts Institute of Technology, 77 Massachusetts Avenue, Cambridge, MA 02139, USA\\
\altaffilmark{12} Scuola Normale Superiore, Piazza dei Cavalieri 7, I-56126 Pisa, Italy}

\begin{abstract}
JWST has extended the search for accreting black holes into the high-redshift, low-mass regime, delivering unprecedented constraints on their formation pathways and seeding mechanisms.
Here we report the discovery of a low-mass black hole candidate in MARTA~3887, a clumpy, metal-poor galaxy at $z=3.7694$.
Combining deep JWST/NIRSpec MSA spectroscopy, HST and JWST imaging, and archival VLT/MUSE integral-field spectroscopy, we characterize the black hole, stellar host, and ionized gas properties of this system over the rest-frame UV-to-optical range.
We identify a broad \ha\ component with $L_{{\rm H}\alpha,{\rm b}}=2.35\times10^{41}$~erg~s$^{-1}$ and ${\rm FWHM}=1270^{+260}_{-250}$~km~s$^{-1}$, with no [O\,\textsc{iii}] counterpart of comparable width.
Under a broad-line-region interpretation, virial calibrations give $\log(M_{\rm BH}/M_\odot)=6.52^{+0.18}_{-0.23}$ and $\lambda_{\rm Edd}=L_{\rm bol}/L_{\rm Edd}=0.07$, or up to $0.38$ for alternative calibrations.
The host is resolved into three compact clumps, with integrated SED fitting yielding $\log(M_\star/M_\odot)=8.16^{+0.09}_{-0.10}$.
The narrow-line spectrum is consistent with highly ionized, metal-poor gas, with $12+\log({\rm O/H})=7.68\pm0.04$ ($\simeq0.10\,Z_\odot$).
The inferred $M_{\rm BH}/M_\star=2.3\%$ lies approximately 2~dex above the extrapolated local relation between black-hole mass and host stellar mass, suggesting that black-hole growth has outpaced the assembly of the observed stellar component.
Illustrative growth tracks show that a lifetime-mean Eddington ratio $\langle\lambda\rangle\simeq5.7$ would be required for a light ($10^2\,M_\odot$) seed if accretion began only when the first 10\% of the observed stellar mass had formed ($t_{10}=92$~Myr in lookback time), whereas an earlier onset of accretion or a more massive seed would relax this requirement.
Within the \textsc{aesopica} framework, a source-matched comparison conditionally favours $10^5\,M_\odot$ seed analogues for the fiducial measurements, while permitting limited overlap with efficiently growing lighter seeds.
In addition, MUSE data reveal extended \lya\ emission ($\simeq6.5$--6.9~proper kpc), offset by $+64^{+10}_{-8}$~km~s$^{-1}$ from systemic, with a provisional escape fraction of $\simeq0.44$.
Together with the narrow Mg~II doublet ratio close to the optically thin limit and a high [O\,\textsc{iii}]/[O\,\textsc{ii}] ratio, these properties make MARTA~3887 a system in which black-hole growth, stellar assembly, and possible ionizing-photon escape channels can be studied jointly.
\end{abstract}

\maketitle

\section{Introduction}
\label{sec:intro}

The origin of the first massive black holes remains a central problem in galaxy evolution.
The existence of ${\sim}10^9\,M_\odot$ quasars within the first billion years requires massive seeds, sustained near-Eddington accretion, super-Eddington episodes, or a combination of these processes \citep{Inayoshi2020}.
Luminous quasars, however, sample only the high-mass tail of the distribution, while black holes with $M_{\rm BH}\sim10^5$--$10^7\,M_\odot$ in lower-mass galaxies provide complementary constraints on the early phases of seeding and black-hole-galaxy co-evolution.

JWST/NIRSpec has extended the search and characterization of early black holes exactly into this regime.
Broad-line AGN with $M_{\rm BH}\sim10^6$--$10^8\,M_\odot$ are now being routinely identified at high redshift, well below the luminosities accessible to pre-JWST quasar surveys \citep{Harikane2023,Kocevski2023,Larson2023,Ubler2023,Matthee2024,MaiolinoSample2024,Greene2024,Juodzbalis2026,Geris2026}.
Many of the newly discovered black holes have large inferred $M_{\rm BH}/M_\star$ ratios relative to local scaling relations \citep{ReinesVolonteri2015,KormendyHo2013,Pacucci2023}, suggesting that black-hole growth might precede the assembly of the dominant stellar component.
Strong selection effects related to flux and line-width sensitivity thresholds favour the detection of massive and/or rapidly accreting black holes at fixed host mass, while measurement scatter further enhances the detected upper envelope \citep{Lauer2007,Li2024}.  Selection-aware analyses indicate that the intrinsic high-redshift $M_{\rm BH}$--$M_\star$ relation may be consistent with the local relation, but with substantially larger scatter capable of producing the observed high-$M_{\rm BH}/M_\star$ tail \citep{Ziparo2026}.  Even if individual detections preferentially sample this upper envelope, the existence and frequency of such overmassive systems remain informative about black-hole formation and early growth.  Models predict different low-mass distributions and scatter depending on their prescriptions for seeding, accretion, and feedback \citep{Habouzit2022,Volonteri2023,Trinca2024}; well-characterized systems at the boundaries of the observed parameter space therefore provide complementary constraints on these processes.

The identification of the faintest candidates requires particular care.
Broad Balmer emission without an associated forbidden-line component of comparable width is a standard signature of a broad-line region (BLR), but at low luminosity it can also arise from interacting supernovae, shocks, outflows, or spatially unresolved galaxy structure \citep{Maiolino2025SN,Geris2026}.
Classical optical narrow-line diagnostics are also poorly discriminating in the metal-poor, highly ionized ISM regime common at high redshift: metal-poor narrow-line regions overlap with the star-forming sequence, while extreme stellar populations can approach classical AGN-classification boundaries \citep{Feltre2016,Gutkin2016,Kewley2013,Mazzolari2024}.

The combined effect of accretion supplying additional hard ionizing photons and radiative or mechanical AGN feedback lowering the neutral-gas covering fraction could open channels through which both AGN- and stellar-produced Lyman-continuum (LyC) photons can escape.
However, this is not an inevitable consequence of hosting an AGN: simulations of high-redshift dwarf galaxies find that supernova feedback can dominate the structure of the ISM and that black-hole activity does not necessarily increase the LyC escape fraction \citep{Trebitsch2018}.  Direct measurements at $z\simeq3$--4 span a broad range, from sub-unity average escape fractions in moderate-luminosity AGN to high values in luminous quasars \citep{Grazian2018,Iwata2022}.
Compactness, high \oiii /\oii, and the presence of strong emission in resonant lines such as \lya\ and Mg~II probing the physical conditions associated with low neutral gas column densities or porous sightlines can therefore test whether active black-hole growth coexists with favourable ionizing radiation escape channels \citep{JaskotOey2013,NakajimaOuchi2014,Verhamme2015,Henry2018,Chisholm2020,Izotov2018,Flury2022,Xu2022,Naidu2022}.
Systems combining these features offer a view of the interplay between black-hole growth, stellar feedback, and ionizing radiation escape.  Each proxy remains degenerate with stellar age, dust, gas geometry, and viewing angle, producing substantial scatter with $f_{\rm esc}^{\rm LyC}$ \citep{Katz2022,Choustikov2024}.

In this paper we present a multi-wavelength analysis of MARTA~3887, a compact, clumpy, and metal-poor galaxy at $z=3.7694$ with robust evidence for a low-mass black hole candidate, combining deep JWST/NIRSpec spectroscopy, HST and JWST imaging, and VLT/MUSE integral-field spectroscopy.
We investigate the origin of the detected \ha\ broad component, place the inferred black hole and host properties on scaling relations, and examine the implications for black-hole-galaxy co-evolution and for LyC escape.
The paper is organized as follows.
Sections~\ref{sec:data}--\ref{sec:muse} describe the data and measurements, Sect.~\ref{sec:discussion} discusses the AGN interpretation, black-hole growth, and ionizing escape, and the final section summarizes our conclusions.
We adopt the Planck 2018 cosmology throughout \citep{Planck2020}.  At $z=3.7694$, $1\arcsec$ corresponds to 7.28~proper kpc.

\begin{figure*}[t]
\centering
\includegraphics[width=0.95\textwidth]{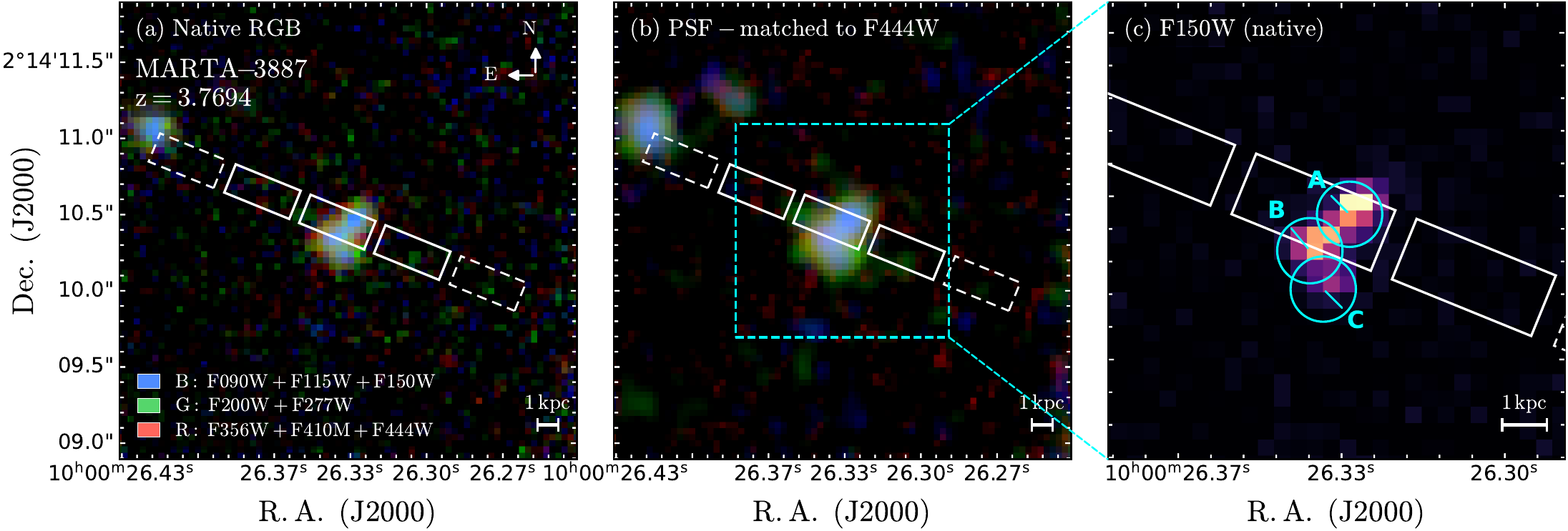}
\caption{JWST/NIRCam view of MARTA-3887 at $z=3.7694$.  Panels (a) and (b) show a $3\arcsec\times3\arcsec$ pseudo-RGB image: blue combines F090W, F115W, and F150W; green combines F200W and F277W; and red combines F356W, F410M, and F444W.  Within each channel, background-subtracted images are combined with inverse-variance weights.  Panel (a) retains their native resolution.  Panel (b) shows the same coadd after STPSF-based matching of every channel to F444W, and the cyan box identifies the region enlarged in panel (c).  Panel (c) is a native F150W zoom; cyan circles and labels mark the north-eastern (A), south-western (B), and southern (C) components of the three-component model (Sect.~\ref{ssec:morphology})  Solid white outlines in all panels mark the three open NIRSpec/MSA shutters in nod~1; dashed outlines in panels (a) and (b) mark the additional sky positions covered by the top and bottom shutters in nods~2 and 3, respectively.  Each scale bar corresponds to 1~proper kpc at the source redshift.
}
\label{fig:rgb-slitlet}
\end{figure*}

\section{Data}
\label{sec:data}

\subsection{MARTA NIRSpec spectroscopy}

MARTA (Measuring Abundances at high Redshift with the $T_{\rm e}$ Approach) is a JWST Cycle~1 GO programme (PID~1879, PI Curti) designed to obtain high-S/N rest-frame optical spectra of galaxies at cosmic noon and directly constrain their chemical abundances through faint auroral lines \citep{Cataldi2025}.
A single JWST/NIRSpec micro-shutter-array (MSA) pointing in the COSMOS field targeted 127 galaxies \citep{Gardner2023,Rigby2023,Jakobsen2022,Ferruit2022}.
The observing strategy combined a very deep G140M/F100LP exposure (31.6~h), optimized for [O\,\textsc{iii}] $\lambda4363$, with G235M/F170LP (7.3~h) coverage of the redder rest-optical spectrum and a shorter G235H/F170LP exposure (2.9~h) for resolved line kinematics.  All configurations used three-shutter slitlets and a three-position nod along the slitlet.

We reduced the data with the pipeline developed by the NIRSpec Instrument Team, adopted by the JADES collaboration, and described by \citet{Scholtz2026}\footnote{Fully reduced data for this programme can be found at \url{https://jades.herts.ac.uk/other_GO/}.}.
Starting from the calibrated count-rate exposures, the workflow identifies contaminated or non-operational shutters, subtracts the background at pixel level using uncontaminated nod positions, extracts each target trace, applies the wavelength and spectrophotometric calibrations, rectifies and combines the exposures, and produces calibrated two- and one-dimensional spectra.  The one-dimensional products use the standard point-source path-loss and slit-loss corrections; we retain the two-dimensional spectra to assess the resolved trace independently in Sect.~\ref{sec:nirspec} and Appendix~\ref{app:ha-tests}.

A relevant feature of this reduction is the extraction and calibration of spectral traces beyond the nominal wavelength limits of each disperser/filter pair, wherever the detector still records non-negligible throughput.
For G235M/F170LP the calibrated range is extended from the nominal red limit near $3.1\,\mu$m to approximately $4.0\,\mu$m \citep{Scholtz2026}.
The final reduced spectra provide therefore usable coverage over $0.97$--$1.89\,\mu$m, $1.70$--$4.00\,\mu$m, and $1.66$--$2.97\,\mu$m, respectively.
G140M contains the rest-UV and blue rest-optical lines, G235M includes \hb, [O\,\textsc{iii}], and recovers the complete \ha+\nii\ complex, while G235H provides higher-resolution coverage of \hb\ and [O\,\textsc{iii}] but does not cover \ha\ for this target (due to the detector gap).

\subsection{HST and JWST imaging}

The imaging data were retrieved from the \href{https://dawn-cph.github.io/dja/imaging/v7/}{version-7 mosaics assembled by the Dawn JWST Archive (DJA; \citealt{Valentino2023,BrammerValentino2025})}.  These mosaics combine all publicly available exposures at the target position on a common astrometric grid; the photometry used in this work is measured directly from the science and inverse-variance images.
The principal broad-band NIRCam coverage \citep{Rieke2023} comes from PRIMER (PID~1837; \citealt{Dunlop2021}) and COSMOS-Web (PID~1727; \citealt{Casey2023}), with additional imaging from Blue Jay (PID~1810; \citealt{Belli2026}), COSMOS-3D (PID~5893), and MINERVA (PID~7814; \citealt{Muzzin2025}).  Together these data provide NIRCam imaging from F090W to F460M.  The DJA products also include archival HST/WFC3 and ACS imaging from F336W to F160W in the COSMOS field \citep{Scoville2007,Grogin2011,Koekemoer2011}, and MIRI imaging \citep{Wright2023} in F770W, F1000W, F1280W, F1500W, F1800W, and F2100W.  Section~\ref{sec:photometry} describes the image-level quality control, PSF treatment, aperture measurements, and the subset adopted for stellar-population fitting.

Figure~\ref{fig:rgb-slitlet} shows the NIRCam imaging and NIRSpec/MSA footprint for MARTA~3887, with the target located in the central shutter and the three-point nodding pattern sampling five effective shutter positions on the sky.  The native RGB image (panel (a)) emphasizes small-scale morphology, while the F444W-PSF-matched rendering (panel (b)) enables a consistent comparison of the light distribution across bands, albeit at reduced angular resolution.
NIRCam imaging depicts MARTA~3887 as a multi-component system, with two knots of comparable brightness intercepted by the MSA spectroscopic aperture, and a fainter, southern knot outside of it; such a three-component nature is better seen at the highest resolution of short wavelength filters, as highlighted by the F150W-view of the system shown in panel (c) of Figure~\ref{fig:rgb-slitlet} (see also Sect.~\ref{ssec:morphology} and the full list of imaging cutouts in Appendix~\ref{app:cutouts}).
We note that, at the MSA slit orientation, the bright clumps have overlapping spatial profiles in the extracted NIRSpec spectra, and we therefore treat the one-dimensional spectrum as an integrated measurement of the two.

\subsection{Archival VLT/MUSE spectroscopy}

MARTA~3887 lies within the public MUSE-Wide pointing \texttt{candels-cosmos-16} \citep{Bacon2010,Herenz2017,Urrutia2019}.
We retrieved the reduced datacube from the ESO Science Archive (programme 097.A-0160(A), dataset ADP.2017-06-19T11:53:53.311).
The fiducial full-depth cube combines two independently reduced observing stacks with different seeing.  It has an effective on-source exposure time of 3867~s and covers $4749.5$--$9349.5$~\AA\ with a spectral sampling of 1.25~\AA\ pixel$^{-1}$ and a spatial sampling of $0.2\arcsec$ pixel$^{-1}$.  We use the full-depth cube for all fiducial measurements and the two constituent stacks only for robustness tests described in Appendix~\ref{app:muse}.
At the redshift of MARTA~3887 this range contains \lya\ and the principal rest-UV emission lines (which are, however, undetected, see Appendix~\ref{app:muse}).
The astrometric correction and data processing of the MUSE datacube are outlined in  Appendix~\ref{app:muse}, while spectral extraction and spatial analysis of the \lya\ emission are discussed in Sect.~\ref{sec:muse}.
We note that no other rest-UV line beyond \lya\ is robustly detected in the MUSE datacube; the apparent He~II $\lambda1640$ feature is identified as a sky-subtraction residual through the independent-exposure tests described in Appendix~\ref{app:muse}.
\FloatBarrier

\section{NIRSpec emission-line results}
\label{sec:nirspec}

\subsection{Broad H$\alpha$}
\label{sec:blrg}
We fit the rest-frame $\lambda \simeq6450$--$6750$~\AA\ region of the G235M spectrum using a local linear continuum and Gaussian emission-line components.
Narrow \ha, \nii\ $\lambda\lambda6548,6584$, and [S\,\textsc{ii}]$\lambda\lambda6718,6732$ share a common velocity and intrinsic dispersion; the [N\,\textsc{ii}] doublet ratio is fixed to its atomic value.
A second Gaussian is included to model any broad component for the (permitted) \ha\ emission.
All components are convolved with the NIRSpec LSF before comparison with the data.

\begin{figure}
\centering
\includegraphics[width=0.99\columnwidth]{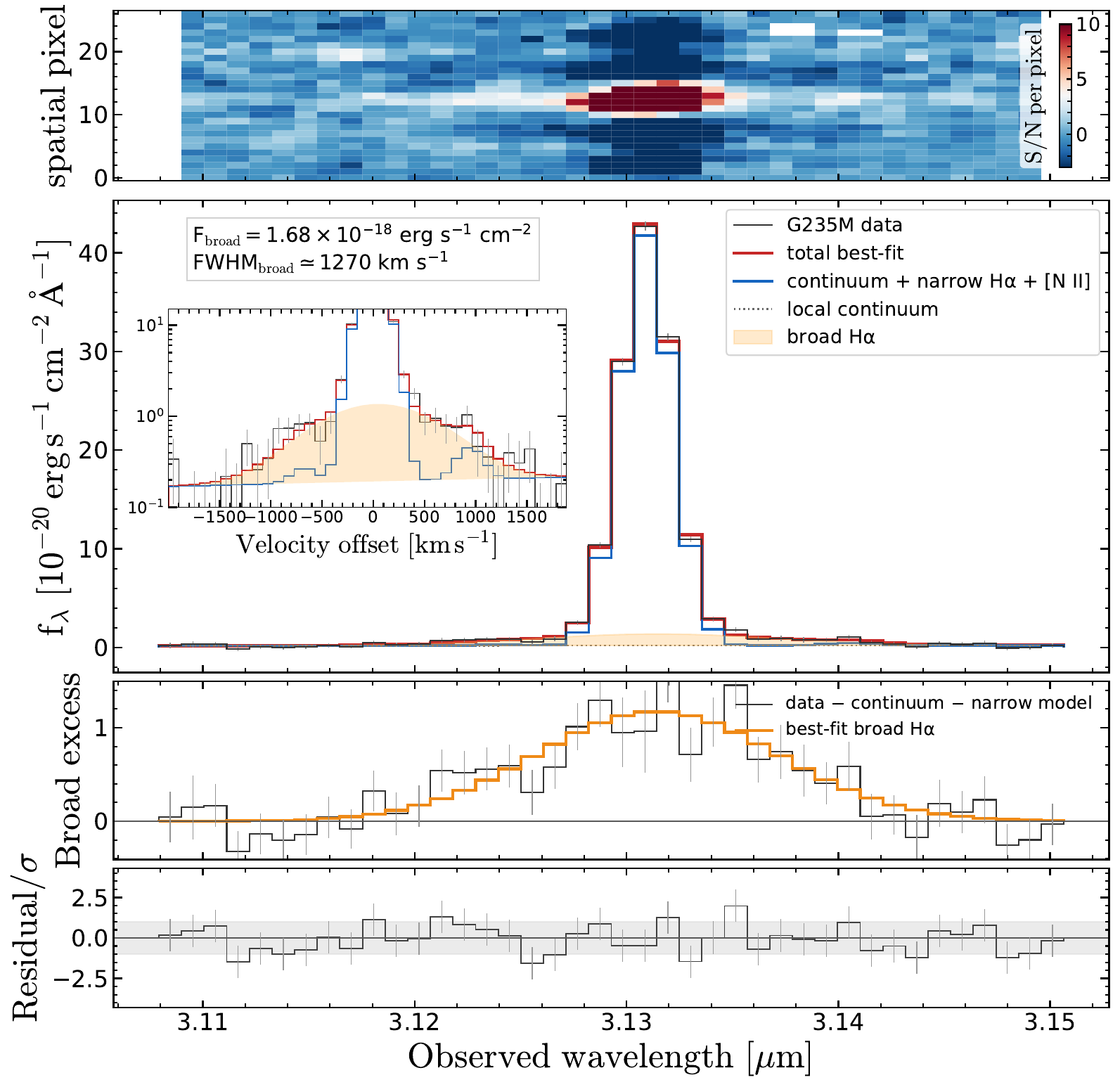}
\caption{Fiducial fit to the \ha+\nii\ complex in the NIRSpec/G235M spectrum.  Narrow permitted and forbidden components share a common kinematic model, while broad \ha\ is fitted independently.  All line components are convolved with the instrumental LSF.
The top panel shows the 2-d spectrum, where each pixel is colour coded by signal-to-noise ratio.
The second panel shows the extracted 1-d spectrum, with the best-fit model and individual components highlighted; the inset panel adopts a log-scale on the $y$ axis to highlight the broad component.
The third panel shows the spectrum after subtracting only the continuum and narrow-line model, together with the inferred broad-\ha\ component; the bottom panel shows the residuals to the complete model.}
\label{fig:halpha-fit}
\end{figure}

The fit in Fig.~\ref{fig:halpha-fit} detects a broad \ha\ component with fiducial flux $F_{{\rm H}\alpha,{\rm b}}=1.69\times10^{-18}\ {\rm erg\ s^{-1}\ cm^{-2}}$, luminosity $L_{{\rm H}\alpha,{\rm b}}=2.35\times10^{41}$~erg~s$^{-1}$, and intrinsic ${\rm FWHM}_{{\rm H}\alpha,{\rm b}}\simeq1270$~km~s$^{-1}$.
Adopting the reference best fit as the fiducial central value, bootstrap refits give $F_{{\rm H}\alpha,{\rm b}}=1.69^{+0.26}_{-0.19}\times10^{-18}$~erg~s$^{-1}$~cm$^{-2}$ and ${\rm FWHM}_{{\rm H}\alpha,{\rm b}}=1270^{+260}_{-250}$~km~s$^{-1}$. The broad model is preferred to a narrow-only model by the Bayesian Information Criterion (BIC; lower values are preferred), with $\Delta{\rm BIC}=-46.8$.

The detection of the broad \ha\ component remains robust against variation in the local continuum and narrow-line assumptions,
is recovered in injection--recovery experiments, and is not produced at comparable significance in null trials.
The two-dimensional G235M spectrum shows the spatially tilted narrow-line trace expected for a resolved source, with the line centroid shifting slightly along the spatial direction; collapsing this trace cannot reproduce the wavelength extent of the broad component.
We additionally test whether blending among intermediate-width \ha\ and both \nii\ lines, all sharing the [O\,\textsc{iii}] non-core kinematics measured below, can account for the broad base, finding that it cannot replace the broad component.
This test is fully described in Appendix~\ref{app:ha-tests}.

\subsection{H$\beta$ and [O\,\textsc{iii}] constraints on the broad component}
\label{sec:oiii_hb}

We next fit only the \hb+[O\,\textsc{iii}] complex in G235M and G235H, without including the \ha\ region in the likelihood.
The two gratings are fitted simultaneously with \textsc{unite} \citep{Hviding2025}, a package for the joint forward modeling of spectra from multiple gratings which includes a proper treatment of the LSFs and a relative flux-calibration nuisance parameter; in our model, we specifically assume shared narrow-line kinematics among the lines, while also including an `intermediate-width' [O\,\textsc{iii}] component to capture any faint non-core emission.
We assume for each grating the LSF for a uniformly illuminated slit \citep{Jakobsen2022}, and allow a shared multiplicative correction factor to account for any deviation from the nominal resolving power;
the fit prefers $R_{\rm scale}=1.086\pm0.007$ for the G235M+G235H fit.

\begin{figure*}
\centering
\includegraphics[width=0.98\textwidth]{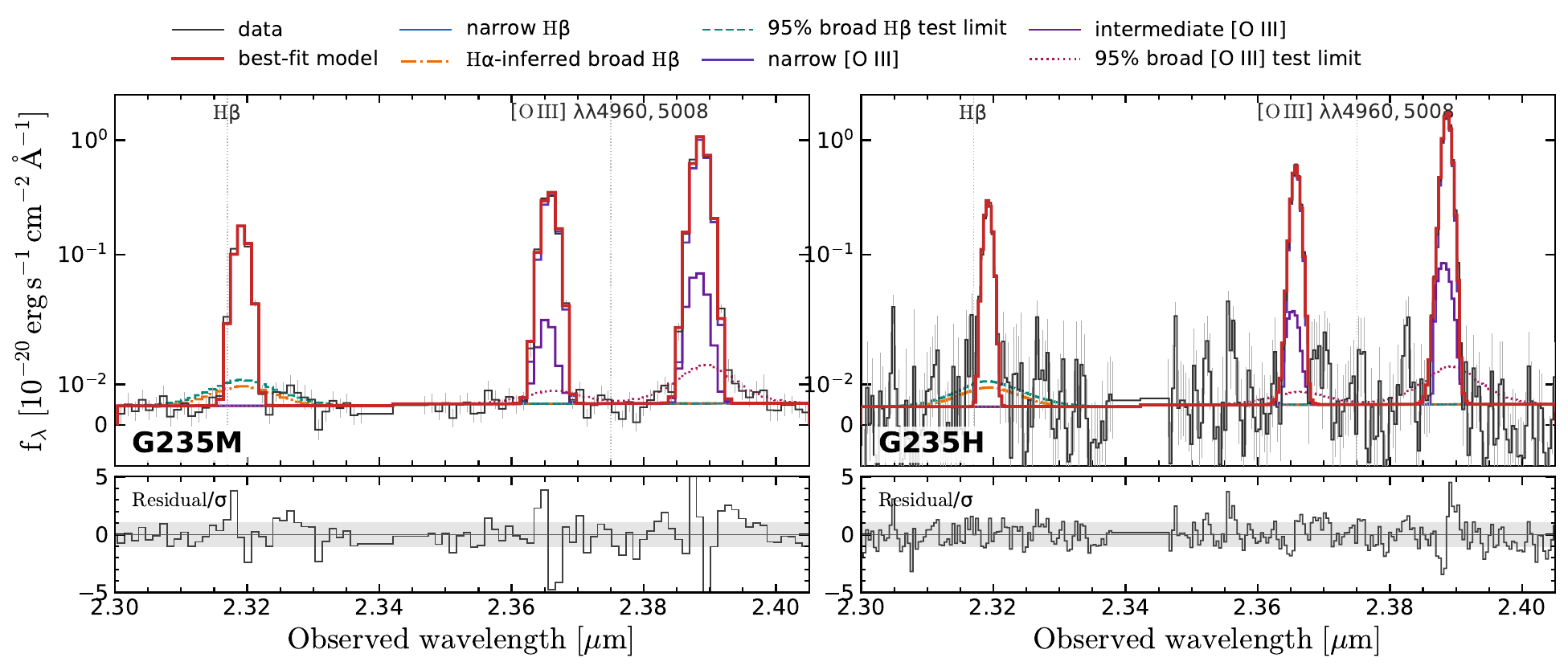}
\caption{Joint G235M (left) and G235H (right) fits to the full \hb+[O\,\textsc{iii}] complex.  Red curves show the best-fit narrow+intermediate model. Solid blue and purple curves show the narrow components and the intermediate [O\,\textsc{iii}] component included in the best-fit model.  The dash-dotted orange curve is the broad-\hb\ template normalized to the broad \ha\ flux for $\ha/\hb=3.1$; it falls below the dashed teal 95\% broad-\hb\ test limit.  The magenta dotted curve shows the 95th-percentile broad [O\,\textsc{iii}] test template.  The broad templates are not selected by BIC and are not included in the fiducial model; a symmetric-logarithmic flux scale is chosen here to highlight the inferred upper limits on their contribution.
Lower panels show residuals in units of the rescaled uncertainty.}
\label{fig:hbeta-oiii-fit}
\end{figure*}

Relative to a narrow-core-only model, including the intermediate [O\,\textsc{iii}] component improves the fit by $\Delta{\rm BIC}=-19.4$.
Such a component has ${\rm FWHM}=299^{+36}_{-32}$~km~s$^{-1}$ and $\Delta v=-61^{+21}_{-35}$~km~s$^{-1}$ relative to the systemic redshift, and contributes at both wavelengths of the [O\,\textsc{iii}] doublet (Fig.~\ref{fig:hbeta-oiii-fit}).
Its width, however, is clearly distinct from that of the $\simeq1270$~km~s$^{-1}$ component detected beneath \ha.
We treat it as a phenomenological non-core component: while its modest blueshift is compatible with a low-velocity ionized-gas outflow, we note that unresolved kinematic structure within the clumpy source or residual mismatch between the true LSF and the uniformly illuminated-slit prescription could also contribute to the overall line profile.

To further test for faint counterparts of the broad \ha, we performed another fit in which we introduce Gaussian templates for broad \hb\ and [O\,\textsc{iii}] whose integrated fluxes are allowed to be either positive or negative.\footnote{Allowing signed amplitudes prevents a weak component from being biased positive by an imposed non-negative boundary.}
Each broad template is assigned an H$\alpha$-informed truncated-Gaussian FWHM prior centred at 1270~km~s$^{-1}$, with a standard deviation of 250~km~s$^{-1}$ and limits of 700--2200~km~s$^{-1}$, together with a uniform $-200<\Delta v<+200$~km~s$^{-1}$ velocity prior.  When broad H$\beta$ and broad [O\,\textsc{iii}] are included simultaneously, they share the same velocity and width.
The corresponding 95th-percentile templates are overplotted across the full \hb+[O\,\textsc{iii}] complex in Fig.~\ref{fig:hbeta-oiii-fit}.  Adding these components gives $\Delta{\rm BIC}=+11.4$, $+2.2$, and $+1.8$, respectively.
Their small positive median amplitudes therefore do not compensate for the additional model parameters, and none is included in the fiducial \hb+[O\,\textsc{iii}] model.
The model-conditional 95\% limits are
$F_{{\rm H}\beta,{\rm b}} < 7.3\times10^{-19}\ {\rm erg\ s^{-1}\ cm^{-2}}$ and
$F_{[\mathrm{O\,III}]5007,{\rm b}} < 9.4\times10^{-19}\ {\rm erg\ s^{-1}\ cm^{-2}}$,
with $F_{[\mathrm{O\,III}]5007,{\rm b}}/F_{[\mathrm{O\,III}]5007,{\rm n}}<0.036$.
For an intrinsic broad-line decrement $\ha/\hb=3.1$, the fiducial broad \ha\ flux predicts instead $F_{{\rm H}\beta,{\rm b}}=5.4\times10^{-19}\ {\rm erg\ s^{-1}\ cm^{-2}}$, below the 95\% limit: thus, the data neither require nor exclude the presence of a broad \hb counterpart.

Because the broad-line amplitudes remain weak, their posteriors depend on the adopted kinematic constraints.  After marginalizing over the H$\alpha$-informed width prior and the uniform $-200<\Delta v<+200$~km~s$^{-1}$ prior adopted above, the broad-H$\beta$ limit provides no positive lower bound on $A_{V,{\rm broad}}$.  Fitting runs in which the broad-line kinematics are fixed or more tightly constrained yield formally stronger though explicitly prior-conditional limits.
We therefore treat the broad H$\beta$ only as an upper limit and apply no extinction correction to the broad H$\alpha$ luminosity.

\subsection{Analysis of the narrow-line spectrum}
\label{ssec:narrow_line_spec}

\begin{figure*}
\centering
\includegraphics[width=0.98\textwidth]{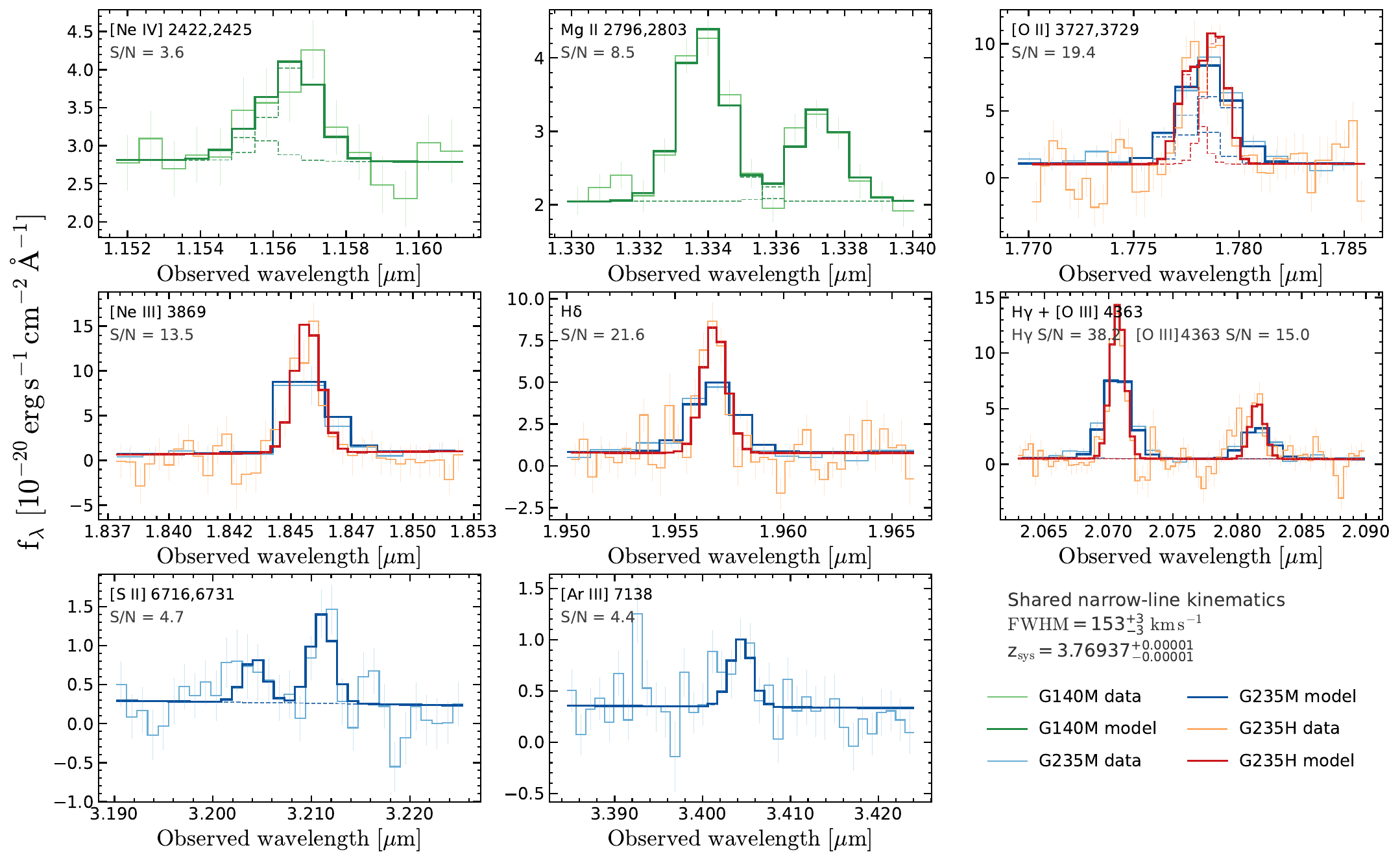}
\caption{Narrow-line fitting across the three available grating spectra.
The panels show the principal rest-UV and optical line complexes in G140M (green), G235M (blue), and G235H (red).
Solid coloured curves are the best-fit LSF-convolved models, and dashed curves isolate the line components for marginally resolved doublets.
}
\label{fig:narrow-inventory}
\end{figure*}

We fit\footnote{Wavelengths adopted in the analysis are vacuum wavelengths; a few conventional shorthand labels retained throughout the paper (for example, [O\,\textsc{iii}] $\lambda5007$) assume air wavelengths but refer to the same transitions.} the full narrow-line spectrum with a separate, LSF-convolved fit that combines G140M, G235M, and G235H.
Non-resonant lines share a common intrinsic narrow-line velocity and width, while the Mg~II$\lambda\lambda2796,2803$ doublet has independent kinematics because resonant scatter can shift and broaden its emergent profile.
The [O\,\textsc{iii}] doublet retains the intermediate component selected in the dedicated fitting described above, while neither \hb\ nor [O\,\textsc{iii}] contain any broad component in this fit.
Each line complex has its underlying continuum modeled in simple linear form, and we include both a multiplicative factor to account for relative flux-calibration offsets between different gratings as well as a correction factor to model variations from the assumed LSF for a uniformly illuminated slit.
Figure~\ref{fig:narrow-inventory} shows the result of the fit.
The shared narrow component has ${\rm FWHM}=153\pm3$~km~s$^{-1}$ and defines the systemic redshift adopted throughout this work, $z_{\rm sys}=3.76937\pm0.00001$ (16th--84th percentile interval).  The formal $\simeq3$~km~s$^{-1}$ difference between its centroid and the initial redshift used to initialize the fit is negligible at the NIRSpec resolution.
The dedicated G235M fit also gives a formal $4.4\sigma$ detection of [Ar\,\textsc{iii}] $\lambda7138$.  Because this wavelength lies close to the onset of the second-order-contaminated region, we report its flux but do not use it in the physical-condition or photoionization fits.
The complete set of measured fluxes, posterior uncertainties, and adopted upper limits is reported in Table~\ref{tab:line-fluxes}.

\subsubsection{Density, temperature, and nebular attenuation}
\label{sec:dust_density}

We infer the electron temperature, low-ionization electron density, and nebular attenuation jointly rather than in a sequence of independent calculations.  The likelihood combines the posterior flux vectors and covariances of the four Balmer lines, [O\,\textsc{iii}] $\lambda4363$ and $\lambda5008$, the G235H [O\,\textsc{ii}] $\lambda\lambda3727,3729$ doublet, and [S\,\textsc{ii}] $\lambda\lambda6716,6731$.  We compare these flux vectors directly with the corresponding PyNeb emissivities \citep{Luridiana2015}, profiling over an independent non-negative normalization for each line family.  This avoids constructing Gaussian approximations to ratios involving marginally detected lines.  For O$^+$, we adopt transition probabilities from \citet{Zeippen1982,Wiese1996} and effective collision strengths from \citet{Kisielius2009}; for O$^{++}$, transition probabilities from \citet{FroeseFischer2004,StoreyZeippen2000} and collision strengths from \citet{Storey2014}; and for S$^+$, transition probabilities from \citet{Rynkun2019} and collision strengths from \citet{TayalZatsarinny2010}.  The H\,\textsc{i} recombination emissivities are from \citet{StoreyHummer1995}.

The Balmer emissivities are evaluated at the sampled temperature and density and attenuated with the \citet{Cardelli1989} curve with $R_V=3.1$.  The [O\,\textsc{iii}] $\lambda4363/\lambda5008$ ratio constrains $T_3\equiv T_{\rm e}({\rm O\,III})$, while [O\,\textsc{ii}] and [S\,\textsc{ii}] are assigned a common representative density for the low-ionization gas.  Because [O\,\textsc{ii}] $\lambda\lambda7320,7330$ is not detected, $T_2\equiv T_{\rm e}({\rm O\,II})$ is linked to $T_3$ through the empirical relation of \citet{Cataldi2025}, including its intrinsic uncertainty.  The ionic abundances are sampled simultaneously and are discussed in Sect.~\ref{sec:metallicity}.

The G235H spectrum gives $F_{3729}/F_{3727}=1.19^{+0.33}_{-0.23}$.  The [S\,\textsc{ii}] ratio, $F_{6716}/F_{6731}=0.51^{+0.26}_{-0.25}$, nominally favours a higher density, although the weaker $\lambda6716$ component is detected at only ${\rm S/N}=2.1$.  Allowing the two ions to have independent densities improves the fit by $\Delta\chi^2=4.2$ relative to the common-density model, corresponding to only a $\simeq2\sigma$ tension.  We therefore retain the common-density description while accounting for the full, non-Gaussian density posterior.

The joint inference gives
\[
T_3=18{,}140^{+770}_{-710}\ {\rm K},\qquad
n_{\rm e}=310^{+600}_{-300}\ {\rm cm^{-3}},
\]
and $E(B-V)_{\rm neb}=0.054\pm0.013$.  The density posterior has a long tail towards the low-density limit and a one-sided 95\% upper bound of $n_{\rm e}<1.5\times10^3$~cm$^{-3}$.  The attenuation constraint is informed jointly by all four Balmer lines: H$\alpha$ comes from the dedicated local fit, whereas H$\gamma$/H$\beta$ and H$\delta$/H$\beta$ provide substantial leverage towards the fitted positive value.  The inferred attenuation remains small under the adopted foreground-screen model and temperature-dependent Case-B emissivities.  Alternative optical attenuation curves change the relevant corrected ratios by amounts well below their measurement uncertainties.

\begin{figure}
\centering
\includegraphics[width=\linewidth]{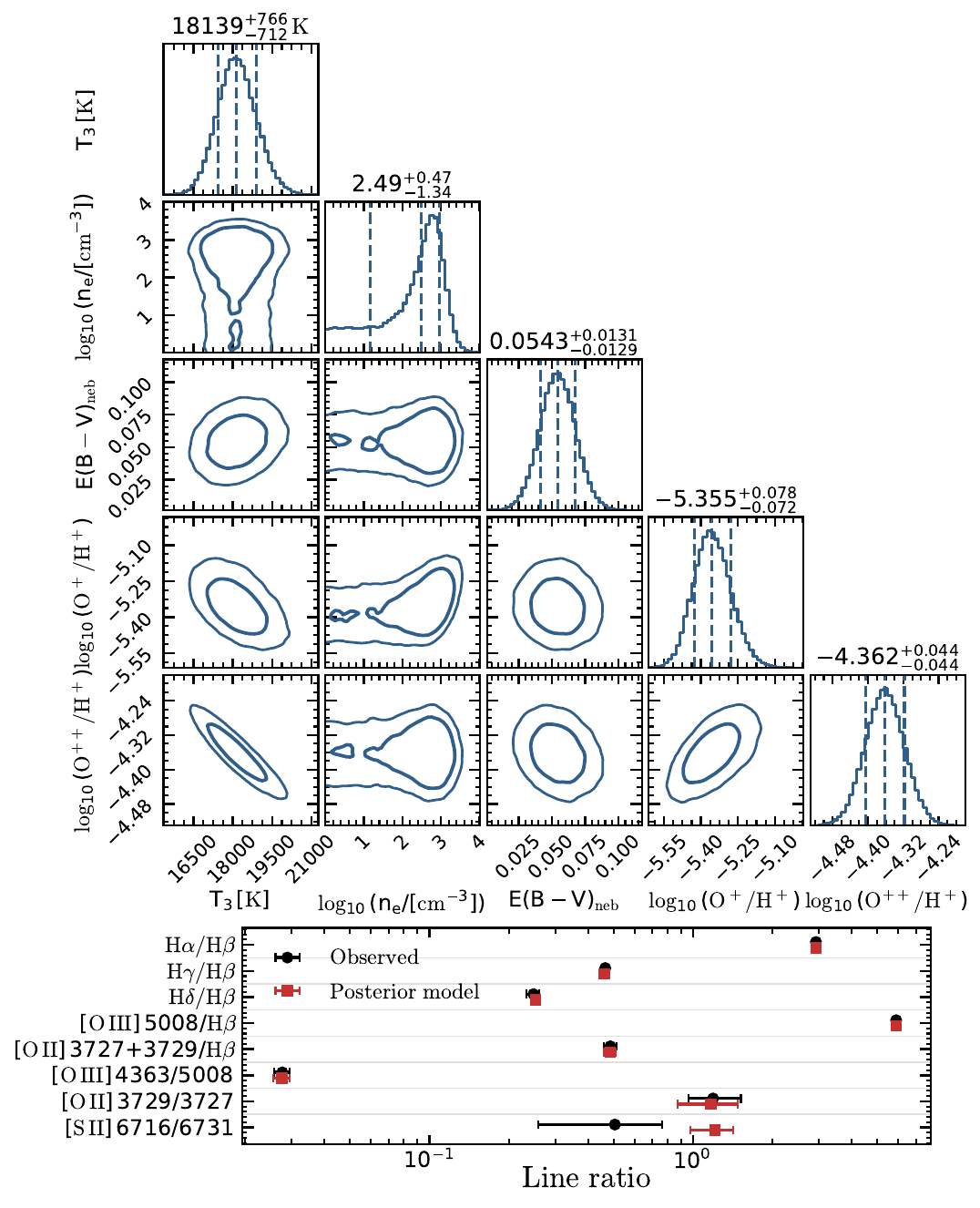}
\caption{Joint constraints on the physical conditions and ionic abundances of the narrow-line gas.  The upper corner plot shows the marginalized posterior distributions and covariances of $T_3$, $n_{\rm e}$, $E(B-V)_{\rm neb}$, O$^+$/H$^+$, and O$^{++}$/H$^+$.  Contours enclose 68\% and 95\% credible regions.  The lower panel compares the measured line ratios with those predicted by the posterior model.  The [O\,\textsc{ii}] density ratio is measured from G235H, whereas the remaining fluxes come from the joint narrow-line fitting of all available gratings (Sect.~\ref{ssec:narrow_line_spec}).  The mild discrepancy for [S\,\textsc{ii}] reflects the $\simeq2\sigma$ tension between the [O\,\textsc{ii}] and [S\,\textsc{ii}] density constraints and the low-S/N ($2.1$) detection of [S\,\textsc{ii}]$\lambda6716$.}
\label{fig:joint-physical-conditions}
\end{figure}

Figure~\ref{fig:joint-physical-conditions} shows the marginalized posterior distributions and covariances, together with the line ratios predicted by the posterior model.  All principal ratios are reproduced within their uncertainties, apart from the mild tension associated with the marginal [S\,\textsc{ii}] $\lambda6716$ measurement.

\subsubsection{Constraints on high-ionization and high-density gas}
\label{sssec:density}

The [Ne\,\textsc{iv}] $\lambda\lambda2422,2425$ doublet probes a more highly ionized phase than [O\,\textsc{ii}] and [S\,\textsc{ii}] and is therefore not included in the common-density inference above.  In the joint spectral fit, $\lambda2425$ is detected at $3.65\sigma$, whereas $\lambda2422$ is not individually detected ($0.87\sigma$; $P[F_{2422}>0]=0.81$).  The resulting posterior gives a one-sided 95\% upper limit $F_{2422}/F_{2425}<0.92$.

We constrain the density from the signed two-line flux vector rather than from a ratio between a detection and an upper limit.  At each trial density, the PyNeb doublet ratio is fixed and a non-negative common normalization is fitted to both transitions, while positive and negative spectral fluctuations are retained in the likelihood.  Marginalizing over the $T_3$ posterior gives $n_{\rm e}({\rm Ne^{3+}})<9.5\times10^3$~cm$^{-3}$ at 95\% confidence; the limit varies by less than 2\% across the allowed temperature range.

\begin{figure}
\centering
\includegraphics[width=0.9\columnwidth]{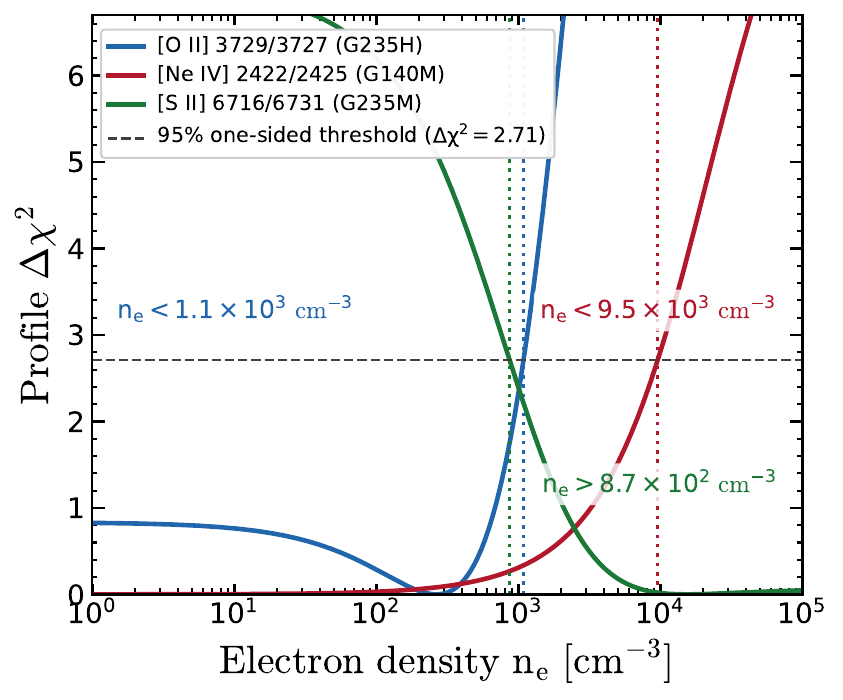}
\caption{Electron-density constraints from gas at different ionization stages.  The curves show the profile likelihoods obtained from the signed flux vectors and posterior covariances of [O\,\textsc{ii}] $\lambda\lambda3727,3729$ in G235H (blue), [S\,\textsc{ii}] $\lambda\lambda6716,6731$ in G235M (green), and [Ne\,\textsc{iv}] $\lambda\lambda2422,2425$ in G140M (red), profiling independently over their normalizations.  Central curves use the median $T_2$ or $T_3$, with shaded bands propagating the corresponding 16th--84th percentile temperature range.  The horizontal dashed line marks the one-sided 95\% threshold ($\Delta\chi^2=2.71$); the bounds are $n_{\rm e}({\rm O^+})<1.1\times10^3$, $n_{\rm e}({\rm S^+})>8.7\times10^2$, and $n_{\rm e}({\rm Ne^{3+}})<9.5\times10^3$~cm$^{-3}$.  The [S\,\textsc{ii}] constraint is tentative because $\lambda6716$ is not individually detected above $3\sigma$.
}
\label{fig:neiv-density}
\end{figure}

Figure~\ref{fig:neiv-density} compares this constraint directly with the individual low-ionization diagnostics.  For each ion, the central curve is evaluated at the median of the relevant temperature posterior, while the shaded band spans its 16th--84th percentiles: $T_2$ for [O\,\textsc{ii}] and [S\,\textsc{ii}], and $T_3$ for [Ne\,\textsc{iv}].  The [Ne\,\textsc{iv}] constraint disfavors a very dense $n_{\rm e}\gtrsim10^4$~cm$^{-3}$ high-ionization component but remains compatible with a moderate-density narrow-line region.  Conversely, the low-ionization density inferred above is typical of high-redshift H\,\textsc{ii} regions \citep[e.g.][]{Sanders2016,Isobe2023Density,Topping2025Density}, but also overlaps the lower-density range of AGN narrow-line regions \citep{Zhang2013}.  The density measurements alone therefore do not discriminate between stellar and AGN excitation.

\subsubsection{Diagnostic diagrams and ionization source}
\label{sssec:diagnostic_diagrams}

We here investigate the position of MARTA~3887 onto classical emission-line-ratio diagnostic diagrams which are commonly adopted to discriminate the dominant source of the ionizing spectrum.
Specifically, we define $\mathrm{N2}=[\mathrm{N\,II}]\lambda6585/\ha$, $\mathrm{S2}=([\mathrm{S\,II}]\lambda6716+[\mathrm{S\,II}]\lambda6731)/\ha$, and $\mathrm{O1}=[\mathrm{O\,I}]\lambda6300/\ha$ for the three classical BPT/VO87 panels \citep{Baldwin1981,VeilleuxOsterbrock1987}.  We further define $R_3=[\mathrm{O\,III}]\lambda5008/\hb$, $\mathrm{O32}=[\mathrm{O\,III}]\lambda5008/([\mathrm{O\,II}]\lambda\lambda3727,3729)$, $R_{23}=([\mathrm{O\,II}]\lambda\lambda3727,3729+[\mathrm{O\,III}]\lambda\lambda4960,5008)/\hb$, $\mathrm{Ne3O2}=[\mathrm{Ne\,III}]\lambda3870/[\mathrm{O\,II}]\lambda\lambda3727,3729$, $\mathrm{O3Hg}=[\mathrm{O\,III}]\lambda4363/\mathrm{H}\gamma$, and $\mathrm{Ne4Hb}=([\mathrm{Ne\,IV}]\lambda2422+[\mathrm{Ne\,IV}]\lambda2425)/\hb$.
[N\,\textsc{ii}] and [O\,\textsc{i}] $\lambda6300$ are undetected and are treated as $3\sigma$ upper limits, giving $\log([\mathrm{N\,II}]/\ha)<-1.76$ and $\log([\mathrm{O\,I}]/\ha)<-1.62$.

\begin{figure*}
\centering
\includegraphics[width=0.99\textwidth]{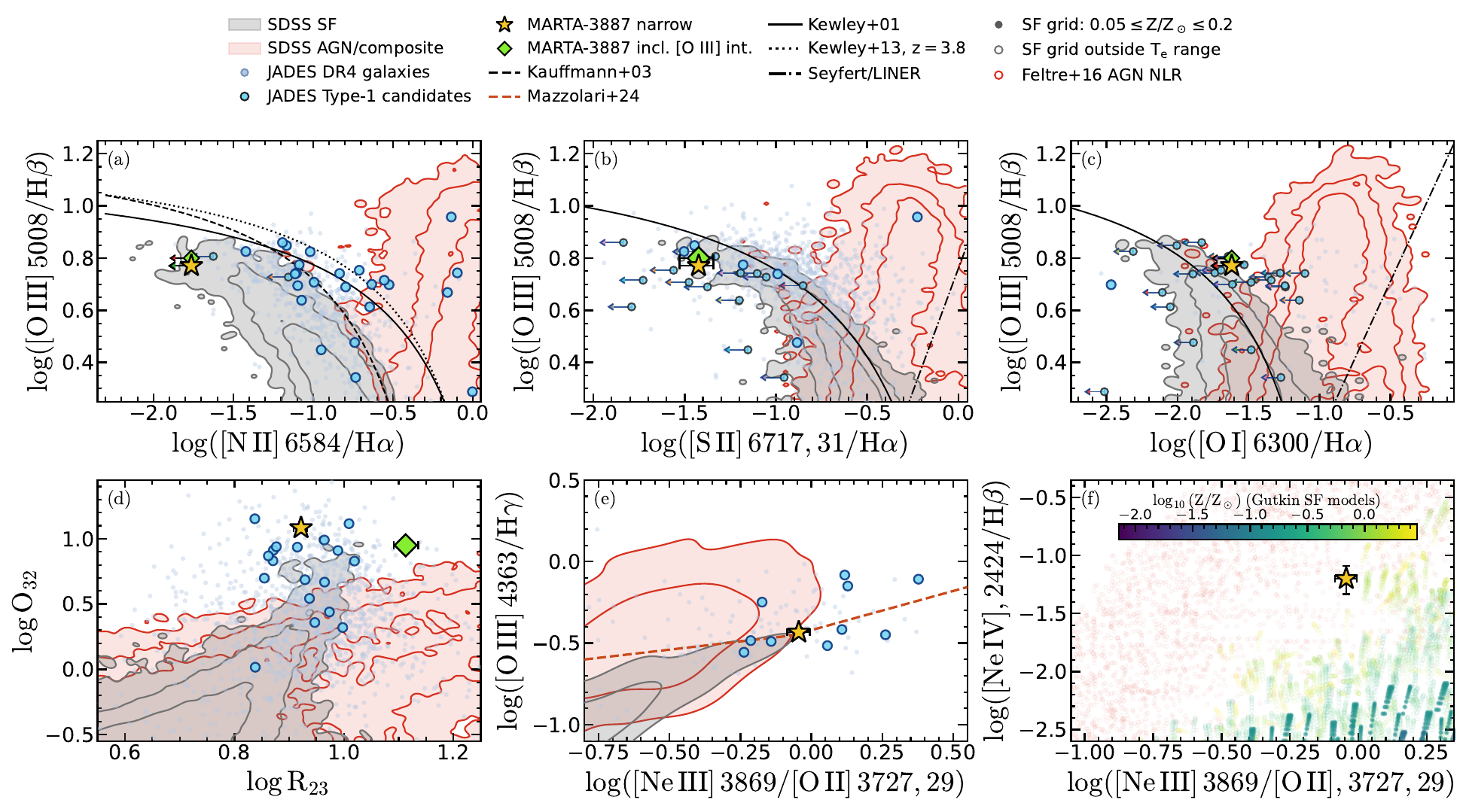}
\caption{Rest-optical/UV emission-line diagnostic diagrams.
Panels (a--c) show [N\,\textsc{ii}]/H$\alpha$, [S\,\textsc{ii}]/H$\alpha$, and [O\,\textsc{i}]/H$\alpha$ against [O\,\textsc{iii}]$\lambda5008$/H$\beta$; panel (d) shows $R_{23}$ versus O32; panel (e) shows the \citet{Mazzolari2024} auroral-line-based diagram using Ne3O2 against [O\,\textsc{iii}]$\lambda4363$/H$\gamma$; and panel (f) shows [Ne\,\textsc{iv}]$\lambda\lambda2422,2425$/H$\beta$ versus Ne3O2.
Grey and red contours represent, respectively, local SDSS galaxies below and above the \citet{Kauffmann2003} [N\,\textsc{ii}]-BPT demarcation; the latter category includes both composite systems and AGN.  The same [N\,\textsc{ii}]-BPT classification is applied across panels (a-e).
Small pale-blue points show the full JADES DR4 line-emitter sample, while cyan circles with dark-blue edges are the JADES Type~1 candidates of \citet{Juodzbalis2026}.
The gold star marks MARTA-3887 using the narrow components; its [N\,\textsc{ii}] and [O\,\textsc{i}] coordinates are $3\sigma$ upper limits.
Green diamonds in panels (a--d) additionally include the intermediate [O\,\textsc{iii}] component (see Sect.~\ref{sec:oiii_hb}).
In panel (f), the \citet{Gutkin2016} star-forming grid is colour-coded by metallicity; filled markers identify models with $0.05\leq Z/Z_\odot\leq0.2$, while open markers show the remaining grid. Open red points show the \citet{Feltre2016} AGN NLR models. Curves show the diagnostic boundaries as listed in the legend \citep{Kauffmann2003,Kewley2001,Kewley2006,Kewley2013,Mazzolari2024}.}
\label{fig:optical-diagnostics}
\end{figure*}

Figure~\ref{fig:optical-diagnostics} compares these ratios with local SDSS DR7 galaxies \citep{Abazajian2009}, using emission-line measurements from the MPA--JHU catalogue \citep{Brinchmann2004,Tremonti2004}.  They are separated a priori into star-forming and AGN/composite populations using their position relative to the \citet{Kauffmann2003} demarcation in the [N\,\textsc{ii}]-BPT diagram.  The same classification is retained in all other optical-line-based panels.  The JADES comparison sample is drawn from DR4 \citep{CurtisLake2026,Scholtz2026,Eisenstein2023}.
Standard diagnostic boundaries \citep{Kauffmann2003,Kewley2001,Kewley2006,Kewley2013} and the JADES DR4 emission-line sample are also shown, with Type~1 AGN identified in \citet{Juodzbalis2026} highlighted.
The measured $\log\mathrm{O32}=1.08$, $\log R_3=0.77$, $\log R_{23}=0.92$, and $\log\mathrm{Ne3O2}=-0.04$ are characteristic of a high-ionization, low-metallicity galaxy, placing MARTA~3887 at the extreme end of the star-forming sequence in the classical BPT and $R_{23}$--O32 diagrams \citep[e.g.][]{Strom2017}.
Its narrow-line ratios are therefore compatible with metal-poor, hard stellar ionizing spectra, and do not independently require an AGN contribution.
In the diagnostic diagram proposed by \citet{Mazzolari2024}, which leverages the detection of the auroral line [O\,\textsc{iii}] $\lambda4363$, ID~3887 lies close to, or nominally just above, the proposed demarcation between AGN and galaxies hosting either star formation, an AGN, or both (panel e of Fig.~\ref{fig:optical-diagnostics}), in line with several other similar JADES type~1 AGN candidates.
Given the measurement uncertainty, proximity to the boundaries, and inferred low metallicity, we treat this primarily as further evidence for a hard stellar ionizing spectrum, while acknowledging the potential for a non-negligible AGN contribution.

The [Ne\,\textsc{iv}] $\lambda\lambda2422,2425$ detection ($3.6\sigma$) provides an additional probe of the ionizing spectrum.  Although [Ne\,\textsc{iv}] is often associated with AGN photoionization at high redshift \citep[e.g.][]{NakajimaMaiolino2022,Cleri2023,Maiolino2024}, hard stellar spectra and stellar end products can also enhance high-ionization lines in metal-poor galaxies \citep{ShiraziBrinchmann2012,Senchyna2017,Schaerer2019}.
From the summed doublet flux we obtain $\log([\mathrm{Ne\,IV}]/\hb)=-1.20\pm0.12$ and $\log([\mathrm{Ne\,IV}]/[\mathrm{Ne\,III}])=-0.84\pm0.12$, while a $3\sigma$ limit on [Ne\,\textsc{v}]$\lambda3426$ gives $\log([\mathrm{Ne\,V}]/[\mathrm{Ne\,IV}])<-0.25$.
The [Ne\,\textsc{iv}]--Ne3O2 panel of Fig.~\ref{fig:optical-diagnostics} compares the detected ratio with the \citet{Gutkin2016} star-forming and \citet{Feltre2016} AGN narrow-line grids.
MARTA~3887 can be reached by the star-forming grid, but mainly at a metallicity of $Z\simeq0.017$--0.020 (i.e. approximately solar), substantially above the conditional `direct' $T_{\rm e}$-based abundance (Sect.~\ref{sec:metallicity}).
Such a mismatch may descend from limitations of the single-zone photoionization models or effectively mark the requirement for an additional hard ionizing component.  It should not, however, be read as a direct inconsistency: an AGN contribution would also invalidate some of the assumptions used to convert the measured temperatures and ionic abundances into a total O/H in classical H\,\textsc{ii} regions.
The low significance of the [Ne\,\textsc{iv}] detection and the absence of [Ne\,\textsc{v}] emission though
suggest only a mild AGN contribution to the narrow-line spectrum, while a proper quantitative decomposition remains difficult: we test this with dedicated photoionization modeling at the end of Sect.~\ref{sec:metallicity}).

\subsubsection{Gas-phase metallicity\label{sec:metallicity}}

The joint inference described in Sect.~\ref{sec:dust_density} also samples the ionic abundances together with $T_3$, $T_2$, $n_{\rm e}$, and $E(B-V)_{\rm neb}$.  At each posterior step, O$^+$/H$^+$ is derived from the [O\,\textsc{ii}] $\lambda\lambda3727,3729$/H$\beta$ ratio using the sampled $T_2$ and low-ionization density.  O$^{++}$/H$^+$ is derived from [O\,\textsc{iii}] $\lambda5008$/H$\beta$ using the sampled $T_3$ and $n_{\rm e}=10^3$~cm$^{-3}$ for the O$^{++}$ zone.  The latter is a representative value consistent with the [Ne\,\textsc{iv}] limit; the [O\,\textsc{iii}] emissivities are insensitive to density over the allowed range.

The resulting low-ionization temperature is $T_2=14{,}610^{+750}_{-740}$~K, where the uncertainty includes the scatter of the adopted $T_2$--$T_3$ relation.  Summing the ionic abundances, $\mathrm{O/H}=\mathrm{O^+/H^+}+\mathrm{O^{++}/H^+}$, gives
\[
12+\log({\rm O/H})=7.68\pm0.04.
\]
The posterior covariance among temperature, density, attenuation, and line fluxes is included directly in this uncertainty.  Approximately 91\% of the inferred oxygen abundance is in O$^{++}$, so the result is only weakly sensitive to the adopted low-ionization density and $T_2$--$T_3$ relation.

The measurement of $T_3$ from [O\,\textsc{iii}] $\lambda4363/\lambda5008$, and the corresponding O$^{++}$/H$^+$ ionic abundance, do not require an assumption about the shape of the stellar ionizing spectrum.  Interpreting the ionic sum as the total oxygen abundance is instead conditional on stellar-dominated narrow-line excitation, the adopted two-zone temperature structure, and a negligible contribution from higher oxygen ionization stages.  The integrated spectrum may also combine regions with different temperatures, densities, and ionizing sources, in which case the inferred quantities are emissivity-weighted effective values.  We assess these systematics below.

\paragraph{Sensitivity to a possible AGN contribution}

Because the detection of broad \ha\ emission (Sect.~\ref{sec:blrg}) suggests the presence of an accreting black hole in the system (Sect.~\ref{sec:agn-candidate}), we investigate whether AGN-powered emission could alter the gas-phase metallicity inference described above.
In fact, if an AGN contributes significantly to the narrow-line ionization, its harder continuum can alter the temperature stratification relative to what is assumed for classical H\,\textsc{ii} regions and star-forming galaxies; while photoionization equilibrium may still hold, converting the measured temperatures and ionic abundances into a `total' O/H would hence require stellar-plus-AGN models that jointly constrain the incident SED, ionization parameter, density, dust, and abundance patterns \citep[e.g.][]{OsterbrockFerland2006,Feltre2016}.
Furthermore, this would place a non-negligible fraction of oxygen in $\mathrm{O^{3+}}$ (in which case an ionization-correction factor would be required and the ionic sum of $\mathrm{O^{+}}$ and $\mathrm{O^{++}}$ would not represent the total oxygen abundance).
However, we detect no significant Si\,\textsc{iv}+O\,\textsc{iv}] $\lambda1400$ emission in the (shallow) MUSE spectrum, with a $3\sigma$ flux limit of $3.9\times10^{-18}$~erg~s$^{-1}$~cm$^{-2}$ for the blended complex, nor secure narrow He\,\textsc{ii} $\lambda4686$ emission in the NIRSpec spectrum.
We therefore find no direct evidence that $\mathrm{O^{3+}}$ contributes substantially and apply no additional ionization correction, while noting that such non-detections do not provide a stringent quantitative limit on $\mathrm{O^{3+}/O}$.

To test the sensitivity to the ionizing spectrum directly, we fitted the MARTA~3887 narrow-line fluxes with the multi-cloud photoionization framework \textsc{homerun} \citep{Marconi2024,Moreschini2026}, adopting either purely stellar or composite stellar+AGN continua.  The model construction and the line-by-line decomposition of the composite solution are presented in Appendix~\ref{app:homerun}.
Both model families reproduce the observed line ratios within the adopted model tolerance; their objective-function values provide a relative ranking within each model family, rather than a formal $\chi^2$ preference between them.
We note however that the stellar-only fit reproduces [Ne\,\textsc{iv}] only through a substantial contribution from gas with $\log U>-1$, or through matter-bounded clouds as proposed by \citet{Moreschini2026}; in the composite fit, the AGN-ionized component instead produces nearly all of the [Ne\,\textsc{iv}] emission.  The attenuation remains negligible in both model families.

For the composite models, we impose a prior on $f_{\rm AGN}({\rm H}\beta)\leq10\%$, motivated by the ionizing-photon budget inferred from broad H$\alpha$ under the preferred BLR interpretation (Sect.~\ref{sec:agn-candidate}).
Given the standard \textsc{Cloudy} \citep{Ferland2017} AGN continuum adopted by \textsc{homerun},\footnote{The adopted SED has $T_{\rm BB}=10^5$~K, $\alpha_{\rm uv}=-0.5$, $\alpha_x=-1$, and $\alpha_{\rm ox}=-1.8$.  It gives $L(h\nu>13.6\,\mathrm{eV})/L_{\rm bol}\simeq0.12$, a mean ionizing-photon energy of $\simeq1.7$~Ryd, and $Q_{\rm AGN}\simeq3.2\times10^9(L_{\rm bol}/{\rm erg\,s^{-1}})$~s$^{-1}$.} the broad-H$\alpha$ luminosity corresponds to $Q_{\rm AGN}\simeq(1$--$2)\times10^{53}$~s$^{-1}$.  The AGN can therefore power a fraction $f_{\rm AGN}({\rm H}\beta)\simeq(0.04$--$0.09)C_{\rm NLR}$ of the narrow Balmer emission, where $C_{\rm NLR}\leq1$ is the narrow-line-region covering factor.  A hotter accretion disc appropriate for a $\sim10^{6.5}\,M_\odot$ black hole can raise this estimate by up to a factor of $\sim2.5$; the 10\% ceiling should therefore be regarded as a fiducial modeling prior rather than a strict physical upper limit.  The best-fitting composite solution has $f_{\rm AGN}({\rm H}\beta)=9\%$ and reproduces all lines within $1\sigma$, except [S\,\textsc{ii}], which is reproduced within $1.5\sigma$.

The predicted, model-based oxygen abundance is $12+\log({\rm O/H})$ =7.79 for the stellar-only fit and 7.81 for the composite fit, a difference of only $0.02$~dex.
In the latter solution, the AGN component contributes $\simeq8\%$ and $\simeq13\%$ of [O\,\textsc{iii}] $\lambda5007$ and $\lambda4363$, respectively: removing these contributions changes $T_{\rm e}({\rm O\,III})$ and O$^{++}$/H$^+$ for the stellar-photoionized gas by approximately $-500$~K and less than 0.03~dex.
We note that the $\sim0.13$~dex offset between the composite \textsc{homerun} oxygen abundance and our fiducial, Te-based estimate persists even in the stellar-only scenario, showing that it is not driven by the inclusion of an AGN continuum but mostly reflects instead the sensitivity of the classical two-zone approach to unresolved density and ionization structures \citep{Moreschini2026}.

As a final, illustrative sensitivity test, replacing the H\,\textsc{ii}-region $T_2$--$T_3$ calibration with the NLR relation of \citet{Dors2020}\footnote{We note such a relation was calibrated for Type~2 AGN narrow-line regions and is mildly extrapolated here towards lower metallicity.} gives $T_{\rm e}({\rm O\,II})\simeq9200$~K and $12+\log({\rm O/H})\simeq7.83$, close to the \textsc{homerun} result.
We hence retain $12+\log({\rm O/H})=7.68\pm0.04$ as the fiducial, conditional H\,\textsc{ii}-region-like oxygen abundance estimate, and allow for an upward systematic of approximately 0.15~dex associated with the unresolved ionization structure; such an uncertainty, however, does not alter the classification of MARTA~3887 as a metal-poor galaxy, nor any of the conclusions of this paper.

\subsubsection{Mg~II emission}
\label{sec:mgii}

The Mg~II $\lambda\lambda2796,2803$ doublet is detected in emission with an intrinsic ${\rm FWHM}=195^{+62}_{-90}$~km~s$^{-1}$ and a centroid offset of $+35^{+13}_{-13}$~km~s$^{-1}$ from systemic.  No broad Mg~II component is required, and the present spectrum shows no evidence for blueshifted Mg~II absorption.
The measured ratio, $R_{\rm MgII}=F_{2796}/F_{2803}=1.90^{+0.27}_{-0.22}$, is close to the optically thin value of two.  In the idealized foreground-screen model, $R_{\rm MgII}=2\exp(-\tau_{2803})$ \citep{Chisholm2020}, and the best-fit ratio corresponds to $\tau_{2803}\simeq0.05$; we note however that the uncertainty extends beyond the physical optically thin limit, and resonant scattering can redirect photons both into and out of the line of sight, hampering a proper quantitative estimate of the optical depth.

The narrow profile rules out Mg~II as an independent BLR virial-mass estimator and instead associates it with gas processed by the host ISM or narrow-line region.  Emission without a detected blueshifted absorption trough does not select a unique origin: it can arise from intrinsically emitted low-optical-depth gas, resonant scattering, or an outflow viewed in an unfavourable geometry \citep[e.g.][]{Erb2012MgII,Martin2013MgII}.  Its doublet ratio is nevertheless consistent with low-optical-depth channels through which Ly$\alpha$ (Sect.~\ref{sec:muse}) and LyC photons may escape, as found e.g. in Green Pea galaxies \citep{Henry2018,Chisholm2020}.
However, in the presence of clumpy geometries the doublet ratio can remain close to two even when optically thick material covers part of the source, while dust and viewing angle alter the emergent fluxes; radiation-hydrodynamic calculations consequently find substantial scatter between Mg~II and LyC escape \citep{Katz2022}.

\section{Host imaging and stellar populations}
\label{sec:photometry}

\subsection{Morphology and photometry}
\label{ssec:morphology}
\begin{figure*}
\centering
\includegraphics[width=0.98\textwidth]{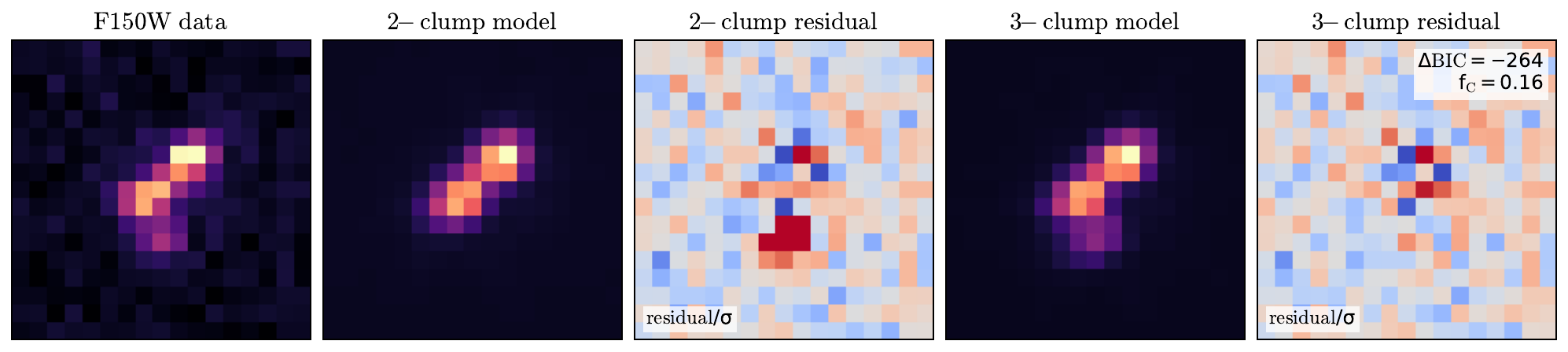}
\caption{PSF-convolved morphological fit on the native F150W image.  From left to right, the panels show the data, the two-component model and its residuals, and the three-component model and its residuals, expressed in units of the pixel uncertainty.  The third component accounts for 16\% of the fitted F150W light and removes the southern excess left by the two-component model.}
\label{fig:f150-morphology}
\end{figure*}

As shown already by Fig.~\ref{fig:rgb-slitlet} (as well as in the full cutouts of Fig.~\ref{fig:cutouts} in Appendix~\ref{app:morphology}), when observed in the rest-frame UV MARTA~3887 appears as constituted by two dominant knots separated by $0.161$--$0.167\arcsec$ ($1.17$--$1.22$ proper kpc) and a fainter southern component $0.115$--$0.125\arcsec$ ($0.84$--$0.91$ proper kpc) from the south-western knot.
Their projected separations establish a multicomponent morphology but do not distinguish star-forming clumps from dynamically distinct merging units.
PSF-convolved modelling favours three components in F115W--F200W, with the southern clump contributing 13--17\% of the light.  Figure~\ref{fig:f150-morphology} illustrates the result in F150W: adding the southern component removes the coherent residual below the two brighter knots and improves the fit by $\Delta\mathrm{BIC}=-264$.
The full multi-band modelling description and assumptions, together with the associated tests, are presented in Appendix~\ref{app:morphology}.
A forced decomposition at longer wavelengths suggests that the south-western component is redder than the north-eastern one, but becomes increasingly prior-dependent as the PSF broadens.
In the following, we therefore adopt only the total, integrated photometry of the system for the purposes of SED-based inference.

For the fiducial SED fitting we adopt a common $0.5\arcsec$ aperture and apply filter-dependent STPSF encircled-energy corrections \citep{Perrin2014}.
NIRCam fluxes are PSF-matched to the F460M reference PSF, whereas HST optical fluxes are retained in native images but measured in the same physical aperture.
The HST F435W and F475W measurements are retained as signed flux constraints on the Lyman break, whereas the F336W non-detection is treated as an upper limit and also constrains direct LyC emission (Sect.~\ref{sec:lyc-leakage}).  NIRCam F300M is excluded from the fiducial SED fit because the H$\alpha$+[N\,\textsc{ii}] complex supplies approximately 48\% of its in-band flux, allowing us to retain the measured H$\alpha$ as an independent check on the recent SFR\footnote{Although BAGPIPES includes nebular emission, fitting
this band would make the inferred recent SFH depend strongly on the adopted stellar ionizing spectrum and other assumptions, such as the presence of ionizing-photon escape and of an AGN contribution to the narrow lines.}; at the same time, we note that the broad H$\alpha$ component contributes only about 5\% of the total F300M flux.
A 5\% fractional systematic floor is added to the detected-band uncertainties.
None of the six available MIRI bands contains a significant detection.  We include each $3\sigma$ upper limit through the cumulative probability of the Gaussian likelihood \citep{Sawicki2012}; this censored treatment preserves the constraint against luminous rest-frame near-IR templates without assigning a positive flux to a non-detection.

\begin{figure}
\centering
\includegraphics[width=0.99\columnwidth]{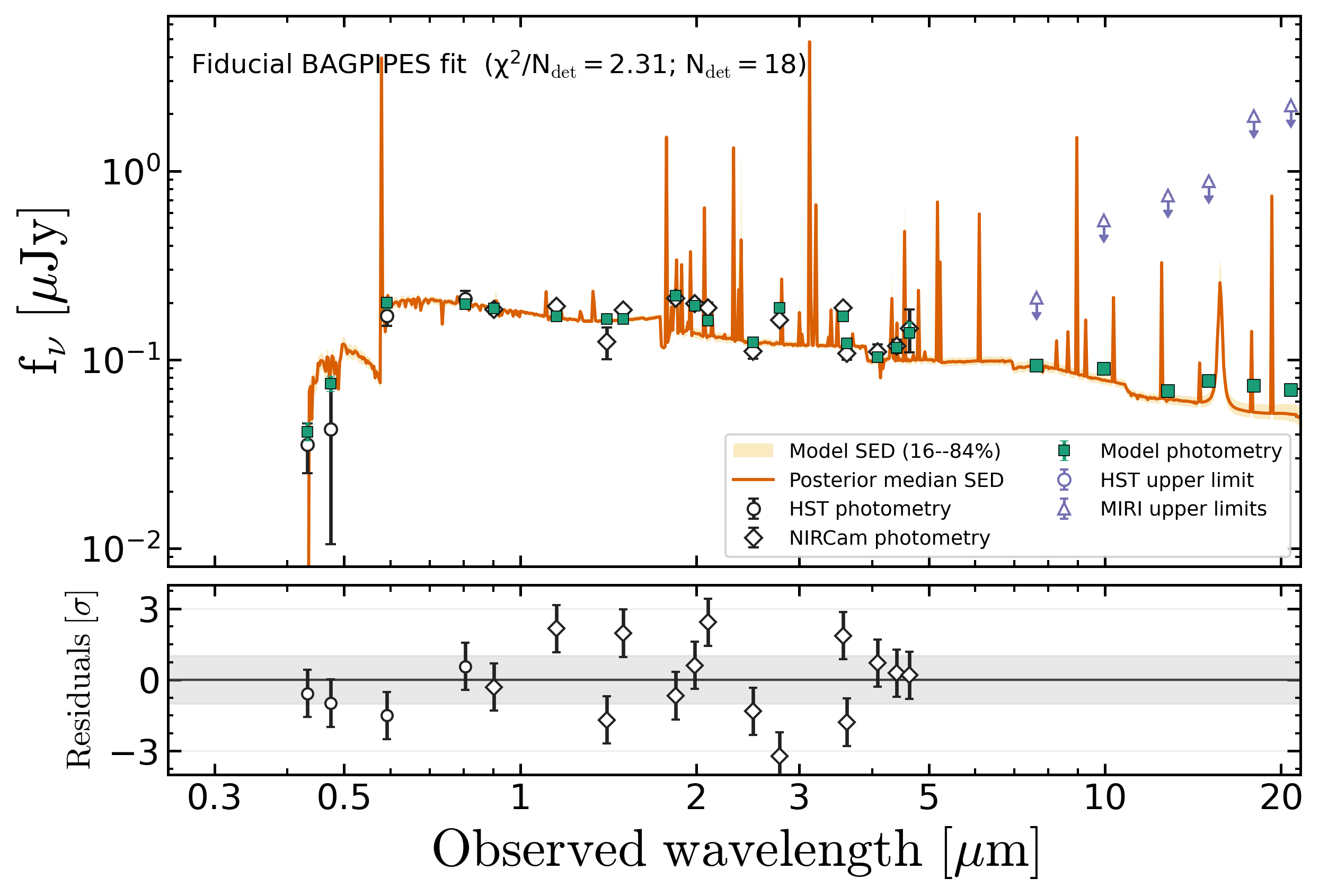}
\caption{Fiducial BAGPIPES SED fitting to the integrated photometry of the galaxy.
Observed HST, NIRCam, and MIRI measurements are distinguished by marker shape and include $1\sigma$ uncertainties.
The orange curve and shaded region show the median and 16th--84th percentile range of the posterior model SED, while green squares show the corresponding posterior synthetic band fluxes.
Purple arrows mark bands treated as upper limits.
The lower panel shows normalized residuals with $1\sigma$ error bars; the quoted $\chi^2/N_{\rm det}$ is computed for detections only and is not a formal reduced $\chi^2$.}
\label{fig:bagpipes-fit}
\end{figure}

\subsection{SED fitting and sensitivity to different SFH priors}
\label{sec:SED_fitting}

We fit the integrated photometry with BAGPIPES \citep{Carnall2018} and sample the posterior with \textsc{nautilus} \citep{Lange2023}; broad-band photometry is input without subtracting the
narrow emission line fluxes, such that their contributions are predicted
self-consistently by the BAGPIPES nebular module, and no spectroscopic line flux is supplied as a separate constraint (with the exception of \ha\ and F300M as described above).
The redshift is fixed from the spectroscopy, and we adopt the BPASS~v2.2.1 binary stellar-population models \citep{StanwayEldridge2018} with a \texttt{imf135\_300} broken-power-law IMF, i.e ${\rm d}N/{\rm d}M\propto M^{-1.30}$ from 0.1 to $0.5\,M_\odot$ and $\propto M^{-2.35}$ from 0.5 to $300\,M_\odot$.  We use a six-bin non-parametric SFH: the logarithmic ratios of adjacent-bin SFRs follow a Student-$t$ continuity prior with scale 0.3, and we repeat the fit with scales 0.6 and 1.0 to test sensitivity to the regularization.
The stellar metallicity has a log-uniform prior over $0.01<Z/Z_\odot<0.3$.  Dust follows a modified Charlot--Fall power law \citep{CharlotFall2000}, $A_\lambda\propto\lambda^{-n}$, with $A_V$ and $n$ fitted; stars younger than 10~Myr and nebular emission receive twice the diffuse-ISM attenuation.
The resulting fit to the global host photometry is shown in Fig.~\ref{fig:bagpipes-fit}.

The fiducial fit gives $\log(M_\star/M_\odot)=8.16^{+0.09}_{-0.10}$, $\log(M_{\star,\rm formed}/M_\odot)=8.22^{+0.11}_{-0.10}$,
and a BAGPIPES-reported 100-Myr-averaged ${\rm SFR}=1.44^{+0.33}_{-0.41}\,M_\odot\,{\rm yr}^{-1}$.
The diffuse attenuation is $A_V=0.021^{+0.037}_{-0.012}$~mag; the fixed birth-cloud factor therefore gives $A_{V,{\rm young}}=0.043^{+0.075}_{-0.025}$~mag for stars younger than 10~Myr.  Expressing the fitted Charlot--Fall curve as the corresponding colour excess, $A_B-A_V$, gives $E(B-V)_{\rm young}=0.012^{+0.008}_{-0.005}$, lower than the independent Balmer-decrement result, $E(B-V)_{\rm neb}=0.054\pm0.013$ (Sect.~\ref{sec:dust_density}), although the two estimates remain consistent within approximately $3\sigma$ and assume different attenuation geometries and curves.
The mass-weighted age is $70^{+180}_{-51}$~Myr.  Broadening the continuity scale to 0.6 and 1.0 gives median surviving masses of 8.15 and 8.11 dex, respectively, but younger and more sharply rising median SFHs: the integrated stellar mass is thus comparatively stable within the regularized continuity family, whereas the detailed early SFH and the onset of the present burst remain prior-sensitive, as expected for non-parametric photometric reconstructions \citep{Leja2019}.
Rising priors can be physically useful at high redshift \citep{Turner2025}, but we do not adopt one as the fiducial model because it would impose the desired early-time behaviour directly.
An unregularized-SFH fit recovers a 0.46-dex lower stellar mass, but provides a poorer photometric fit ($\Delta\chi^2=7.5$; Appendix~\ref{app:max-old}).  We retain the continuity result with scale=0.3 as the fiducial value and propagate this downward excursion as an asymmetric host-model systematic in Fig.~\ref{fig:bh-scaling} and the related discussion.

From the fiducial posterior, the lookback times at which 10\% and 50\% of the final formed mass were already in place are
$t_{10}=92^{+891}_{-61}\ {\rm Myr}$ and $t_{50}=18^{+8}_{-5}\ {\rm Myr}$; the long upper tail of $t_{10}$ is relevant, as the data clearly identify a recent dominant episode but do not determine whether a low-level older component existed.
In this sense, we also perform a `maximally-old stress test', as described in Appendix~\ref{app:photometry}, limiting any hidden old component to $M_{\rm old,formed}<4.6\times10^7\,M_\odot$ at 95\% posterior probability (Fig.~\ref{fig:max-old}).

The fiducial BAGPIPES model contains stellar and nebular emission but no AGN continuum.  We test for possible systematics on the stellar mass inference by fitting the photometry simultaneously with the same host model and an additional AGN power law, spanning $-1\leq\alpha_\nu\leq0$.  The compact broad-\ha\ contribution is first integrated through and subtracted from each filter ---the only non-negligible
corrections being in F277W and F356W--- while the narrow nebular lines remain included self-consistently in the model.
The data do not require a positive AGN component: the host+AGN fits have $\Delta\ln\mathcal{Z}=-3.7$ to $-3.4$ relative to the host-only fit, and the envelope over the range of adopted slopes gives an AGN fraction below 10.3\% at rest-frame 5100~\AA\ (95\% posterior probability).  The corresponding median stellar masses span $8.14\leq\log(M_\star/M_\odot)\leq8.16$, i.e. offset only by $-0.05$ to $-0.02$~dex relative to the fiducial fit.
We note that such constraints apply only to the emergent AGN continuum in the model and do not limit an intrinsic nuclear continuum that might be subject to independent attenuation.

\subsection{Aperture-matched H$\alpha$ star-formation rate}
\label{sec:halpha-sfr}

The narrow \ha\ luminosity provides an independent tracer of star formation on a $\lesssim10$~Myr timescale, provided that its excitation is dominated by massive stars.
To perform a meaningful comparison with the SED-based SFR estimate, we aperture-correct the measured \ha\ flux by mapping the NIRSpec point-source-calibrated narrow-line spectrum to the $0.5\arcsec$-radius aperture used for the SED fitting.
We define $C_{\rm ap}\equiv F_{0.5\arcsec}/F_{\rm NIRSpec,ps}$, where $F_{\rm NIRSpec,ps}$ is the flux returned by the point-source-calibrated one-dimensional extraction.  At the observed wavelength of \ha, F300M gives $C_{\rm ap}=1.76$, with a conservative range 1.64--2.10 that brackets the PSF-transfer and direct synthetic-photometry tests (Appendix~\ref{app:slitloss}).

The resulting, `total' narrow-line flux is $F_{\rm H\alpha,n}=2.28\times10^{-17}$~erg~s$^{-1}$~cm$^{-2}$, or $L_{\rm H\alpha,n}=3.18\times10^{42}$~erg~s$^{-1}$ before attenuation correction, which becomes $3.61\times10^{42}$~erg~s$^{-1}$ assuming the joint Balmer-decrement constraint from Sect.~\ref{sec:dust_density}.
To derive the \ha-based SFR, we use the high-redshift, sub-solar-metallicity conversion of \citet{Clarke2024}, also adopted by \citet{Khostovan2026},
$\log\left(\frac{{\rm SFR}_{\rm H\alpha}}{M_\odot\,{\rm yr}^{-1}}\right)=\log\left(\frac{L_{\rm H\alpha}}{\rm erg\,s^{-1}}\right)-41.59$.
This gives ${\rm SFR}_{\rm H\alpha}=9.3\,M_\odot\,{\rm yr}^{-1}$, with an aperture-mapping range of $8.6$--$11.1\,M_\odot\,{\rm yr}^{-1}$ and an additional $\simeq0.15$-dex conversion systematic associated with the poorly constrained stellar iron abundance.
The SED-based posterior gives ${\rm SFR}_{0-10\,{\rm Myr}}=5.42^{+0.77}_{-0.52}\,M_\odot\,{\rm yr}^{-1}$ and predicts an attenuated \ha\ flux of $1.98^{+0.19}_{-0.18}\times10^{-17}$~erg~s$^{-1}$~cm$^{-2}$; the aperture-matched observed flux is approximately 15\% higher.  The larger difference between ${\rm SFR}_{0-10\,{\rm Myr}}$ and the narrow-H$\alpha$ estimate below also reflects the different BPASS and metal-poor H$\alpha$--SFR conversions and their attenuation prescriptions.  We therefore regard the two estimates as consistent within the calibration, aperture, and excitation systematics and the short-timescale variability expected for this bursty system.

\section{MUSE Ly$\alpha$ emission}
\label{sec:muse}

\begin{figure*}
\centering
\includegraphics[width=0.9\textwidth]{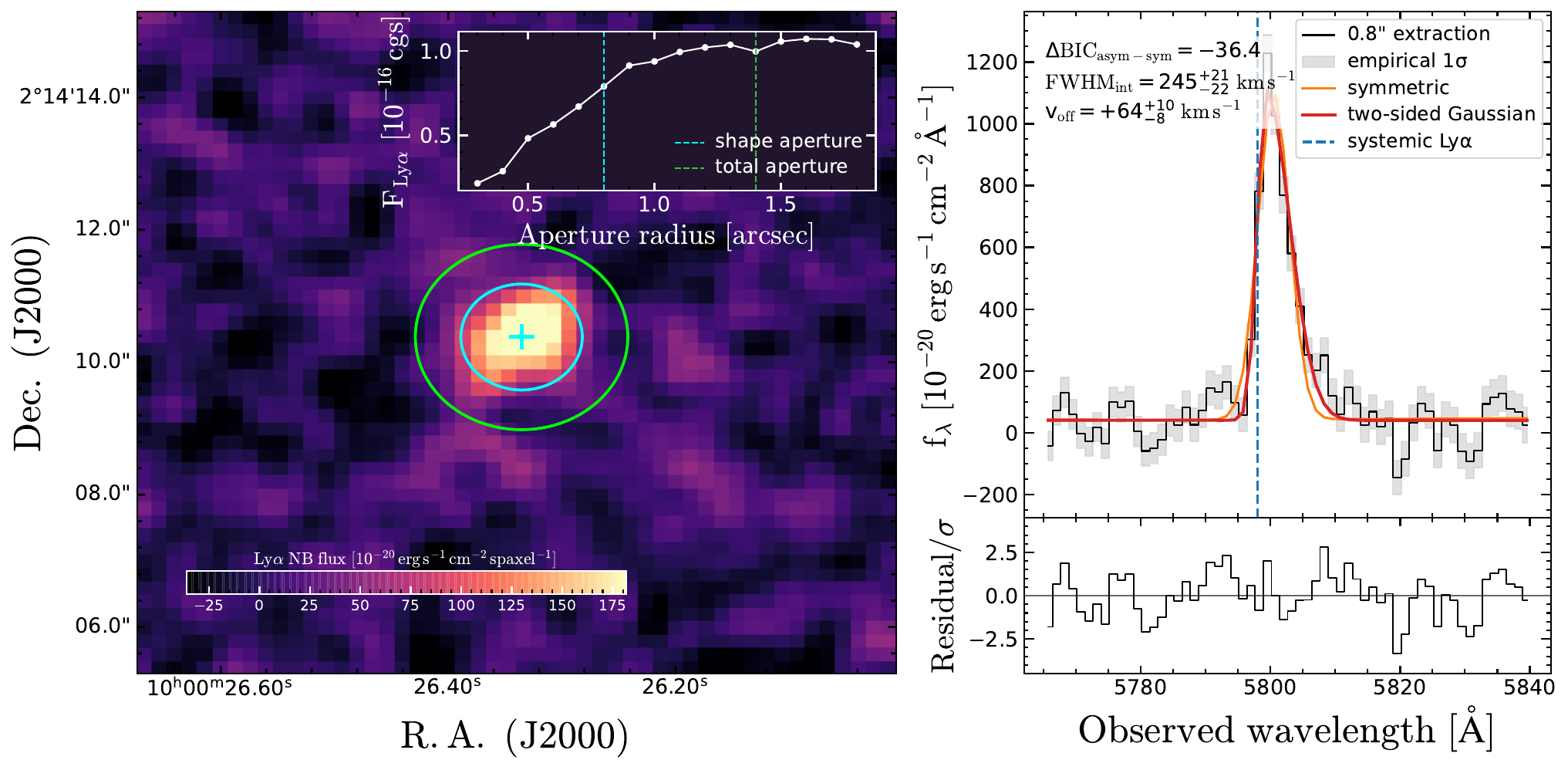}
\caption{\lya\ emission in the MUSE datacube.  The left panel shows the continuum-subtracted narrow-band map, with the $0.8\arcsec$ profile aperture in cyan and the $1.4\arcsec$ total-flux aperture in green; the inset shows the corresponding curve of growth.  The right panel shows the $0.8\arcsec$-extracted spectrum (maximizing the S/N) and the symmetric and two-sided-Gaussian fits, while the lower-right panel shows the normalized residuals to the latter.  The spectrum is not sufficiently sensitive to robustly detect any unattenuated broad-\lya\ flux as predicted from the broad \ha\ (Sect.~\ref{sec:muse}).}
\label{fig:muse-lya-profile}
\end{figure*}

We here analyze VLT/MUSE observations (Sect.~\ref{sec:data}) of MARTA~3887, probing its rest-frame UV spectrum.
After registering the MUSE WCS (see Appendix~\ref{app:muse}), clear emission is seen in the datacube at $0.037\arcsec$ from the NIRCam position---consistent with the astrometric uncertainty---which we identify as \lya.
The left-hand panel of Figure~\ref{fig:muse-lya-profile} shows the `pseudo narrow-band' surface brightness map obtained by collapsing the datacube over the line, the $r=0.8\arcsec$ aperture maximizing the SNR and used to study the line profile, and the $r=1.4\arcsec$ aperture adopted to recover the total flux, as set by the curve of growth approaching its plateau (inset panel).

The emission is spatially resolved and elongated.
An elliptical-Gaussian fit gives an observed FWHM of $1.24\arcsec\times0.84\arcsec$, compared with a $0.85\arcsec\times0.76\arcsec$ PSF, and a deconvolved FWHM of approximately $0.95\arcsec\times0.23\arcsec$ ($6.9\times1.7$ proper kpc) at a major-axis position angle of $\simeq133^\circ$.  The \lya\ emission encompasses the multiclump UV source, extends beyond it, and probes the inner ISM or CGM without isolating an individual knot.  The measured deconvolved major-axis FWHM of $\simeq7$ proper kpc and the curve-of-growth plateau at $\simeq10$ proper kpc are comparable to the compact \lya\ haloes commonly detected around high-redshift star-forming galaxies \citep{Wisotzki2016,Leclercq2017,Kusakabe2022}.  They are substantially smaller than the $\sim50$--$100$ proper-kpc nebulae routinely recovered around optically luminous quasars in MUSE surveys \citep{Borisova2016,ArrigoniBattaia2019,Lusso2019,Fossati2021}.  This comparison is however purely descriptive, since both the ionizing luminosity and the surface-brightness sensitivity differ markedly between the samples; the present observations therefore do not exclude a fainter large-scale halo around MARTA~3887.

The right-hand panel of Fig.~\ref{fig:muse-lya-profile} shows the extracted \lya\ profile and its best-fit.
The line-profile models are convolved with the wavelength-dependent MUSE line-spread function measured by \citet{Bacon2017}; at the observed \lya\ wavelength this corresponds to ${\rm FWHM}_{\rm LSF}=2.69$~\AA\ ($139$~km~s$^{-1}$).
The total flux is $F_{\rm Ly\alpha}=9.97\times10^{-17}$~erg~s$^{-1}$~cm$^{-2}$, or $L_{\rm Ly\alpha}=1.39\times10^{43}$~erg~s$^{-1}$.
A two-sided Gaussian profile is preferred over a symmetric Gaussian by $\Delta{\rm BIC}_{\rm asym-sym}=-36.4$.
The intrinsic FWHM is $245^{+21}_{-22}$~km~s$^{-1}$, the peak offset is $+64^{+10}_{-8}$~km~s$^{-1}$ with respect to the systemic redshift, and the red-to-blue width ratio is $2.85^{+0.43}_{-0.46}$.

We use the aperture-corrected narrow \ha\ flux, $F_{\rm H\alpha,n}(r<0.5\arcsec)=2.28\times10^{-17}$~erg~s$^{-1}$~cm$^{-2}$ (Sect.~\ref{sec:halpha-sfr}), to infer a $F_{\rm Ly\alpha}/F_{\rm H\alpha,n}=4.38$; this assumes that the non-resonant \ha\ emission is compact and enclosed by the $0.5\arcsec$ aperture, while any additional extended \ha\ emission would lower the inferred escape fraction.
The observed \lya\ flux is not dust-corrected: absorption by dust, enhanced by resonant scattering, is one of the losses that $f_{\rm esc}^{\rm Ly\alpha}$ is intended to measure.
The \ha\ flux is corrected because it provides the non-resonant estimate of the intrinsic recombination luminosity.  Applying a foreground-screen correction to \lya\ would remove part of the attenuation being measured and would not capture its non-local resonant transfer.
For Case~B recombination at the measured gas temperature and density (Sect.~\ref{sec:dust_density}), the intrinsic ratio is $(\mathrm{Ly}\alpha/\ha)_{\rm int}=8.82$.
We therefore define
$f_{\rm esc}^{\rm Ly\alpha}=F_{\rm obs}(\mathrm{Ly}\alpha)/[8.82\,F_{\rm int}(\ha)]$, which gives $f_{\rm esc}^{\rm Ly\alpha}\simeq0.44$ with an allowed range 0.36--0.48 reflecting the conservative $C_{\rm ap}$ envelope for \ha\ aperture correction and the uncertainty on $E(B-V)_{\rm neb}$.
We note that the apertures remain imperfectly matched, as the MUSE \lya\ flux is integrated to $r=1.4\arcsec$, whereas the \ha\ mapping is tied to the $r=0.5\arcsec$ aperture used in the SED fitting; furthermore, the relative MUSE/NIRSpec absolute calibration remains uncertain.
Additional \lya\ production through fluorescence or collisional excitation would also bias this simple recombination-based estimate high.

At face value, the measured escape fraction implies that a substantial fraction of the \lya\ photons produced by recombinations escape into the $\simeq10$~kpc aperture despite resonant scattering, consistent with a low effective dust opacity and low-column or porous H\,\textsc{i} geometry.
Small \lya\ peak offsets or separations and high $f_{\rm esc}^{\rm Ly\alpha}$ are empirically associated with direct LyC emitters, although with substantial scatter and a strong dependence on viewing angle and neutral-gas geometry \citep{Verhamme2015,Verhamme2017,Izotov2018,Gazagnes2020,Flury2022,Naidu2022}.  High O32 increases the probability of LyC leakage but is not sufficient on its own \citep{JaskotOey2013,NakajimaOuchi2014,Flury2022,Choustikov2024}.  Recent empirical applications therefore combine ionization-sensitive line ratios with UV slope and compactness \citep{Steidel2018,Pahl2021,Chisholm2022,Mascia2023,Mascia2024,Saxena2024}.  The Mg~II doublet provides a complementary resonant-line probe of low-column escape paths, subject to radiative-transfer and geometric degeneracies \citep{Xu2022,Henry2018,Chisholm2020,Katz2022}.
Taken together, these properties are consistent with ionizing escape channels in MARTA~3887 but do not yield a quantitative estimate of the LyC escape fraction.
\lya\ photons can scatter around optically thick material, whereas LyC photons are absorbed, and both escape processes are anisotropic; moreover,  LyC photons that escape before producing recombinations also reduce \ha\ and \lya\ together and are not counted by their ratio.
We consequently regard MARTA~3887 as a multi-proxy, though indirect LyC-leaker candidate.

Finally, the presence of the broad \ha\ component naturally motivates a search for a broad \lya\ counterpart.
We find no reproducible detection of such a component in the MUSE data: adding a direct, symmetric Gaussian with the broad-\ha\ kinematics to the $0.8\arcsec$-extracted spectrum gives $F_{\rm Ly\alpha,b}<3.4\times10^{-17}$~erg~s$^{-1}$~cm$^{-2}$ (one-sided 95\% confidence).  The mild preference for a non-zero component in the full-depth cube is however not reproduced in either of the two independent observing subsets and is therefore not interpreted as a detection.
For comparison, the observed broad-\ha\ flux predicts $F_{\rm Ly\alpha,b}\simeq(0.6$--$1.5)\times10^{-17}$~erg~s$^{-1}$~cm$^{-2}$ for intrinsic Ly$\alpha$/H$\alpha$ ratios ranging from a representative observed quasar-BLR value ($\simeq3.3$; \citealt{Kriss1984}) to Case~B ($\simeq8.8$), respectively, and before accounting for the partial IGM transmission of the blue wing at $z=3.77$.
The present data are thus not sensitive enough to identify an unattenuated counterpart.

At the same time, we note that the galaxy-integrated properties inferred for MARTA~3887, including the low $E(B-V)_{\rm neb}$ and high $f_{\rm esc}^{\rm Ly\alpha}$, do not necessarily apply to the sightline towards the compact nucleus.
A modest nuclear dust column, to which the present broad-\hb\ limit is insensitive (Sect.~\ref{sec:oiii_hb}), would attenuate Ly$\alpha$ by roughly three--four times more than \ha\.
Moreover, in the case of a static, fully covering $T\simeq10^4$~K H\,\textsc{i} screen, $N_{\rm HI}\sim10^{20}$~cm$^{-2}$ gives an optical depth of order unity even at $\simeq600$~km~s$^{-1}$ from line center.
Resonant scattering can therefore increase the path length and effective dust attenuation of a compact nuclear Ly$\alpha$ component \citep[e.g.][]{Neufeld1990,Dijkstra2014}.
The extended and asymmetric \lya\ emission seen in the MUSE data therefore most likely traces galaxy-scale radiative transfer and does not directly probe the sub-pc BLR.
We therefore treat both the observed profile and the absence of a reproducible broad component as neutral with respect to the AGN interpretation discussed below.

\section{Discussion}
\label{sec:discussion}

\begin{figure*}
\centering
\includegraphics[width=0.99\textwidth]{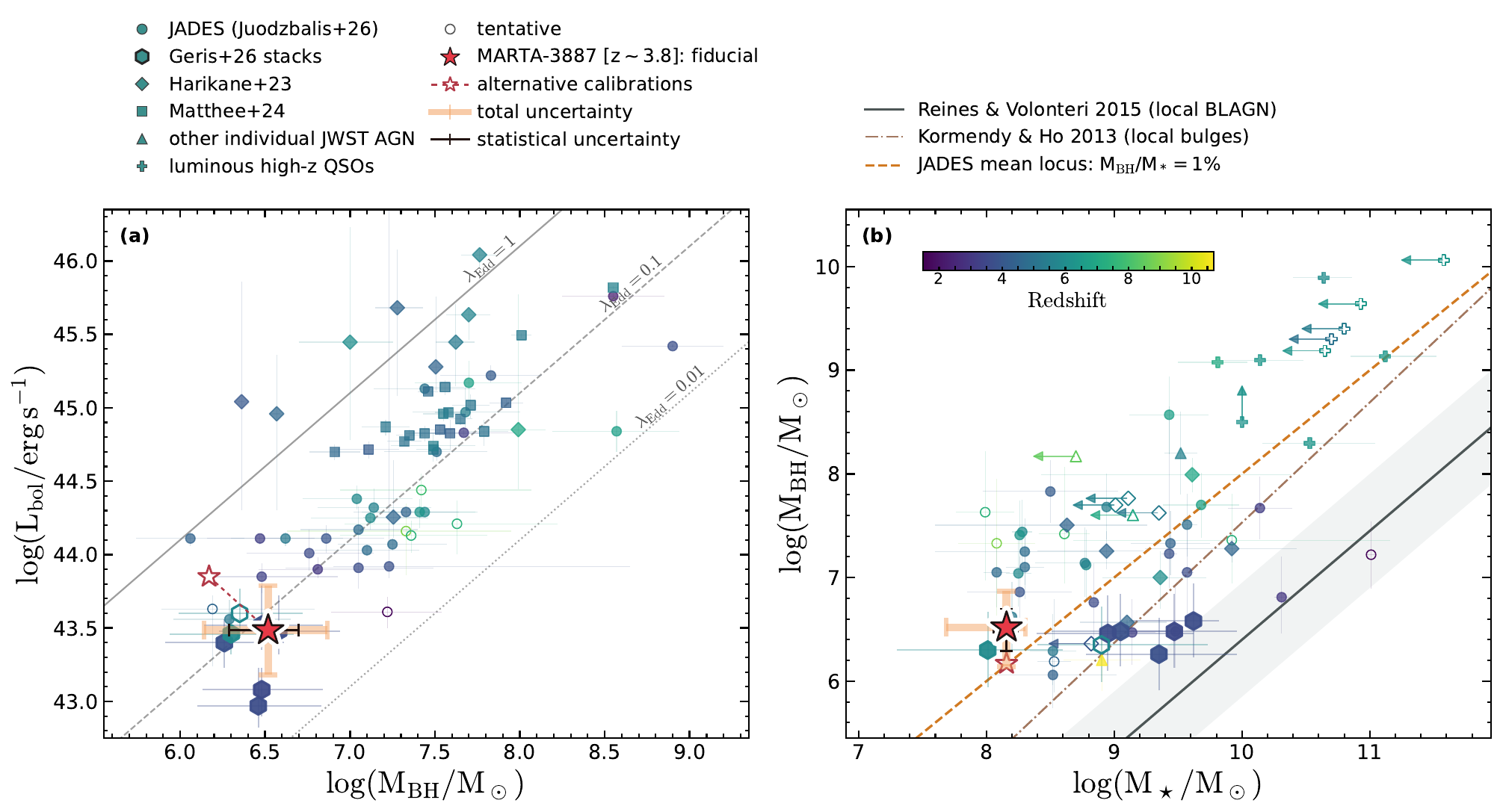}
\caption{Black-hole accretion and host scaling relations.
The filled red star marks MARTA~3887 using the \citet{ReinesVolonteri2015} broad-\ha\ mass, the \citet{SternLaor2012} bolometric correction, and the integrated-host stellar mass fiducial measurements. Black inner bars show statistical uncertainties, while pale outer bars add the 0.30-dex virial and bolometric-calibration terms; in panel (b) they additionally include the asymmetric host-SED systematic from the unregularized-SFH and maximally-old stress tests described in Appendix~\ref{app:max-old}.
The connected open star shows the \citet{GreeneHo2005} virial mass together with the broad-\ha-inferred $L_{5100}$ and \citet{Richards2006} bolometric correction.
Panel (a) shows bolometric luminosity versus single-epoch black-hole mass, comparing MARTA~3887 to samples from JADES \citep{Juodzbalis2026}, the stacked measurements of \citet{Geris2026}, \citet{Harikane2023}, and \citet{Matthee2024}; diagonal lines mark constant Eddington ratio.  Panel (b) shows black-hole mass versus total host stellar mass, and includes also individual JWST AGN \citep{Ubler2023,Kokorev2023,Furtak2024,Maiolino2024}, and luminous quasars \citep{Ding2023,Stone2024,Yue2024QSO}.
The curves in panel (b) show the local broad-line AGN relation and scatter \citep{ReinesVolonteri2015}, the local classical-bulge relation \citep{KormendyHo2013}, and the descriptive $M_{\rm BH}/M_\star=1\%$ JADES locus.
Hexagons identify stack averages, while other marker shapes identify individual-object samples;  points are colour-coded by redshift, open symbols denote tentative broad-line identifications, and arrows retain published mass limits. }
\label{fig:bh-scaling}
\end{figure*}

\subsection{A broad-line AGN candidate}
\label{sec:agn-candidate}

The robust detection of a broad \ha\ component (Sect.~\ref{sec:blrg}) provides the primary observational evidence for a low-mass accreting black hole in MARTA~3887.
We adopt the BLR interpretation for the quantitative inferences below, while retaining such scenario as conditional because the present single-epoch data cannot exclude every transient or compact high-density alternative.

The broad component is confined to the permitted \ha\ emission, is centered close to systemic, and is substantially wider than narrow and outflow-like components.
No comparably broad forbidden-line counterpart is required by the data (Sect.~\ref{sec:oiii_hb}), while any constraint on the broad \hb\ remains weak ($\mathrm{S/N}\lesssim2$).
No X-ray counterpart is present in the Chandra COSMOS-Legacy or XMM-COSMOS point-source catalogues \citep{Cappelluti2009,Civano2016,Marchesi2016}; however, the more constraining Chandra soft-band depth corresponds to $L_{2-10\,\mathrm{keV}}\lesssim(3.5$--$5)\times10^{43}$~erg~s$^{-1}$.  Adopting the luminosity-dependent $2$--$10$~keV bolometric correction of \citet{Duras2020}, including its intrinsic scatter, predicts an X-ray flux more than an order of magnitude below the catalogue depth for the fiducial broad-\ha-based $L_{\rm bol}$.  The non-detection therefore provides no meaningful constraint on the BLR interpretation, particularly because X-ray weakness is common among faint JWST broad-line AGN \citep{Ananna2024,Yue2024Xray,Maiolino2025SN,Geris2026}.

Under the BLR interpretation, we estimate the black-hole mass with a single-epoch virial relation.
For consistency with recent literature compilations such as that analysing the JADES sample \citep{Juodzbalis2026}, we adopt the \citet{ReinesVolonteri2015} broad-\ha\ calibration,
\begin{equation}
 \begin{split}
 \log\frac{M_{\rm BH}}{M_\odot}={}&6.60+0.47\log\frac{L_{{\rm H}\alpha,{\rm b}}}{10^{42}\ {\rm erg\ s^{-1}}}\\
 &+2.06\log\frac{{\rm FWHM}_{{\rm H}\alpha,{\rm b}}}{10^3\ {\rm km\ s^{-1}}},
 \end{split}
\end{equation}
which gives $\log(M_{\rm BH}/M_\odot)=6.52^{+0.18}_{-0.23}$ from the statistical line uncertainties, while we retain a separate 0.30-dex systematic uncertainty term for the single-epoch calibration.
The \citet{GreeneHo2005} prescription instead gives $\log(M_{\rm BH}/M_\odot)=6.17$, and we use it to illustrate the calibration systematics.

Alternatives to the black-hole scenario can be assessed directly against the properties of MARTA~3887.
The non-core [O\,\textsc{iii}]-emitting component has an intrinsic ${\rm FWHM}=299$~km~s$^{-1}$, substantially narrower than broad \ha.
Assigning its kinematics simultaneously to \ha\ and both \nii\ lines does not reproduce the observed \ha+\nii\ base: this intermediate-complex model is disfavoured relative to broad \ha\ by $\Delta{\rm BIC}=41.8$ (Appendix~\ref{app:ha-tests}).
Together with the absence of a forbidden-line component of comparable width and the spatial-profile tests, this result rules out aperture smearing and disfavours an ordinary low- or moderate-density galaxy-scale outflow.
A separate compact phase with densities approaching or exceeding the [O\,\textsc{iii}] $\lambda5008$ critical density could suppress its forbidden emission and cannot be excluded by the present data.

A luminous interacting core-collapse SN remains the principal transient alternative.
The measured $L_{{\rm H}\alpha,{\rm b}}=2.35\times10^{41}$~erg~s$^{-1}$ and ${\rm FWHM}\simeq1270$~km~s$^{-1}$ are attainable by Type~IIn-like events.
For the measured ${\rm SFR}_{\rm H\alpha}=9.3\,M_\odot\,{\rm yr}^{-1}$, the time-averaged scaling adopted in recent assessments of JWST broad-\ha\ sources predicts
$L_{{\rm H}\alpha,{\rm b}}^{\rm SN}\simeq5.6\times10^{40}$~erg~s$^{-1}$, approximately 4.2 times below the observed luminosity \citep{Maiolino2025SN,Geris2026}.
We stress that this comparison concerns the mean contribution of the underlying SN population and is not an upper limit on a single event: an individual luminous interacting SN could in fact reach the observed luminosity, provided it is caught during the comparatively brief phase in which broad \ha\ is strong \citep{Kiewe2012,Taddia2013,Smith2017}.
Nonetheless, the nearly systemic and approximately symmetric profile, without an obvious P-Cygni feature nor the large late-time blueshift often produced by dust in SN ejecta, still favours the BLR interpretation \citep{Gall2014}.

\begin{figure}
\centering
\includegraphics[width=\columnwidth]{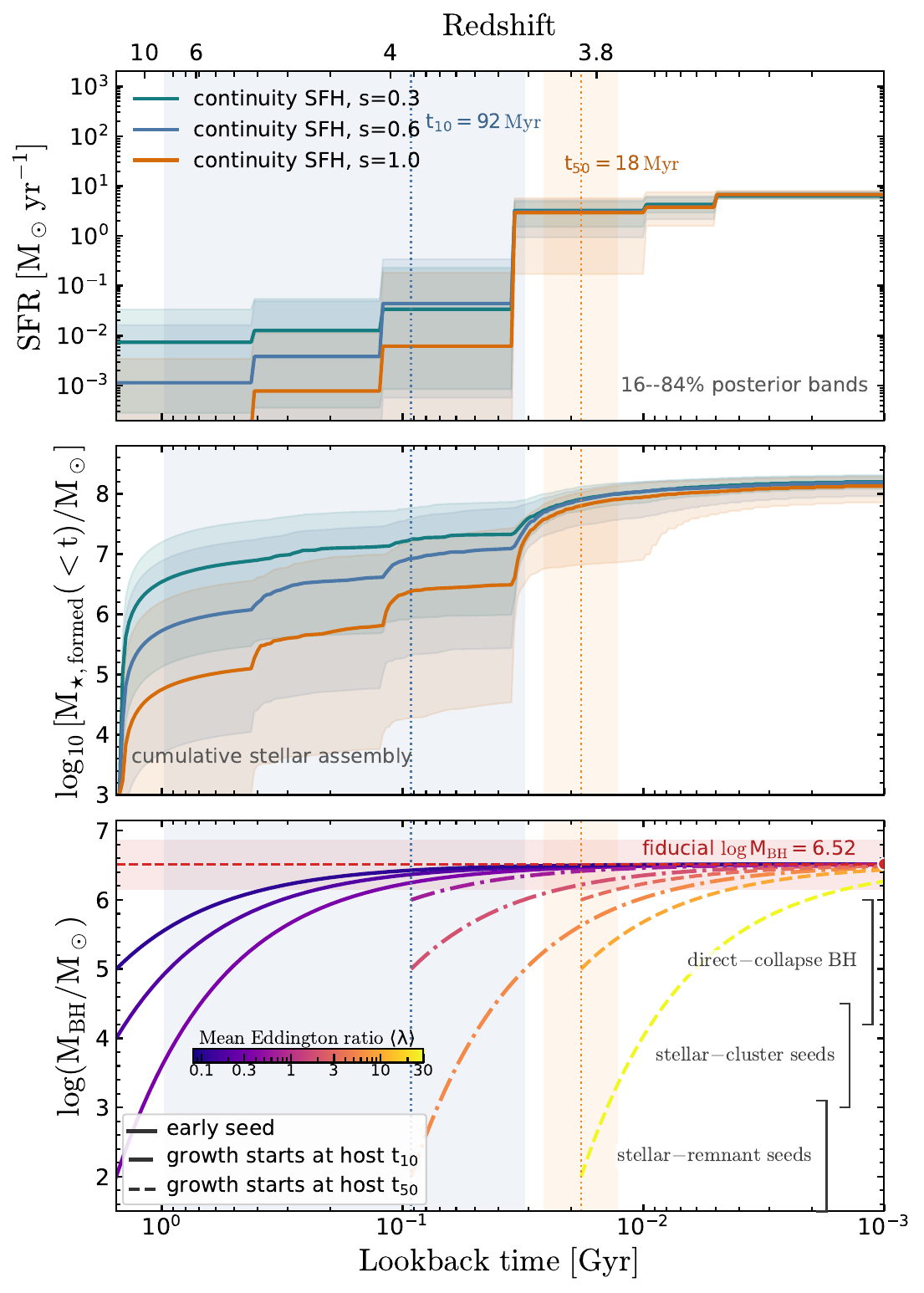}
\caption{Host galaxy and black-hole growth are compared on a common (logarithmic) lookback-time axis.  Top: BAGPIPES SFH posterior medians and 16th--84th percentile intervals for three continuity-prior scales.  Middle: cumulative formed stellar mass.  Bottom: illustrative black-hole tracks normalized to the fiducial broad-\ha-based mass.  The vertical dotted lines mark the median host $t_{10}$ and $t_{50}$; solid curves begin at cosmic age 0.1~Gyr, dash-dotted curves begin at $t_{10}$, and dashed curves begin at $t_{50}$.  The latter two need to be considered as limiting `late-start' scenarios: the stellar assembly percentiles do not measure the seed epoch or the onset of black-hole accretion.  Track colours encode the required constant mean Eddington ratio.
}
\label{fig:bh-growth}
\end{figure}

\begin{figure*}[t]
\centering
\includegraphics[width=0.99\textwidth]{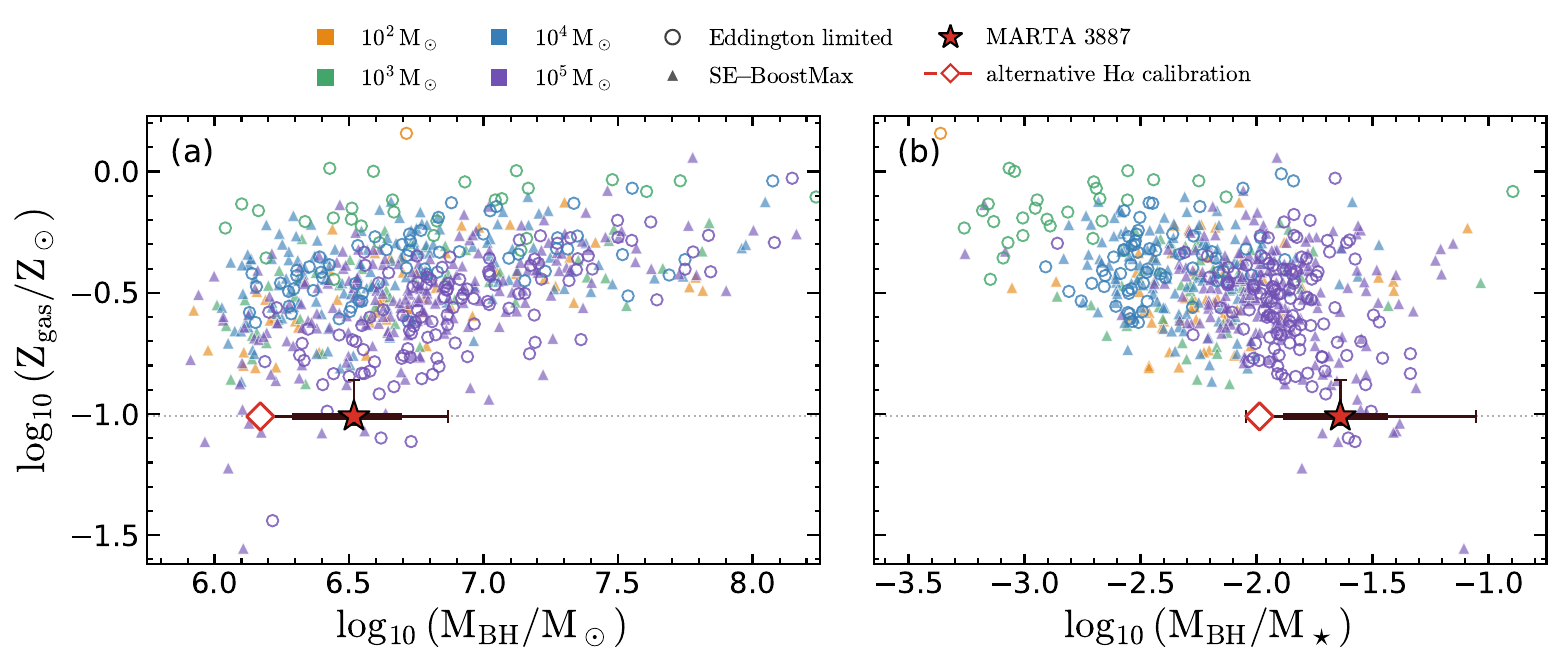}
\caption{Selection-aware comparison between MARTA~3887 and the $z=3.9$ \textsc{aesopica} populations.  The simulated objects shown satisfy the \textsc{balmersopica} broad-H$\alpha$ selection calculated for the NIRSpec/G235M 7.3-h MARTA observations at $z=3.75$.  Panel (a) shows gas metallicity against black-hole mass, and panel (b) shows gas metallicity against the black-hole to stellar-mass ratio.  Colours identify the imposed seed mass; open circles show the fiducial Eddington-limited runs and filled triangles the `SE--BoostMax' runs, which permit boosted and super-Eddington accretion.  The red star marks MARTA~3887 using the fiducial broad-H$\alpha$ virial calibration.  Thick horizontal bars show the formal uncertainty while thin bars additionally include the adopted virial and host-SED systematics;  the connected open diamond shows instead the alternative broad-H$\alpha$ calibration.  The thick vertical bar marks the formal uncertainty on the direct-method oxygen abundance, while the thin one-sided extension shows the $+0.15$-dex systematic allowed by the \textsc{homerun} models.}
\label{fig:aesopica-marta}
\end{figure*}

\subsection{Black-hole accretion and host scaling relations}
\label{sec:bh-scaling}

We compare MARTA~3887 with published JWST broad-line AGN and luminous high-redshift quasars in Fig.~\ref{fig:bh-scaling}.
The left-hand panel shows the $L_{\rm bol}$ versus $M_{\rm BH}$ relationships, and includes the JADES census from \citet{Juodzbalis2026}, the stacked measurements of \citet{Geris2026}, and the samples of \citet{Harikane2023} and \citet{Matthee2024}, with lines marking constant-Eddington-ratio loci.
The right-hand panel places MARTA~3887 on the $M_{\rm BH}$--$M_{\star}$ diagram, together with the \citet{Geris2026} stacks, individual JWST AGN \citep{Ubler2023,Kokorev2023,Furtak2024,Maiolino2024}, and quasars with resolved or constrained host light \citep{Ding2023,Stone2024,Yue2024QSO}.
Using $L_{\rm bol}=130L_{{\rm H}\alpha,{\rm b}}$ \citep{SternLaor2012} gives $\log(L_{\rm bol}/{\rm erg\ s^{-1}})=43.49$ and
$\lambda_{\rm Edd}\equiv L_{\rm bol}/L_{\rm Edd}=0.074$, corresponds to a radiatively active, moderately accreting black hole.
The systematic uncertainty in the bolometric conversion is comparable to the statistical uncertainty, placing MARTA~3887 at the low-mass, low-luminosity boundary of current spectroscopic JWST samples.

For comparison, we also follow the procedure adopted by \citet{Harikane2023}, using the broad-\ha--continuum relation calibrated by \citet{GreeneHo2005}:
we first convert the measured broad-line luminosity into $\lambda L_\lambda(5100)=10^{44}[L_{{\rm H}\alpha,{\rm b}}/(5.25\times10^{42})]^{1/1.157}=6.8\times10^{42}$~erg~s$^{-1}$, and the \citet{Richards2006} factor $L_{\rm bol}=10.33\lambda L_\lambda(5100)$ then gives $\log(L_{\rm bol}/{\rm erg\ s^{-1}})=43.85$.
Combined with the \citet{GreeneHo2005} virial mass, $\log(M_{\rm BH}/M_\odot)=6.17$, this yields $\lambda_{\rm Edd}\simeq0.38$.  The factor of $\approx5$ between the two $\lambda_{\rm Edd}$ estimates arises from combining the alternative bolometric conversion ---raising $L_{\rm bol}$ by 0.36~dex--- with the alternative virial relation which lowers $M_{\rm BH}$ by 0.35~dex; changing either calibration alone gives $\lambda_{\rm Edd}\simeq0.17$.
The \citet{Harikane2023}-like conversion implies an AGN continuum of 39.7~nJy at observed $2.432\,\mu$m, well above the slope-dependent 95\% limits of 8.3--12.2~nJy from the simultaneous host+AGN fits (Sect.~\ref{sec:photometry}).
Both bolometric chains, like the single-epoch virial relations, are calibrated primarily on lower-redshift Type-1 AGN and are extrapolated here to a low-mass $z=3.77$ source.
We also note that in both chains the quantities entering $\lambda_{\rm Edd}$ inherit the broad-\ha\ luminosity and are therefore highly covariant.
We therefore leave the full $\lambda_{\rm Edd}\simeq0.07$--0.38 bracket as a conservative systematic uncertainty, and
show the results from alternative calibration chains as connected open symbols in Fig.~\ref{fig:bh-scaling}.

From the fiducial host stellar mass we infer $M_{\rm BH}/M_\star=2.3\%$.  The adoption of an unregularized-SFH would raise this to $\simeq6.6\%$, whereas the simultaneous host+AGN fits change the stellar mass by less than 0.05~dex.
This is 0.36~dex above the $M_{\rm BH}/M_\star=1\%$ locus of the JADES sample \citep{Juodzbalis2026}, and lies well above the local broad-line AGN relation of \citet{ReinesVolonteri2015}, although that relation is only weakly anchored at the stellar mass of MARTA~3887.
The \citet{KormendyHo2013} relation is shown only for visual reference, since it is calibrated against classical-bulge mass and is not directly applicable to the total stellar mass of a clumpy, star-forming galaxy.
The alternative virial calibration yields $M_{\rm BH}/M_\star\simeq1.0\%$, close to the JADES mean locus, without bringing the source onto the local broad-line AGN relation.
In Fig.~\ref{fig:bh-scaling} we separate statistical errors from the 0.30-dex virial and bolometric systematics; the host-mass outer uncertainty is asymmetric and incorporates the explicitly explored SFH and old-population tests (see also Appendix~\ref{app:max-old}).
We also note that our comparison adopts the stellar mass of the integrated system, whereas the morphology of MARTA~3887 is dominated by at least two bright components that are both captured by the NIRSpec slit (Fig.~\ref{fig:rgb-slitlet} and Sect.~\ref{ssec:morphology}).
The NIRSpec spectra do not localize the broad component to either of the two clumps; should it be associated with only one of them, the physically relevant host mass would be smaller and the local $M_{\rm BH}/M_\star$ ratio correspondingly higher.

Taken at face value, the offset measured for MARTA~3887 implies that its black hole assembled rapidly relative to the presently observed stellar host, compatible with a phase in which black-hole growth leads stellar assembly.
A massive seed---including a direct-collapse black hole (DCBH)---merger-triggered or otherwise bursty super-Eddington growth, and subsequent feedback-regulated evolution can all produce such a state \citep[e.g.][]{Inayoshi2020,Pacucci2024,Trinca2024}.

The stacked analysis of \citet{Geris2026} provides relevant population context, suggesting the presence of an average population at $3<z<7$ with $\log(M_{\rm BH}/M_\odot)\simeq6.2$--6.6 and $\lambda_{\rm Edd}\simeq0.02$--0.1 sharing the same low-mass, moderately accreting regime probed by our fiducial MARTA~3887 estimates.
At $3<z<5$, their inferred active black-hole mass function at a few times $10^6\,M_\odot$ is $(3.1$--$8.7)\times10^{-3}\,{\rm Mpc}^{-3}\,{\rm dex}^{-1}$.
The inferred black-hole mass function is well reproduced by the active subset of the \textsc{cat} framework \citep{Trinca2022} in which black holes grow through short, episodic super-Eddington phases \citep{Trinca2024,Geris2026}.  In the specific model comparison presented by \citet{Geris2026}, heavy-seed realizations also provide a closer match than models restricted to light seeds and Eddington-limited growth.  More generally, however, seeding and accretion remain strongly degenerate: efficient growth can allow both light and heavy seeds to populate the observed region and progressively erase their initial mass differences \citep{Habouzit2022,Hu2025,Koudmani2026,Dayal2026}.
By $z\simeq3.8$, subsequent accretion may have erased much of the information about the initial conditions, so the position of MARTA~3887 alone cannot identify its black-hole seed (we further explore some of the possibilities in Sect.~\ref{sec:bh-growth}).

The same population comparison cautions against interpreting individual offsets as direct evidence for evolution of the normalization of the entire $M_{\rm BH}$--$M_\star$ relation.
Broad-line flux and line-width thresholds preferentially select massive, actively accreting black holes at fixed host mass, while measurement scatter moves objects into the detectable upper envelope \citep{Li2024}.
Stacking analyses recover lower-mass and less actively accreting black holes whose mean locations lie closer to the local relation than those of individual detections. Nevertheless, the stacked measurements remain, on average, above the local relation and do not establish whether a corresponding population exists below it. This is consistent with individually detected AGN preferentially sampling the upper-\(M_{\rm BH}\) tail at fixed host mass.
Furthermore, host detectability introduces an additional selection effect, and stellar masses depend on properly separating nuclear from stellar light, while the resolved NIRCam morphology and $<10.3\%$ conditional limit on the emergent AGN continuum at 5100~\AA\ reduce this ambiguity for MARTA~3887.

\subsection{Co-evolution of black-hole and galaxy host}
\label{sec:bh-growth}

Figure~\ref{fig:bh-growth} compares the posterior host SFH and cumulative formed mass with illustrative black-hole growth tracks\footnote{These tracks are not inferred black-hole growth histories} on the same lookback-time axis.
The stellar-mass assembly times $t_{10}$ and $t_{50}$ do not constrain the formation time of the seed or the onset of accretion; we use them only to construct limiting scenarios in which black-hole growth begins as late as the dominant observed episode of stellar assembly.

To build the latter, we assume a single seed branch with no black-hole mergers after $t_{\rm seed}$,
\begin{equation}
 M_{\rm BH}(t)=M_{\rm seed}\exp\left[\frac{\langle\lambda_\mathrm{Edd}\rangle(t-t_{\rm seed})}{t_{\rm Sal}}\right],
\end{equation}
with constant mean Eddington ratio $\langle\lambda_\mathrm{Edd}\rangle$, radiative efficiency $\epsilon=0.1$, and Salpeter time $t_{\rm Sal}=0.05$~Gyr \citep{Salpeter1964}.  Each track is normalized to reach the fiducial $\log(M_{\rm BH}/M_\odot)=6.52$ at the observed epoch.
Seeds of $10^2$, $10^4$, and $10^5\,M_\odot$ placed at an illustrative cosmic age of 0.1~Gyr (corresponding to z$\sim30$, i.e. the left edge of Fig.~\ref{fig:bh-growth}) require $\langle\lambda_\mathrm{Edd}\rangle\simeq0.34$, 0.19, and 0.11, respectively.  Imposing the median $t_{10}=92$~Myr as the start of accretion would instead require $\langle\lambda_\mathrm{Edd}\rangle\simeq5.7$ for a $10^2\,M_\odot$ seed, rising to $\simeq29$ if growth began at $t_{50}=18$~Myr.
Repeating the calculation with the lower \citet{GreeneHo2005} mass changes the early-seed values to 0.31, 0.16, and 0.087.

We stress that the inferred lifetime-mean Eddington ratios depend on the adopted final mass but not on the estimate of the instantaneous $\lambda_{\rm Edd}$, as they assume a constant radiative efficiency of $\epsilon=0.1$ and parameterize the time-averaged mass growth.
As such, they should not be interpreted as luminosities known to have been reached at any particular epoch, but in our no-merger scenario they only describe the accretion required along one progenitor branch; mergers with other seed-bearing progenitors would reduce that requirement.
Relatedly, the resolved clumpy morphology of MARTA~3887 is compatible with, though does not firmly establish, assembly through a galaxy merger.
In particular, the NIRSpec spectrum contains no spectroscopic evidence for two BLRs or for a dual active nucleus.
Any black-hole merger relevant to the mass assembly of MARTA~3887 would therefore have occurred earlier in its merger tree and cannot be reconstructed from the current morphology or integrated spectra.
Spatially resolved spectroscopy is required to distinguish genuine dual AGN from non-AGN clumps, chance alignments, or lensing in multiple systems \citep{Perna2023,Scialpi2024,Ubler2024}.

The three adopted seed masses are intended to bracket the main families of proposed formation pathways.  The $10^2\,M_\odot$ track is representative of light seeds left by massive Pop~III stars \citep{MadauRees2001}; the $10^4\,M_\odot$ track represents intermediate-mass seeds assembled through runaway stellar collisions, repeated mergers, and gas accretion in dense primordial or nuclear star clusters \citep{PortegiesZwart2004,DevecchiVolonteri2009,Boekholt2018}; and the $10^5\,M_\odot$ track corresponds to the characteristic mass scale of heavy seeds formed through the collapse of supermassive stars in atomic-cooling haloes \citep{BrommLoeb2003,Begelman2006,LodatoNatarajan2006,Inayoshi2020}.

In the classical DCBH picture, rapid inflow in an atomic-cooling halo forms a $\sim10^4$--$10^6\,M_\odot$ seed before ordinary star formation fragments the gas; this generally requires pristine or only very weakly enriched gas, together with conditions that suppress H$_2$ cooling \citep{Omukai2008,Inayoshi2020}.
In this sense, the low metallicity measured in MARTA~3887, $12+\log({\rm O/H})=7.68$ ($\simeq0.10~Z_\odot$), is qualitatively consistent with a chemically young host and, while being already far above the metallicities normally required for the classical DCBH channel, it refers to the ionized gas observed at $z=3.7694$, long after any plausible seed-formation epoch.
The seed could moreover have formed in a different, initially pristine branch of the merger tree before entering the presently observed, mildly enriched system.


The constant-$\langle\lambda_\mathrm{Edd}\rangle$ curves further compress a potentially intermittent accretion history into one time-averaged parameter.  Models developed for the JWST broad-line population commonly predict feedback-regulated cycles in which short super-Eddington episodes can provide a substantial fraction of the integrated mass growth. For instance, the semi-analytic framework of \citet{Trinca2024} produces repeated 0.5--3~Myr bursts with duty cycles of 1--4\%, while the cosmological \textsc{aesopica} simulations find model-dependent bursts lasting approximately 10~Myr and super-Eddington fractions of only about 0.2--2\% at $z=4$ \citep{Koudmani2026}.  The difference in timescale reflects the adopted accretion and feedback prescriptions, but both results illustrate that a black hole observed at the fiducial instantaneous $\lambda_{\rm Edd}=0.07$ may have experienced much more rapid earlier episodes.
Efficient accretion can also erase much of the seed-mass imprint by $z\sim4$ \citep{Koudmani2026}, further limiting what the current mass and host SFH alone can reveal about the initial seeding.

Taken together, the proposed analysis does not allow us to date the formation of the black-hole seed precisely, nor to establish whether it pre-dated the galaxy.
The exercise nevertheless excludes a restricted but physically informative scenario: under the adopted no-merger, $\epsilon=0.1$ assumptions, a light seed whose growth began only during the dominant recent episode of stellar assembly could not reach the observed mass through sustained sub-Eddington accretion.  At least one of the following is therefore required: an earlier onset of black-hole growth, a more massive seed, intermittent episodes of rapid accretion, or growth through black-hole mergers.  The comparison thus identifies a mismatch between the rapid recent assembly of the surviving stellar host and the time required to build the inferred black-hole mass, while remaining unable to determine whether black-hole growth preceded, accompanied, or followed the full assembly history of the galaxy.

\subsection{Insights on black-hole seeding from cosmological simulations}
\label{sec:aesopica-seeding}

To place MARTA~3887 in the context of a self-consistent cosmological model, we compare its black-hole and host properties with the $z=3.9$ \textsc{aesopica} populations of \citet{Koudmani2026}. All runs evolve the same comoving $40\,h^{-1}$ Mpc cosmological volume with the same Planck cosmology and numerical resolution. Black holes are inserted dynamically into previously unseeded haloes once they exceed $2\times10^9\,h^{-1}\,M_\odot$; therefore they do not all begin growing at a common redshift, with the first seeds appearing at $z\sim18$.
Within each run, all newly inserted black holes are assigned the same fixed initial mass; the four seed masses, $10^2$, $10^3$, $10^4$, and $10^5\,M_\odot$, are explored in separate simulations. The suite includes both the fiducial, Eddington-limited Bondi prescription and the `SE--BoostMax' model, in which boosted accretion and short super-Eddington episodes up to $f_{\rm Edd,max}=10$ are allowed.
We forward-model the observational selection with \textsc{balmersopica}, using a dedicated NIRSpec/G235M realization matching the 7.3-h MARTA exposure at $z=3.75$, the closest tabulated redshift to MARTA~3887. The forward model predicts the broad-H$\alpha$ flux and width from the simulated $M_{\rm BH}$ and $L_{\rm bol}$, propagates them through the instrumental sensitivity, and applies the same broad-line detectability criteria used by \citet{Koudmani2026}.
Figure~\ref{fig:aesopica-marta} shows only the subset of simulated black holes satisfying this matched selection: panel~(a) compares gas metallicity with black-hole mass, while panel~(b) replaces black-hole mass with the black-hole to stellar-mass ratio. In both panels, colours identify the seed mass and marker shapes distinguish the fiducial Eddington-limited and `SE--BoostMax' accretion prescriptions.
When interpreting this figure, it is important to recall that the simulated metallicity is the SFR-weighted mean gas metallicity within the stellar half-mass radius, whereas the MARTA~3887 point adopts the nebular oxygen abundance derived in Sect.~\ref{sec:metallicity} as an observational proxy for $Z_{\rm gas}/Z_\odot$. MARTA~3887 also lies close to the low-host-mass resolution limit of the simulations.

Under the Eddington-limited prescription, the location of MARTA~3887 is reached only by the most massive seeds, whereas sufficiently efficient, bursty growth in the `SE--BoostMax' case allows lighter seeds to approach the same region. We quantify this comparison by requiring consistency with the expanded uncertainty intervals shown by the thin error bars in Fig.~\ref{fig:aesopica-marta}, including the adopted systematic uncertainties in $M_{\rm BH}$, $M_{\rm BH}/M_\star$, and metallicity. For the fiducial virial calibration, 11 of the 112 MARTA-detectable objects in the fiducial $10^5\,M_\odot$-seed run ($\sim10$\%) and 7 of 217 in its `SE--BoostMax' counterpart ($\sim3$\%) satisfy the joint constraints. None of the detectable objects in the $10^2$--$10^4\,M_\odot$-seed runs does so, even after including the systematic uncertainty associated with the $T_{\rm e}$-based metallicity estimate discussed in Sect.~\ref{sec:metallicity}.
Considering the alternative virial calibration produces limited overlap with lighter seeds: 2 of 55 detectable objects in the `SE--BoostMax' $10^3\,M_\odot$ run ($\sim4$\%) and 1 of 133 in the corresponding $10^4\,M_\odot$ run ($\sim1$\%), compared with 4 of 112 ($\sim4$\%) and 6 of 217 ($\sim3$\%) in the fiducial and `SE--BoostMax' $10^5\,M_\odot$ runs, respectively. No $10^2\,M_\odot$-seed analogue satisfies the joint constraints. Thus, although the alternative calibration opens a small region of overlap with efficiently growing lighter seeds, the closest and most numerous analogues remain associated with $10^5\,M_\odot$ seeds.

We stress that these numbers should not be interpreted directly as seed-mass probabilities: the denominators include only simulated black holes that would satisfy the broad-H$\alpha$ detectability criteria in the matched 7.3-h G235M observations, rather than the complete simulated populations.
Moreover, each run assigns the same fixed seed mass to every eligible halo crossing a common halo-mass threshold, rather than modelling the channel-dependent formation criteria and relative occurrence rates of light, intermediate, and massive seeds. The fractions derived above therefore measure only how frequently each assumed model produces a detectable `MARTA-like' galaxy and cannot be converted directly into probabilities for the corresponding formation channels.

In conclusion, within the \textsc{aesopica} framework the combined low metallicity and high $M_{\rm BH}/M_\star$ of MARTA~3887 conditionally favour a massive seed; a lighter seed remains possible, but would require some of the most efficient early-growth histories explored by the simulations.

\subsection{Low-column-density gas and the possible link to LyC leakage}
\label{sec:lyc-leakage}

The provisional \lya\ escape fraction, the small \lya\ velocity offset, the Mg~II ratio near its optically thin value, the high O32 value, and low metallicity consistently point toward a low effective H\,I column or porous, anisotropic gas geometry.
As already noted in Sect.~\ref{sec:muse}, MUSE cannot measure the LyC directly: at $z=3.7694$ the Lyman limit is observed at 4349~\AA, below its blue cutoff, and MARTA~3887 should therefore be considered as an indirect LyC-leaker candidate.

The existing HST imaging nevertheless provides some direct observational constraints (see Figure~\ref{fig:cutouts}).
In fact, WFC3/F336W probes the SED below the redshifted Lyman limit (its pivot wavelength corresponds to $\simeq703$~\AA\ in the rest frame) and gives a $3\sigma$ upper limit of $f_\nu<0.46$~nJy ($m_{\rm AB}>32.24$), delivering no direct LyC detection; ACS/F435W is detected at $3.4\sigma$, but the filter straddles 4349~\AA\ and includes transmission redward of the Lyman edge.
Converting the F336W limit into an absolute $f_{\rm esc}^{\rm LyC}$ would require the intrinsic stellar-plus-AGN ionizing spectrum and marginalization over the strongly stochastic IGM+CGM transmission \citep{Inoue2014,Bassett2021,Steidel2018,Pahl2021}; this is especially important because F336W probes the far LyC near 700~\AA, where the line-of-sight transmission at this redshift can be very small.  Direct LyC surveys at $z\simeq3$ consequently infer escape fractions statistically over ensembles or explicitly forward-model sightline transmission rather than treating a single non-detection as an intrinsic constraint \citep{Steidel2018,Pahl2021}.
The photometry therefore constrains the product of internal escape and IGM transmission and cannot isolate $f_{\rm esc}^{\rm LyC}$; the F336W non-detection neither confirms nor rules out the low-column channels suggested by the emission line diagnostics.

The presence of an accreting black hole in MARTA~3887 does not imply that escaping ionizing photons are AGN-powered; the full budget is $Q({\rm H})=Q({\rm H})_\star+Q({\rm H})_{\rm AGN}$, with potentially distinct stellar and AGN escape fractions.
The narrow, extended \lya\ emission probes transfer through host/NLR-scale gas and cannot identify the ionizing source.  The available data cannot determine whether any low-covering-factor geometry is produced by stellar feedback, AGN activity, or the intrinsic structure of the multi-clump system; an AGN contribution is possible but not required.
At $z=3.7694$, the stochastic IGM transmission at the rest-frame wavelengths sampled by F336W and the uncertain AGN EUV contribution make any absolute $f_{\rm esc}^{\rm LyC}$ estimate weak and strongly model-dependent.  We therefore do not infer an intrinsic LyC escape fraction from the available HST imaging.

\section{Summary and conclusions}

We have presented a multi-wavelength study of MARTA~3887, a compact, clumpy galaxy at $z=3.7694$ possibly hosting a low-mass accreting black hole, combining deep JWST/NIRSpec spectroscopy, HST and JWST imaging, and archival VLT/MUSE data.  Our analysis was designed to establish the origin of its broad \ha\ emission, characterize the stellar and ionized gas properties of the galaxy, and place the inferred black hole properties in the context of the low-mass AGN population now emerging from JWST surveys.  Our principal results and conclusions are summarized below.

\begin{itemize}
\item JWST/NIRSpec detects broad \ha\ emission with $L_{{\rm H}\alpha,{\rm b}}=2.35\times10^{41}$~erg~s$^{-1}$ and ${\rm FWHM}=1270^{+260}_{-250}$~km~s$^{-1}$ (Sect.~\ref{sec:blrg}, Fig.~\ref{fig:halpha-fit}).  Such a component is robust against the adopted continuum, line-spread function, and narrow-line model, is spatially coincident with the narrow-line emission, and has no comparably broad [O\,\textsc{iii}] counterpart (Sect.~\ref{sec:oiii_hb}, Fig.~\ref{fig:hbeta-oiii-fit}; Appendix~\ref{app:ha-tests}, Fig.~\ref{fig:halpha-robustness}).  These properties favour a broad-line region origin; an interacting supernova or a compact, high-density outflow are less likely, though they cannot be fully excluded with single-epoch data (Sect.~\ref{sec:agn-candidate}).

\item The narrow-line spectrum is consistent with high-excitation, metal-poor gas.  Assuming stellar-dominated photoionization, a joint inference of the physical conditions gives $E(B-V)_{\rm neb}=0.054\pm0.013$, $T_{\rm e}({\rm O\,III})=18{,}140^{+770}_{-710}$~K, and $12+\log({\rm O/H})=7.68\pm0.04$ (Sects.~\ref{sec:dust_density} and \ref{sec:metallicity}, Fig.~\ref{fig:joint-physical-conditions}).
The detection of [Ne\,\textsc{iv}] points to an additional high-ionization phase, for which the doublet ratio limits the density to $n_{\rm e}<9.5\times10^{3}$~cm$^{-3}$ (Sect.~\ref{sssec:density}, Fig.~\ref{fig:neiv-density}); together with the position of MARTA~3887 on diagnostic diagrams (Sect.~\ref{sssec:diagnostic_diagrams}, Fig.~\ref{fig:optical-diagnostics}), this is compatible with a modest contribution from the NLR of the AGN.
Multi-cloud photoionization modeling reproduce the spectrum with an energetically consistent AGN component powering $\lesssim10\%$ of the narrow H$\beta$ flux; the model-based abundances ($12+\log({\rm O/H})\simeq7.8$) imply a systematic upward uncertainty of $\simeq0.15$~dex on the $T_{\rm e}$-based value, independent of the AGN contribution (Sect.~\ref{sec:metallicity}; Appendix~\ref{app:homerun}, Fig.~\ref{fig:homerun-contributions}).

\item Morphologically, MARTA~3887 comprises at least three components, well resolved in the rest-frame UV imaging and separated by $0.8$--$1.2$ proper kpc (Sect.~\ref{ssec:morphology}, Figs.~\ref{fig:rgb-slitlet} and \ref{fig:f150-morphology}; Appendix~\ref{app:morphology}).
SED fitting to the integrated photometry gives $\log(M_\star/M_\odot)=8.16^{+0.09}_{-0.10}$ and a rising recent SFH, with a 100-Myr-averaged ${\rm SFR}=1.44^{+0.33}_{-0.41}\,M_\odot\,{\rm yr}^{-1}$ and ${\rm SFR}_{0-10\,{\rm Myr}}=5.42^{+0.77}_{-0.52}\,M_\odot\,{\rm yr}^{-1}$ (Sect.~\ref{sec:SED_fitting}, Fig.~\ref{fig:bagpipes-fit}), the latter consistent with the aperture- and dust-corrected narrow-H$\alpha$ estimate of $9.3\,M_\odot\,{\rm yr}^{-1}$ within the systematics (Sect.~\ref{sec:halpha-sfr}).

\item Under the broad-line region interpretation, the fiducial virial calibration gives $\log(M_{\rm BH}/M_\odot)=6.52^{+0.18}_{-0.23}$ and $\lambda_{\rm Edd}=0.07$, with alternative calibration chains spanning $\lambda_{\rm Edd}=0.07$--$0.38$ (Sects.~\ref{sec:agn-candidate} and \ref{sec:bh-scaling}).
The resulting $M_{\rm BH}/M_\star=2.3\%$ lies approximately 2~dex above the (extrapolated) local relation for broad-line AGN, placing MARTA~3887 at the edge of the low-mass end of current JWST samples (Sect.~\ref{sec:bh-scaling}, Fig.~\ref{fig:bh-scaling}).
In an illustrative no-merger scenario with constant radiative efficiency, a light seed whose growth began only with the dominant recent episode of stellar assembly could reach the inferred mass only through sustained super-Eddington accretion ($\langle\lambda_{\rm Edd}\rangle\simeq5.7$).  Earlier onset of accretion, a more massive seed, intermittent rapid episodes, or black-hole mergers would relax this requirement.  
A source-matched comparison with \textsc{aesopica} simulations conditionally favours $10^5\,M_\odot$ seeds for the fiducial measurements, while the alternative virial calibration permits limited overlap with an efficiently growing lighter-seed population (Sects.~\ref{sec:bh-growth} and \ref{sec:aesopica-seeding}; Figs.~\ref{fig:bh-growth} and \ref{fig:aesopica-marta}).

\item MUSE resolves asymmetric \lya\ emission over a 6.5--6.9 proper-kpc major-axis FWHM, with $L_{\rm Ly\alpha}=1.39\times10^{43}$~erg~s$^{-1}$, an intrinsic FWHM of $245^{+21}_{-22}$~km~s$^{-1}$, and a peak offset of $+64^{+10}_{-8}$~km~s$^{-1}$ (Sect.~\ref{sec:muse}, Fig.~\ref{fig:muse-lya-profile}).  The provisional $f_{\rm esc}^{\rm Ly\alpha}\simeq0.44$, the small \lya\ velocity offset, the Mg~II doublet ratio ($R_{\rm MgII}=1.90^{+0.27}_{-0.22}$; Sect.~\ref{sec:mgii}), and the high O32 consistently point to low-column-density or porous sightlines.
Altogether, they make MARTA~3887 a strong, indirect LyC-leaker candidate, although the available data cannot measure the LyC escape fraction nor establish any direct causal link with the AGN (Sect.~\ref{sec:lyc-leakage}).
\end{itemize}

In conclusion, MARTA~3887 extends the individually detected population of accreting black holes into the metal-poor, $M_\star\simeq10^8\,M_\odot$ host regime, offering a view of black-hole growth outpacing stellar assembly in a young, low-mass galaxy beyond cosmic noon.

\section*{Acknowledgments}
MC, MB, CG, ID, LV, and AT acknowledge financial support from the
ASI--INAF Agreement No. 2025-6-HH.0 for the project ``Attività scientifiche propedeutiche alla partecipazione alla missione PRIMA del JPL/NASA''.
SK acknowledges support from the Royal Society under grant number URF~R1~251867.
FM acknowledges support from the INAF Fundamental Astrophysics programme 2022 and 2023.  FM, AM and GC acknowledge support from the European Union through the NextGenerationEU plan, Mission~4, under PRIN-MUR project ``PROMETEUS'' (202223XPZM; CUP C53D2300080-006).  AM acknowledges INAF funding through the ``Ricerca Fondamentale 2023'' programme (mini-grant 1.05.23.04.01).  GC acknowledges financial support from INAF under the Large Grant 2022 ``The metal circle: a new sharp view of the baryon cycle up to Cosmic Dawn with the latest generation IFU facilities'' and the COSMOWEBB project ``A new era for cosmology: exploiting the JWST revolution'', funded by the Italian National Institute for Astrophysics (INAF) through the PRORIS programme, CUP C83C26000300005.  AF acknowledges support from project ``VLT-MOONS'' CRAM 1.05.03.07.  LS acknowledges financial support from the PhD grant funded through PNRR funds, Notice No.~3264 of 28 December 2021, PNRR M4C2, reference IR0000034 STILES, Investment~3.1 (CUP C33C22000640006).  AT acknowledges support from the ``INAF Ricerca Fondamentale 2023 \& 2024 -- Large GO'' grant.
EB acknowledges funding from the INAF Ricerca Fondamentale 2024 (GO grant ``A JWST/MIRI MIRACLE: Mid-IR Activity of Circumnuclear Line Emission'' and MiniGrant RSN1 1.05.24.07.01).
GV acknowledges financial support from the Italian National Institute for Astrophysics (INAF) under the IAF - Astrophysics Fellowships in Italy grant CUP C59J21034720001 - ``AD MAJORA''.

This work is based on observations made with the NASA/ESA/CSA James Webb Space Telescope.  The data were obtained from the Mikulski Archive for Space Telescopes at the Space Telescope Science Institute, which is operated by the Association of Universities for Research in Astronomy, Inc., under NASA contract NAS~5-03127 for JWST.  These observations are associated with JWST programmes 1727, 1810, 1837, 1879, 5893, and 7814.  Some of the imaging products presented herein were retrieved from the Dawn JWST Archive (DJA).  DJA is an initiative of the Cosmic Dawn Center (DAWN), which is funded by the Danish National Research Foundation under grant DNRF140.  The DJA imaging products were reduced with \textsc{grizli} \citep{Brammer2023grizli}.  This work is also based on observations made with the NASA/ESA Hubble Space Telescope and obtained from the Mikulski Archive for Space Telescopes, and on observations collected at the European Southern Observatory under ESO programme 097.A-0160(A).  This research made use of \textsc{Astropy}, \textsc{NumPy}, \textsc{SciPy}, and \textsc{Matplotlib} \citep{Astropy2013,NumPy2020,SciPy2020,Matplotlib2007}.
Generative-AI tools (OpenAI GPT-5.6 Sol) were used to assist with code development, figure preparation, and language editing; all the scientific analysis and interpretation, numerical results, references to literature work, and final text have been validated and remains under the full responsibility of the authors.

\bibliographystyle{aa}
\bibliography{references}

\FloatBarrier
\onecolumngrid
\clearpage
\twocolumngrid
\appendix
\setcounter{figure}{0}
\renewcommand{\thefigure}{\thesection\arabic{figure}}
\makeatletter
\@addtoreset{figure}{section}
\makeatother
\section{Aperture corrections on the NIRSpec spectrum}
\label{app:slitloss}

The reduced one-dimensional NIRSpec products include the standard point-source path-loss and slit-loss correction.  MARTA~3887 is nevertheless resolved into several compact knots, so this calibration does not recover the flux that would be measured by the $0.5\arcsec$-radius aperture used for the integrated SED fitting.
We here define the aperture-correction factor as $C_{\rm ap}\equiv F_{0.5\arcsec}/F_{\rm NIRSpec,ps}$, where the denominator is the flux returned by the point-source-calibrated one-dimensional extraction.  We use this imaging-based correction only when the comparison is explicitly referred to the same $0.5\arcsec$ aperture and a nearby NIRCam band provides a plausible morphological proxy for the quantity of interest, as F300M does e.g. for narrow \ha.

For each NIRCam band, we projected the full central shutter onto the native image using the WCS and every MSA configuration that enters the extracted spectrum (39 G140M and 12 G235M integration/nod footprints, equal-weighted).  We measured the admitted fraction of the $0.5\arcsec$ aperture light and divided the corresponding NIRSpec point-source throughput by it.  The three nod positions are repeated across the exposures, but retaining every footprint ensures that the adopted geometry follows the actual data combination.  At the observed wavelength of \ha, the direct F300M image proxy gives $C_{\rm ap}=1.76^{+0.06}_{-0.09}$.

We tested the imaging/spectroscopy PSF mismatch separately.
We fitted the F300M morphology with three intrinsic Gaussian clumps convolved with the NIRCam STPSF, re-convolved the best-fit intrinsic model with the NIRSpec PSF at the wavelength of interest, and repeated the shutter calculation.  This gives $C_{\rm ap}=1.67^{+0.05}_{-0.03}$.  An independent comparison of synthetic G235M photometry with the measured F300M $0.5\arcsec$ flux gives $1.83^{+0.27}_{-0.21}$.  We adopt the direct-image value for narrow \ha\ and use the envelope 1.64--2.10 as a conservative aperture-mapping systematic.

We note that, although the inferred factor varies across the NIRCam bands, we do not interpolate it and rescale the complete spectrum.
In fact, the imaging supplies a broad-band continuum proxy, while the true cross-dispersion morphology of every individual emission line remains unknown; a line-by-line application would therefore assume that all lines follow the nearby continuum morphology.
Indeed, applying such broad-band factors to the Balmer series produces inconsistent decrements.  The correction is consequently used only for the integrated narrow \ha\ flux in Sect.~\ref{sec:halpha-sfr}, for which the target $0.5\arcsec$ aperture and the F300M proxy are explicitly defined.  For the \lya\ comparison in Sect.~\ref{sec:muse}, we use this aperture-corrected \ha\ flux under the additional assumption that the non-resonant recombination emission is compact (the larger $r=1.4\arcsec$ MUSE aperture preventing a strictly aperture-matched escape fraction calculation).
All other narrow-line ratios, and associated diagnostics, retain the values inferred directly from fitting the pipeline-calibrated NIRSpec spectrum.
Finally, the broad \ha\ component is treated as a point-like source and retains the native pipeline calibration, with no further correction applied.

\section{Broad H$\alpha$ robustness tests}
\label{app:ha-tests}

Because the broad \ha\ component is detected at modest contrast against a much brighter narrow line, we here test its width recovery against different choices in the local spectral model and the resolved morphology of the source.
Starting from the fiducial fit described in Sect.~\ref{sec:blrg}, we repeated the analysis while changing the fitted wavelength interval, the polynomial order of the continuum, the inclusion of the adjacent [S\,\textsc{ii}] lines, and the constraints on the narrow component.
Furthermore, we also replaced the fiducial scaled uniformly illuminated-slit LSF with the nominal uniform-slit and point-source prescriptions \citep{deGraaff2024}, and allowed a Gauss--Hermite rather than Gaussian broad profile.
In all cases, we found that including a broad component is favoured by the data.  For the cases in which the narrow-line kinematics is left free, the inferred broad flux lies within $(1.53$--$1.77)\times10^{-18}$~erg~s$^{-1}$~cm$^{-2}$, the intrinsic FWHM within 1200--1480~km~s$^{-1}$, and $-52<\Delta{\rm BIC}<-41$.
The inferred broad component properties are therefore mostly insensitive to the continuum and LSF choices explored here.

The [O\,\textsc{iii}] fit independently identifies a non-core component with ${\rm FWHM}=299$~km~s$^{-1}$ and $\Delta v=-61$~km~s$^{-1}$ (Sect.~\ref{sec:oiii_hb}).
We test whether the apparent broad base could instead arise from the blending of intermediate-width \ha\ and \nii\ emission by assigning these three lines the measured [O\,\textsc{iii}] kinematics, while fixing the \nii\ doublet ratio.  The same component is also allowed for the [S\,\textsc{ii}] doublet over the wider fitted window.  This intermediate-complex model is disfavoured relative to the fiducial broad-\ha\ model by $\Delta{\rm BIC}=41.8$.  Adding broad \ha\ to the intermediate-complex model improves its BIC by 28.2, whereas retaining the intermediate components in addition to broad \ha\ is not supported.  In the combined fit, the broad component has $F_{{\rm H}\alpha,{\rm b}}=1.56\times10^{-18}$~erg~s$^{-1}$~cm$^{-2}$ and ${\rm FWHM}=1246$~km~s$^{-1}$, consistent with the fiducial measurements.  Thus, the detected [O\,\textsc{iii}] wing cannot account for the broad \ha\ base.  Compact shocks or high-density outflows remain possible because their forbidden emission may be collisionally suppressed, but they are not favoured over the broad-line model by the present data.

\begin{figure*}
\centering
\begin{minipage}[t]{0.39\textwidth}
\makebox[\linewidth][l]{\textbf{(a)}}\\[-0.5ex]
\includegraphics[width=\linewidth]{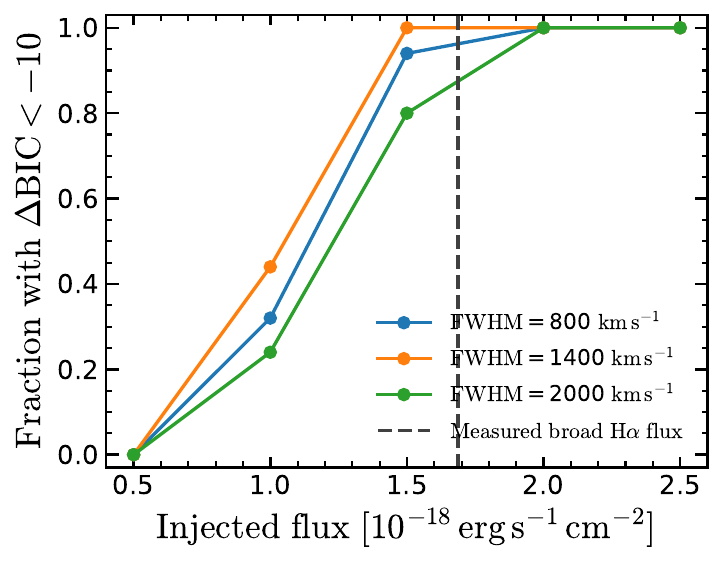}
\end{minipage}
\begin{minipage}[t]{0.59\textwidth}
\makebox[\linewidth]{\makebox[0.49\linewidth][l]{\textbf{(b)}}\makebox[0.49\linewidth][l]{\textbf{(c)}}}\\[-0.5ex]
\includegraphics[width=\linewidth]{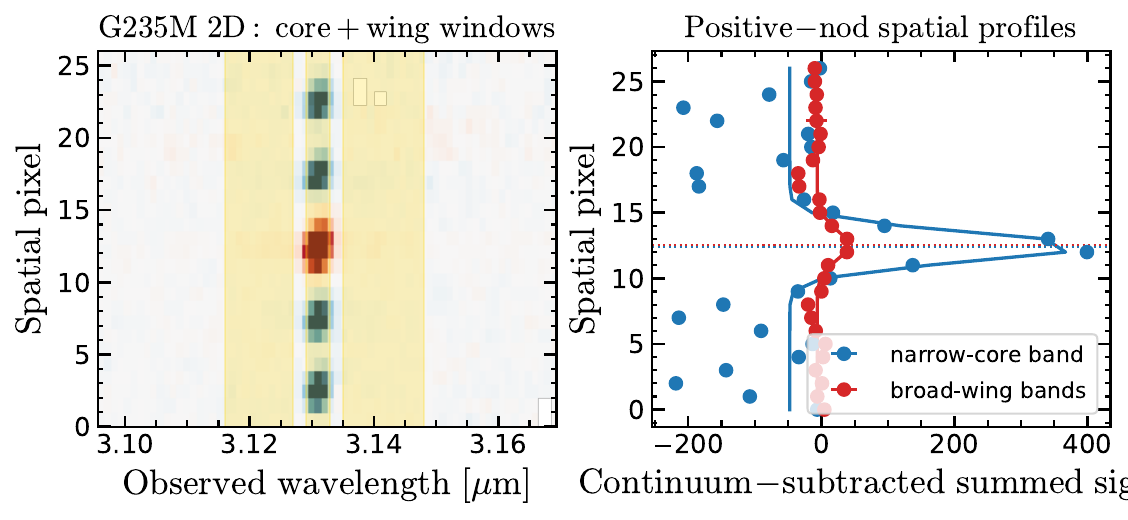}
\end{minipage}
\caption{Robustness tests for the broad \ha\ component.  Panel (a): fraction of injection--recovery realizations in which the broad model is decisively preferred ($\Delta{\rm BIC}<-10$), as a function of injected flux and width; the vertical dashed line marks the measured broad-line flux.  Panel (b): continuum-subtracted G235M two-dimensional spectrum around \ha; shaded regions mark the narrow-core and symmetric broad-wing integration windows.  Panel (c): spatial profiles obtained from those windows for all three nod positions, with the fits performed on the positive nod containing MARTA~3887.  The wing emission is co-spatial with the narrow core, while the tilt of the resolved narrow-line trace is insufficient to reproduce the broad spectral base.}
\label{fig:halpha-robustness}
\end{figure*}

Next, we quantified the statistical recovery of the broad \ha\ component with three complementary resampling experiments.
First, 350 parametric bootstrap realizations generated from the fiducial broad-line model have a median $F_{{\rm H}\alpha,{\rm b}}=1.76\times10^{-18}$~erg~s$^{-1}$~cm$^{-2}$.  We retain the reference-fit value, $1.69\times10^{-18}$~erg~s$^{-1}$~cm$^{-2}$, as the fiducial centre and recenter the bootstrap percentile widths on it, giving the $1.69^{+0.26}_{-0.19}\times10^{-18}$ interval quoted in Sect.~\ref{sec:blrg}.  The same bootstrap gives ${\rm FWHM}_{{\rm H}\alpha,{\rm b}}=1270^{+260}_{-250}$~km~s$^{-1}$.
Second, we injected components with FWHM of 800, 1400, and 2000~km~s$^{-1}$ over a range of line fluxes and refitted each realization with both the narrow-only and broad-line models.
At $1.5\times10^{-18}$~erg~s$^{-1}$~cm$^{-2}$, i.e. close to the actual measured flux, 80--100\% of the trials meet the conservative $\Delta{\rm BIC}<-10$ criterion; the recovery fraction reaches unity for every tested width by $2.0\times10^{-18}$~erg~s$^{-1}$~cm$^{-2}$ (Fig.~\ref{fig:halpha-robustness}(a)). The median recovered fluxes and widths track the injected values without evidence for a preferential positive-flux bias.
Finally, none of 250 spectra simulated from the best-fitting narrow-only model produces a BIC improvement as large as the observed $\Delta{\rm BIC}=-46.8$; the 1st--99th percentile interval of the null distribution is $6.8<\Delta{\rm BIC}<14.5$.

Finally, the two-dimensional G235M spectrum provides a separate test of morphological broadening.
Because MARTA~3887 is resolved within the shutter, the narrow-line trace is slightly tilted along the dispersion direction.
Figure~\ref{fig:halpha-robustness}(b) shows the core and symmetric wing intervals used to construct the continuum-subtracted profiles in Figure~\ref{fig:halpha-robustness}(c).  The wing emission is centred on the positive-nod signal: its centroid differs from that of the narrow core by only $0.14\pm0.11$ spatial pixels, and its fitted spatial width, $\sigma=1.17\pm0.16$ pixels, is consistent with the narrow-core value of $1.145\pm0.009$ pixels.  Thus the spectral wings are spatially associated with MARTA~3887 rather than with a displaced contaminant.  Moreover, collapsing templates based on the resolved narrow trace or on the adjacent continuum does not reproduce the wavelength extent of the broad base, showing that the tilted narrow-line morphology alone cannot account for the flux excess in the wings as seen in the one-dimensional spectrum.

In conclusion, the spectral decomposition, resampling experiments, and two-dimensional tests demonstrate that the measured broad component is not an artefact of the adopted continuum, LSF, noise realization, or resolved narrow-line trace, corroborating the scenario at the base of the interpretation presented in Sect.~\ref{sec:agn-candidate}.

\section{HOMERUN photoionization decomposition}
\label{app:homerun}

\textsc{homerun} represents an observed emission-line spectrum as a non-negative weighted linear combination of constant-density, single-cloud \textsc{Cloudy} photoionization models \citep{Marconi2024,Ferland2017}.  At a given chemical composition and incident continuum, the library spans electron density and ionization parameter; the cloud weights are determined by minimizing the \textsc{homerun} loss function.  This construction allows different transitions to arise predominantly from different regions of the $(n_{\rm e},U)$ grid rather than imposing a single characteristic density and ionization parameter on the integrated spectrum.  The minimum-loss solution is identified separately for each metallicity and ionizing continuum, as described in detail by \citet{Marconi2024} and \citet{Moreschini2026} (see also \citet{Castellano2026} for another application to a high-z galaxy).

We fitted the narrow-line measurements and limits listed in Table~\ref{tab:line-fluxes}.  The stellar-only family contains clouds illuminated by the adopted stellar continuum, while the composite family combines independent stellar- and AGN-ionized cloud ensembles.  To test whether the hard detected lines can be reproduced without an AGN, the stellar grid is extended to $\log U=+0.5$, beyond the $\log U\leq-1$ range normally adopted for the standard H\,\textsc{ii}-region grid.  For the composite family we impose the fiducial $f_{\rm AGN}({\rm H}\beta)\leq10\%$ prior described in Sect.~\ref{sec:metallicity}.  We explored separate attenuation configurations for the two components and recovered only small fitted extinctions.  The comparison in the main text is based on the objective-function minima within each model family; these values provide a relative model ranking and are not interpreted as formal $\chi^2$ probabilities.

\begin{figure*}
\centering
\includegraphics[width=0.95\textwidth]{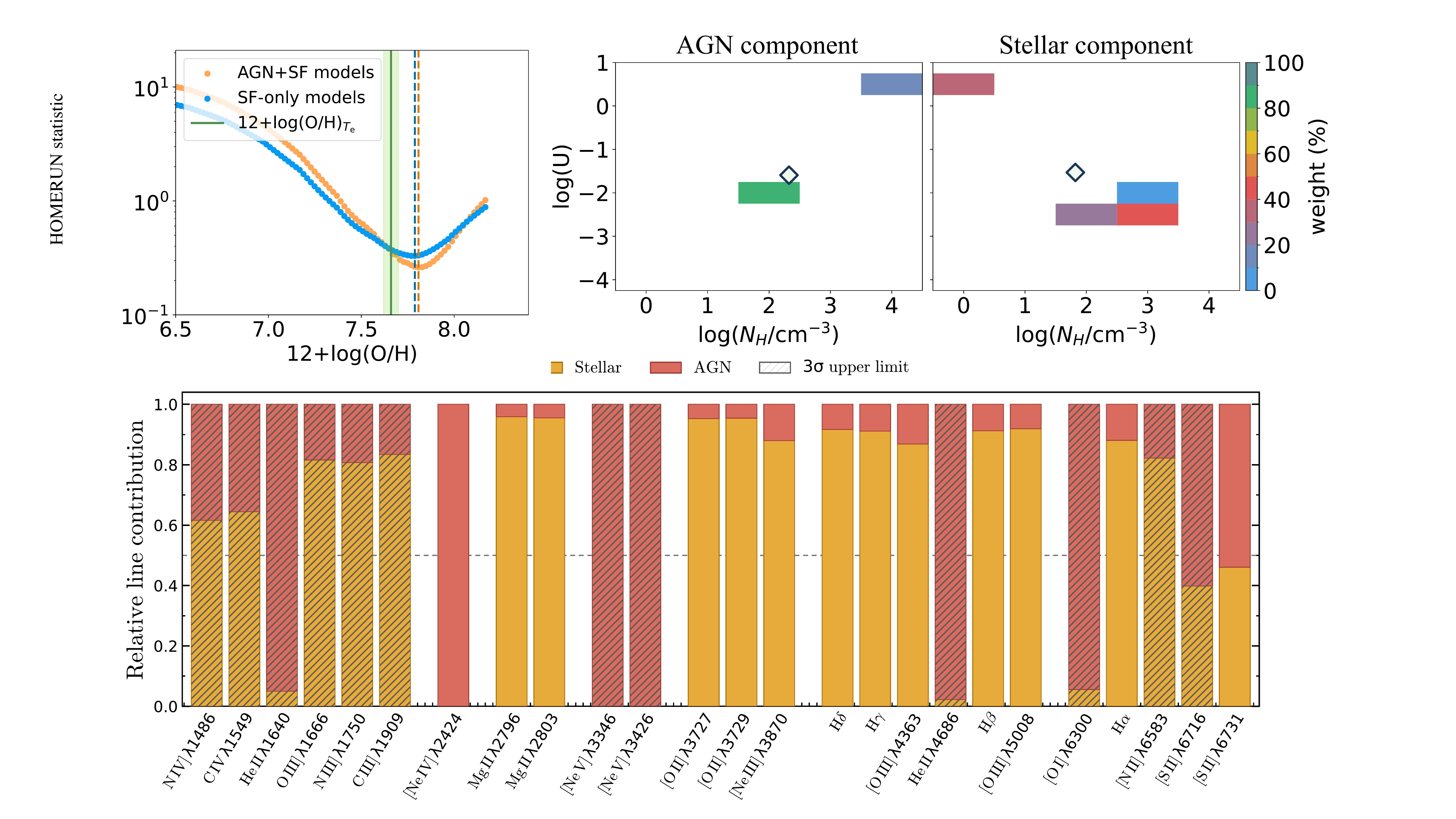}
\caption{Summary of the \textsc{homerun} photoionization modeling.  Upper left: value of the \textsc{homerun} goodness-of-fit statistic as a function of oxygen abundance for the stellar-only (blue) and stellar+AGN (orange) model families.  Dashed lines mark their respective minima at $12+\log({\rm O/H})=7.79$ and 7.81.  The green line and band show the fiducial direct-method abundance and its statistical uncertainty.  Because the statistic includes the adopted model tolerance, it is used to rank solutions within each family rather than as a formal $\chi^2$ probability.  Upper middle and right: non-zero cloud weights in the $(n_{\rm H},U)$ plane for the AGN and stellar components, respectively, of the best-fitting composite solution; colours encode the percentage weight assigned to each grid cell and diamonds mark the weight-averaged locations.  Lower panel: relative stellar and AGN contributions to each predicted line flux.  Every stacked bar sums to unity.  Hatched bars identify transitions entering the fit as one-sided $3\sigma$ upper limits, for which the partition describes a model prediction below the limit rather than a detected flux.  The Balmer and principal oxygen lines remain stellar dominated, whereas the detected [Ne\,\textsc{iv}] emission is produced almost entirely by AGN-ionized clouds.}
\label{fig:homerun-contributions}
\end{figure*}

Figure~\ref{fig:homerun-contributions} summarizes the metallicity search and the structure of the best-fitting composite solution.  The stellar-only and stellar+AGN families reach their minimum loss at $12+\log({\rm O/H})=7.79$ and 7.81, respectively, compared with the fiducial direct-method value of $7.68\pm0.04$.  The shallow, nearly coincident minima show directly that adding the adopted AGN continuum has little effect on the model-based abundance.  The non-zero cloud weights span more than one density and ionization parameter in both components: the stellar ensemble is weighted mainly towards $\log U\simeq-2$, whereas the AGN ensemble includes both low-density, high-ionization gas and denser clouds with lower $U$.  These weights describe an unresolved emission-line mixture and should not be interpreted as distinct spatial zones.

The lower panel of Fig.~\ref{fig:homerun-contributions} shows how the composite solution partitions the predicted flux of each independent line constraint between the stellar and AGN components.  The stellar component supplies 91\% of H$\beta$, 88\% of H$\alpha$, 87\% of [O\,\textsc{iii}] $\lambda4363$, and 92\% of [O\,\textsc{iii}] $\lambda5008$; it also contributes more than 95\% of both [O\,\textsc{ii}] and Mg\,\textsc{ii}.  By contrast, the detected [Ne\,\textsc{iv}] emission is almost entirely assigned to AGN-ionized gas.  The predicted He\,\textsc{ii}, [Ne\,\textsc{v}], and [O\,\textsc{i}] fluxes are likewise AGN dominated, but these transitions enter only as upper limits and therefore do not constitute additional detections of an AGN component.  The [S\,\textsc{ii}] doublet receives comparable contributions from the two components and is also the least accurately reproduced detected complex.

The oxygen lines that set $T_{\rm e}({\rm O\,\textsc{iii}})$ and dominate the abundance estimate therefore remain primarily associated with the stellar component.  This explains why the inferred O/H changes by only 0.02~dex between the stellar-only and composite solutions.  The detailed fractions remain conditional on the adopted model grids and on the H$\beta$ prior; they should not be interpreted as an observational separation of spatially distinct emitting regions.

\section{Photometric and morphological properties of the host galaxy}
\label{app:photometry}

\subsection{Image cutouts and integrated photometry}
\label{app:cutouts}

\begin{figure*}
\centering
\includegraphics[width=\textwidth]{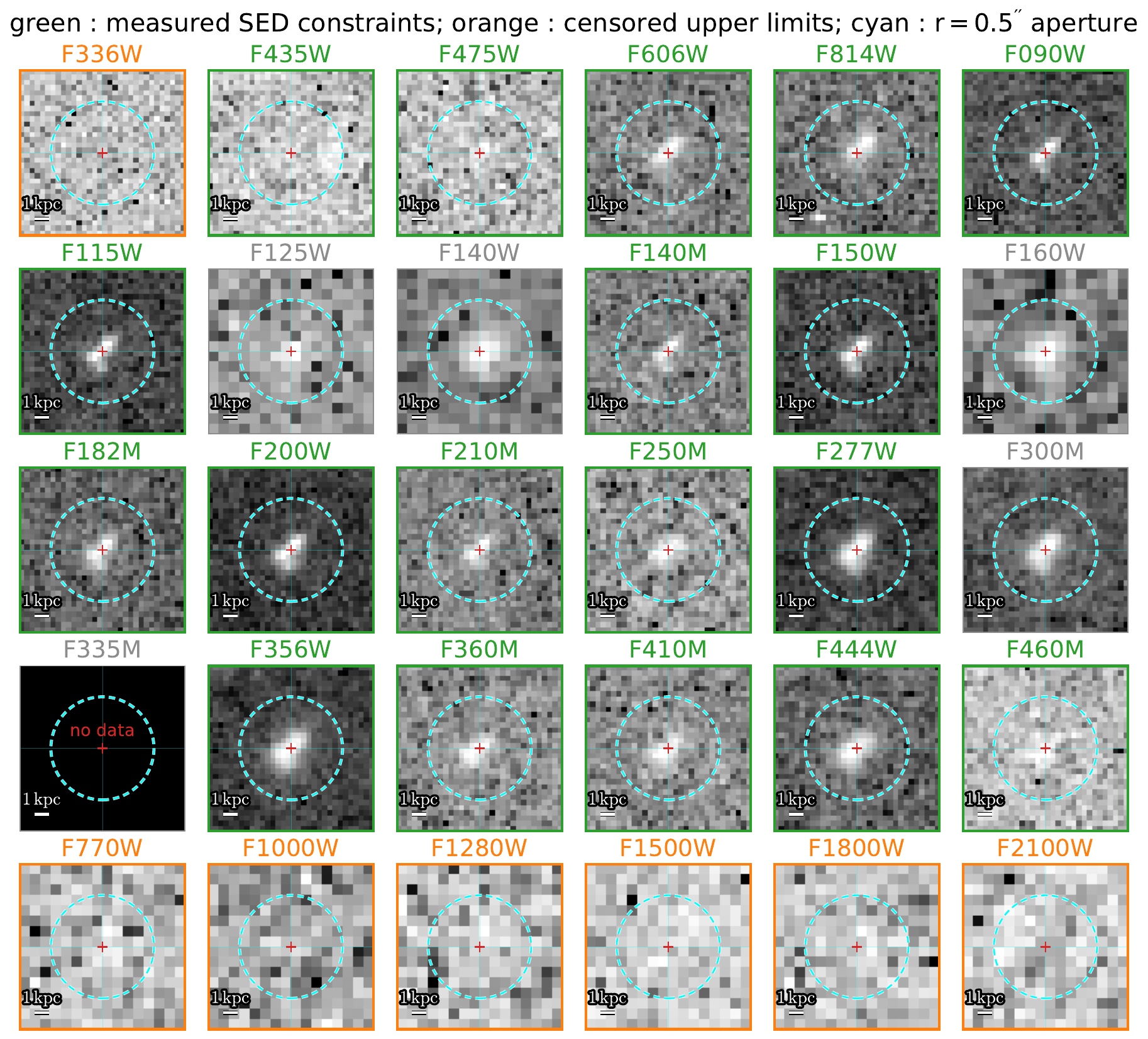}
\caption{HST, NIRCam, and MIRI cutouts in a common $1.6\arcsec$ field.  The cyan dashed circle marks the $r=0.5\arcsec$ aperture used for the integrated photometry.  The scale bar in each panel denotes 1 proper kpc at $z=3.7694$ ($1\arcsec=7.28\,\mathrm{kpc}$).  Green frames mark bands entering the SED fit as measured flux constraints, including the low-S/N F475W measurement.  Orange frames mark the HST F336W and MIRI bands treated as censored upper limits.  Grey labels denote excluded or missing bands.}
\label{fig:cutouts}
\end{figure*}

Figure~\ref{fig:cutouts} presents all of the imaging data considered for MARTA~3887 on the same $1.6\arcsec$ field.  The HST panels in the first two rows establish the Lyman-break constraints discussed in Sect.~\ref{sec:lyc-leakage}; in particular, F336W is a non-detection, whereas F435W contains flux from both sides of the redshifted Lyman limit.  The NIRCam panels from F090W to F460M show that the two principal knots and the fainter southern component are contained within the $0.5\arcsec$ aperture used for the integrated photometry.  F300M is displayed for completeness but excluded from the SED because it contains broad \ha, while the black F335M panel records missing coverage.  The six MIRI panels in the bottom row contain no significant source at the target position and enter the fit only as upper limits.

We measured the integrated fluxes in a common $0.5\arcsec$ aperture, which includes all internal knots while avoiding the nearest external neighbour, and estimated the local background in a clipped $0.70\arcsec<r<1.05\arcsec$ annulus.  Before measurement, the NIRCam images were matched to the F460M PSF so that the aperture samples a consistent light fraction across filters; filter-dependent STPSF encircled-energy corrections were then applied.  HST images were retained at their native resolution but measured with the same sky aperture.  The MIRI images were likewise kept at their native PSFs, and their signed aperture measurements were converted into conservative total-flux limits using the appropriate encircled energies.

The BAGPIPES likelihood treats detected HST and NIRCam bands as Gaussian measurements.  For each MIRI band, it instead uses the Gaussian cumulative probability below the quoted $3\sigma$ limit \citep{Sawicki2012}.
This censored treatment preserves the information that bright rest-frame near-IR templates are disfavoured without assigning a positive flux to a non-detection, as shown in the `hidden-old-population test' below.

\subsection{Multi-component morphology}
\label{app:morphology}

\begin{figure*}
\centering
\includegraphics[width=0.98\textwidth]{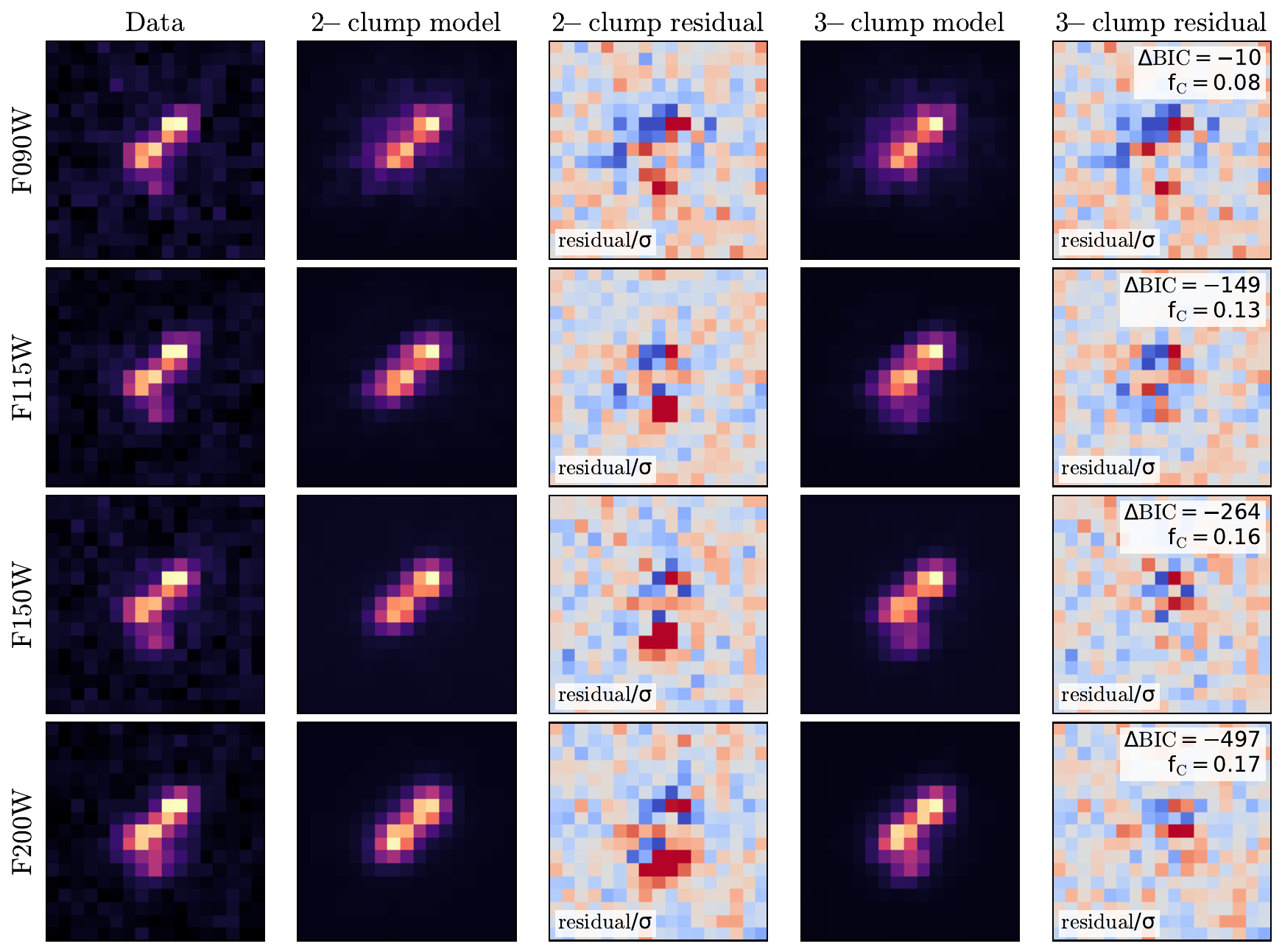}
\caption{PSF-convolved morphology fit in the native F090W, F115W, F150W, and F200W images.  From left to right, each row shows the data, the two-component model and its residuals, and the three-component model and its residuals, normalized by the pixel uncertainty.  The final panels report the BIC difference between the three- and two-component fits and the fitted light fraction of component C.  The southern component is strongly preferred in F115W--F200W, where it removes the coherent residual below the two brighter knots.}
\label{fig:three-clump}
\end{figure*}

We fitted the native F090W, F115W, F150W, and F200W images with PSF-convolved Gaussian components.  F200W defines the relative positions and intrinsic shapes, after which each filter has independent non-negative component amplitudes and a local sky plane.  Each row of Figure~\ref{fig:three-clump} follows the five-panel decomposition used for the F150W fit in the main text: data, two-component model and residuals, and three-component model and residuals.  Residuals are expressed in units of the pixel uncertainty.
The third component is only marginal in F090W but is consistently recovered from F115W through F200W; over these three bands it contributes approximately 13--17\% of the total light and removes the coherent southern residual left by the two-component model.

To test for colour gradients across the system, we also fitted the fluxes of the three components in the remaining NIRCam images, while keeping their F200W positions and intrinsic shapes fixed and adopting the appropriate STPSF for each band.
In the blue bands the clumps remain sufficiently well separated that their fitted fluxes are only weakly correlated ($|\rho|\lesssim0.25$).  At 3.5--4.6\,$\mu$m, the broader PSF blends them more strongly: flux can be exchanged between neighbouring components with little change in the total model, and the largest pairwise correlations increase to $|\rho|=0.24$--$0.45$.
These correlations describe only the statistical degeneracy within the adopted model.  They do not include systematic uncertainties from assuming that the intrinsic F200W morphology is unchanged at all wavelengths, from using a detector-centre STPSF rather than the effective position- and exposure-weighted PSF of the mosaic, or from genuine wavelength-dependent structures.
Accordingly, we retain only the qualitative indication that the south-western component is redder than the north-eastern one, and we do not adopt the inferred fluxes to perform independent SED fitting at the clump level; all physical properties of the host remain inferred from the integrated photometry.

\subsection{Sensitivity to SFH priors and `maximally-old' stress tests}
\label{app:max-old}

As a weakly regularized alternative to the fiducial continuity SFH model presented in Sect.~\ref{sec:SED_fitting}, we retain the same six time bins and parameter bounds but replace the Student-$t$ continuity terms with independent flat priors on the logarithmic adjacent-bin SFR ratios.  This fit converges on a strongly rising, $\lesssim5$~Myr solution with $\log(M_\star/M_\odot)=7.70^{+0.08}_{-0.02}$, 0.46~dex below the fiducial mass.  Its best photometric likelihood is poorer than that of the scale-0.3 continuity fit by $\Delta\chi^2=7.5$.  The experiment demonstrates the residual sensitivity of the host mass to the SFH prior; we retain the regularized solution as fiducial while including the full downward excursion in the asymmetric host-model systematic.

Furthermore, the fiducial non-parametric SFH identifies the host galaxy growth as dominated by a recent star-formation episode, but cannot exclude the presence of a faint, high-mass-to-light-ratio population.
We therefore constructed a `maximally old' two-component model consisting of a 5--100~Myr burst, required to reproduce the young light, and an independent exponentially declining component whose onset is restricted to ages older than 200~Myr.  This stress test attempts to quantify hidden mass and is not treated as an alternative preferred SFH.

\begin{figure*}
\centering
\includegraphics[width=0.92\textwidth]{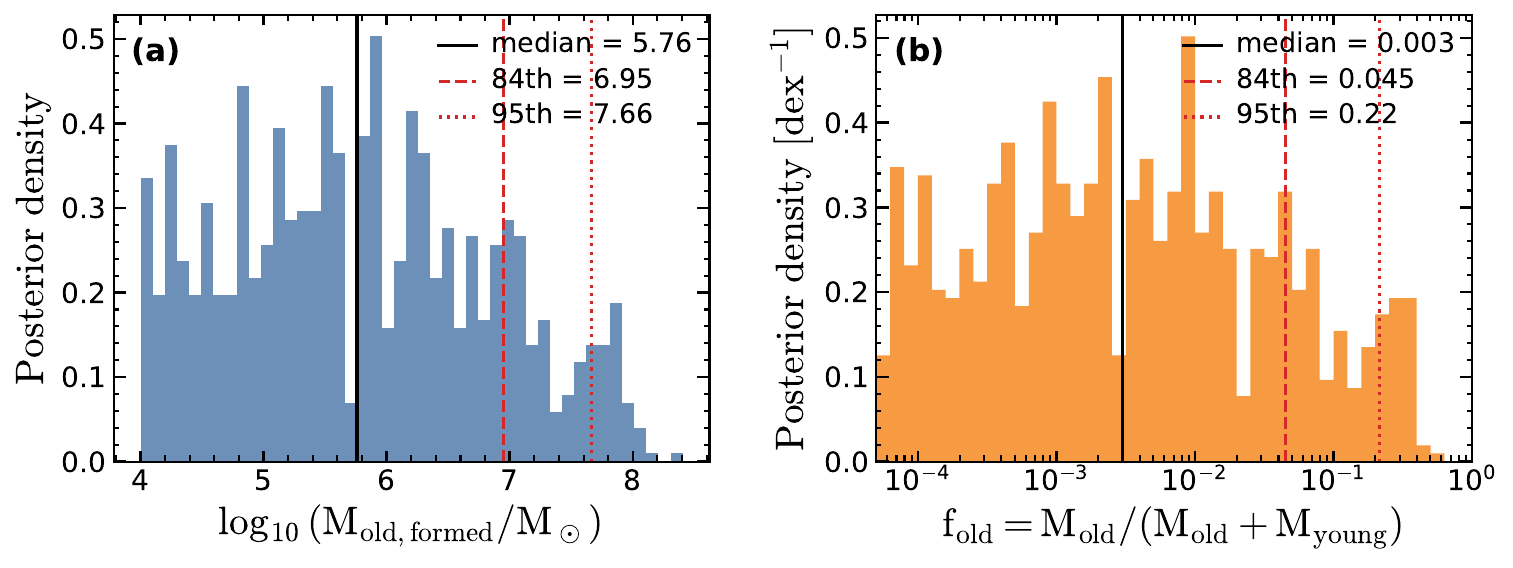}
\caption{Posterior for the maximally-old test including all MIRI upper limits.  Panel (a) shows the formed mass of the independent old component; panel (b) shows its fraction of the combined old-plus-young formed mass on a logarithmic axis.  Solid black, dashed red, and dotted red lines mark the median, 84th percentile, and 95th percentile, respectively.}
\label{fig:max-old}
\end{figure*}

Figure~\ref{fig:max-old}(a) shows the posterior formed mass of the old component.  It peaks at a negligible contribution, with a median of $5.7\times10^5\,M_\odot$, but has a long prior-sensitive tail reaching $8.9\times10^6\,M_\odot$ and $4.6\times10^7\,M_\odot$ at the 84th and 95th percentiles.  Figure~\ref{fig:max-old}(b) expresses the same result as a fraction of the combined old-plus-young formed mass and gives $f_{\rm old}<0.22$ at 95\% posterior probability.  Repeating the experiment without the F1000W--F2100W limits relaxes the corresponding bounds to $M_{\rm old,formed}<2.0\times10^8\,M_\odot$ and $f_{\rm old}<0.62$.  The ensemble of MIRI non-detections therefore limits, but does not exclude, an older population.

\section{MUSE data processing and emission lines}
\label{app:muse}

\subsection{Astrometric registration and noise rescaling}

The archive MUSE world-coordinate system is displaced from the HST/JWST reference frame.  We measured a rigid translation from three bright, isolated reference sources detected in both the MUSE white-light image and HST/F814W.
The mean correction is $(\Delta\alpha\cos\delta,\Delta\delta)_{\rm HST-MUSE}=(-0.348\arcsec,-1.892\arcsec)$, with an rms scatter of approximately $0.03\arcsec$ among the three matches.  No additional rotation or scale term is required by this small reference set.

\begin{figure*}
\centering
\includegraphics[width=0.96\textwidth]{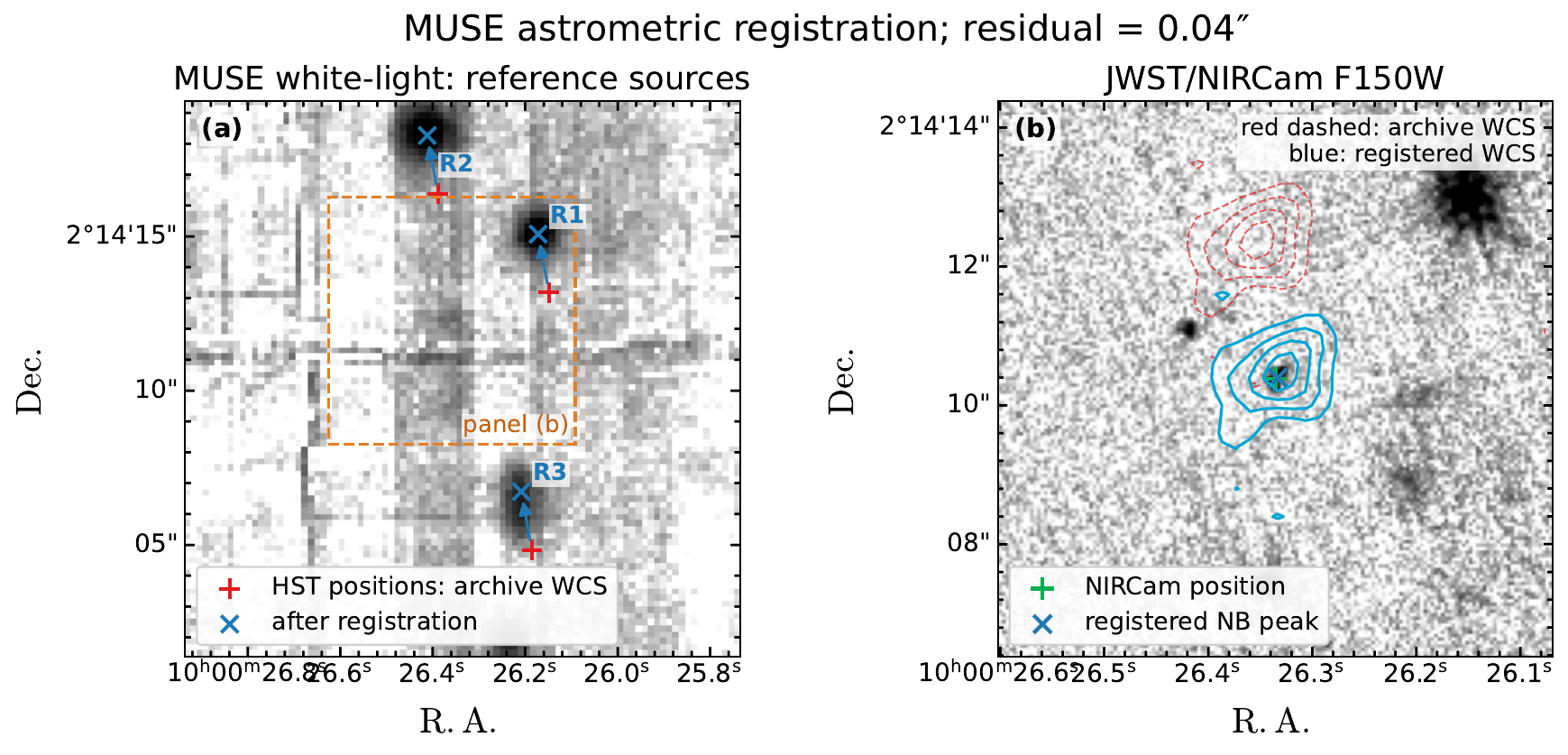}
\caption{Registration of the MUSE cube to the HST/JWST astrometric frame.  Panel (a) shows the MUSE white-light image and the three reference sources R1--R3.  Red pluses mark their HST/F814W positions projected through the archive MUSE WCS; arrows connect them to the blue crosses obtained after applying the rigid $1.92\arcsec$ translation.  The dashed orange box identifies the target-centred field shown in panel (b).  Panel (b) shows the JWST/NIRCam F150W image with the \lya\ narrow-band contours before (red dashed) and after (blue solid) the correction.  The green plus marks the NIRCam position and the blue cross the registered \lya\ peak, separated by $0.037\arcsec$.}
\label{fig:muse-astrometry}
\end{figure*}

Figure~\ref{fig:muse-astrometry} shows the three reference matches and the effect of the correction.  When the HST positions are projected using the archive MUSE WCS, they are displaced from the corresponding white-light sources by the common $1.92\arcsec$ offset (red pluses in Fig.~\ref{fig:muse-astrometry}a).  The rigid translation moves them to the blue crosses, which coincide with the MUSE sources.  The same correction moves the \lya\ narrow-band contours from the dashed red to the solid blue position in Fig.~\ref{fig:muse-astrometry}(b).  The registered \lya\ peak lies only $0.037\arcsec$ from the NIRCam position, comparable to the registration rms.  This corrected frame is used for the apertures, centroid, and spatial measurements reported in Sect.~\ref{sec:muse}.

To account approximately for resampling covariance and residual sky structure not captured by the formal variance cube, we measure the wavelength-dependent scatter among 16 blank apertures around the source and rescale the spectral uncertainties accordingly.  The same procedure is applied to the two constituent MUSE stacks (good and bad seeing).
As an additional spatial check, the stack with better seeing gives a deconvolved major-axis FWHM of $0.89\arcsec$ ($6.5$ proper kpc), consistent with the $0.95\arcsec$ ($6.9$ proper kpc) measurement from the full-depth cube; the minor-axis size is less robust.

\subsection{Rest-UV lines (non-)detections}

We searched the $0.8\arcsec$ MUSE extraction for the principal rest-UV emission-line complexes covered at the systemic redshift.  For each feature we fitted a local linear continuum together with an LSF-convolved template of intrinsic width 100~km~s$^{-1}$; doublet ratios were fixed to the values stated in the extraction model and are used only to define integrated limits.
For all spectral fits we adopt the wavelength-dependent MUSE line-spread function measured from sky lines by \citet{Bacon2017}, ${\rm FWHM}_{\rm LSF}(\lambda)=5.866\times10^{-8}\lambda^2-9.187\times10^{-4}\lambda+6.040$~\AA, where $\lambda$ is the observed wavelength in \AA.
Formal cube errors were multiplied by the wavelength-dependent correction inferred from blank apertures, as described above.  We evaluated each template at systemic velocity and also scanned $-400<\Delta v<400$~km~s$^{-1}$ in 25~km~s$^{-1}$ steps as a check for displaced emission.

\begin{figure*}
\centering
\includegraphics[width=0.98\textwidth]{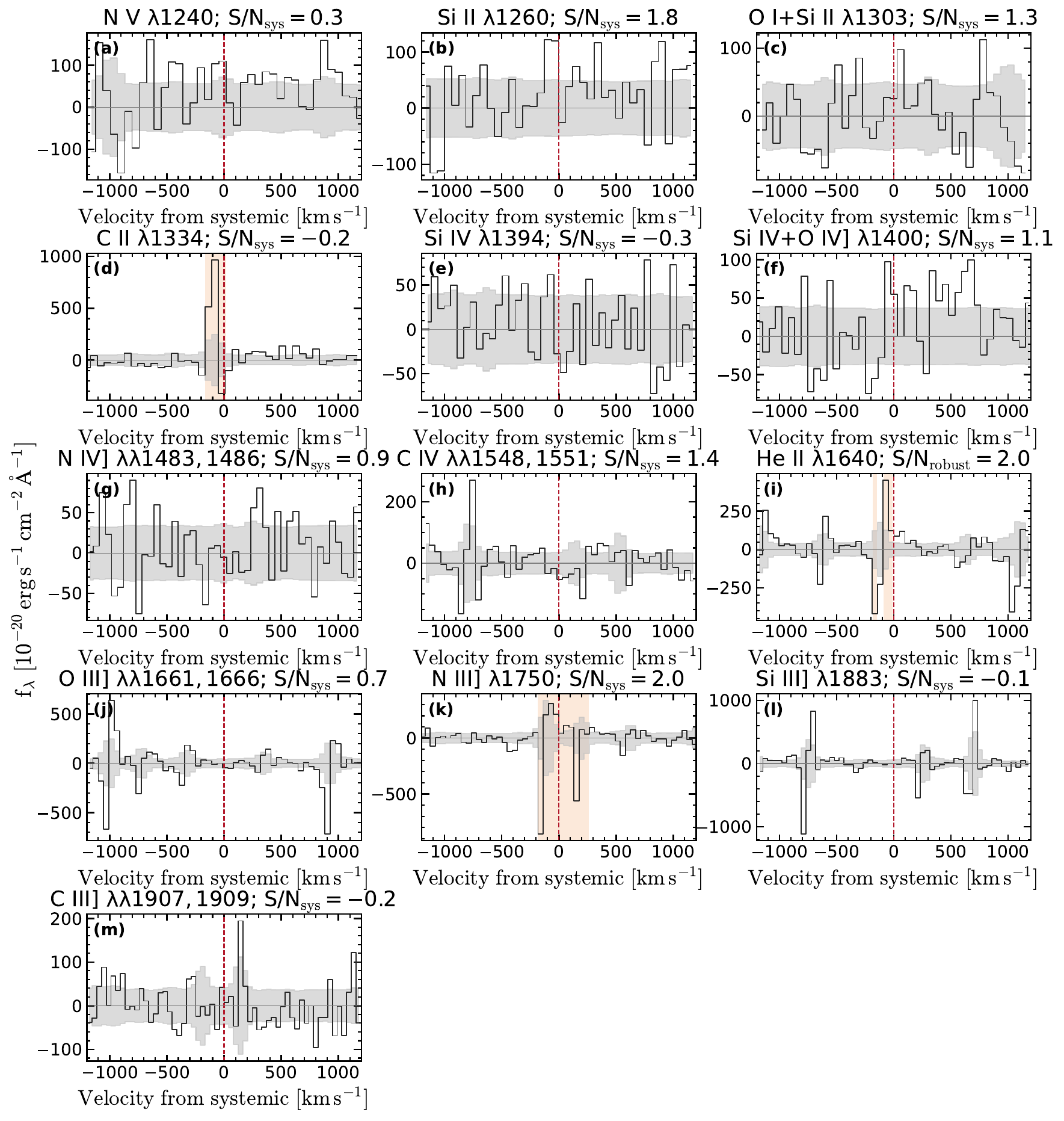}
\caption{MUSE rest-UV line inventory for MARTA~3887.  Panels (a)--(m) show the continuum-subtracted $0.8\arcsec$-aperture spectrum around each transition or line complex; velocities are measured relative to $z=3.7694$.  Grey bands denote the empirically rescaled $1\sigma$ uncertainty and red dashed lines mark systemic velocity.  Pale-orange vertical bands identify the channels affected by sky residuals or stack inconsistency in the C~II, He~II, and N~III] panels.  The quoted He~II significance includes the channel mask and stack-difference systematic described in the text.}
\label{fig:muse-uv-nondetections}
\end{figure*}

Figure~\ref{fig:muse-uv-nondetections} displays every tested emission line complex in velocity space.  Panels (a)--(c), (e)--(h), (j), (l), and (m) show no clear feature above $3\sigma$.  The large spike near C~II in panel (d) coincides with a narrow sky residual, while N~III] in panel (k) lies in a similarly compromised spectral region; neither feature is considered astrophysical.  The shaded bands show the empirically rescaled $1\sigma$ uncertainties, and the systemic matched-filter S/N is printed above each panel.

He~II $\lambda1640$ requires a separate qualification.  Three spectral samples near the expected line centre show strong sign reversals between the two constituent stacks and are therefore masked independently of their value in the combined spectrum.  The masked fit has a formal S/N of 3.2.  We additionally take half the difference between the amplitudes measured in the two stacks as a stack-dependent systematic uncertainty; adding this term in quadrature reduces the robust significance to S/N$=2.0$, as reported in Figure~\ref{fig:muse-uv-nondetections}(i).  The residual structure is also spatially stripe-like rather than source-like.  We therefore classify it as a sky-subtraction residual and retain only an upper limit on He~II.

\section{Emission-line measurements}
\label{app:line-fluxes}

Table~\ref{tab:line-fluxes} reports the emission-line measurements used throughout this work.  Detections are listed with asymmetric posterior uncertainties, while non-detections are represented by the adopted one-sided $3\sigma$ limits.  Derived doublet sums are included where they enter relevant diagnostic ratios.
\begin{deluxetable*}{@{}lllr@{}}
\tabletypesize{\scriptsize}
\setlength{\tabcolsep}{2.2pt}
\renewcommand{\arraystretch}{0.95}
\tablecaption{\MakeUppercase{Emission-line measurements and upper limits for MARTA~3887}\label{tab:line-fluxes}}
\tablehead{\colhead{Line} & \colhead{Component} & \colhead{Instrument} & \colhead{Flux [or $3\sigma$ limit]}}
\startdata
{Ly$\alpha$ 1216} & broad & MUSE 0.8\arcsec & $<41$ \\
{Ly$\alpha$ 1216} & total & MUSE 1.4\arcsec & $99.7\,\substack{+6.9\\-6.9}$ \\
{Ly$\alpha$ 1216} & total & MUSE 0.8\arcsec & $66.6\,\substack{+2.8\\-2.6}$ \\
{N\,\textsc{v} 1240} & narrow & MUSE 0.8\arcsec & $<8.63$ \\
{Si\,\textsc{ii} 1260} & narrow & MUSE 0.8\arcsec & $<5.13$ \\
{O\,\textsc{i}+Si\,\textsc{ii} 1303} & narrow & MUSE 0.8\arcsec & $<5.05$ \\
{C\,\textsc{ii} 1334} & narrow & MUSE 0.8\arcsec & $<7.77$ \\
{Si\,\textsc{iv} 1394} & narrow & MUSE 0.8\arcsec & $<4.07$ \\
{Si\,\textsc{iv}+O\,\textsc{iv}] 1400} & narrow & MUSE 0.8\arcsec & $<3.93$ \\
{N\,\textsc{iv}] 1483+1486} & narrow & MUSE 0.8\arcsec & $<5.78$ \\
{C\,\textsc{iv} 1548+1551} & narrow & MUSE 0.8\arcsec & $<5.68$ \\
{He\,\textsc{ii} 1640} & narrow & MUSE 0.8\arcsec & $<10.7$ \\
{O\,\textsc{iii}] 1661+1666} & narrow & MUSE 0.8\arcsec & $<10$ \\
{N\,\textsc{iii}] 1750} & narrow & MUSE 0.8\arcsec & $<8.81$ \\
{Si\,\textsc{iii}] 1883} & narrow & MUSE 0.8\arcsec & $<6.88$ \\
{C\,\textsc{iii}] 1907+1909} & narrow & MUSE 0.8\arcsec & $<8.51$ \\
{C\,\textsc{ii}] 2324} & narrow & NIRSpec G140M & $<1.37$ \\
{C\,\textsc{ii}] 2325} & narrow & NIRSpec G140M & $<0.618$ \\
{C\,\textsc{ii}] 2324+2325+2327} & sum & NIRSpec G140M & $<0.815$ \\
{C\,\textsc{ii}] 2327} & narrow & NIRSpec G140M & $<0.837$ \\
{[Ne\,\textsc{iv}] 2422} & narrow & NIRSpec G140M & $<0.257$ \\
{[Ne\,\textsc{iv}] 2422+2425} & sum & NIRSpec G140M & $0.286\,\substack{+0.081\\-0.076}$ \\
{[Ne\,\textsc{iv}] 2425} & narrow & NIRSpec G140M & $0.229\,\substack{+0.066\\-0.06}$ \\
{Mg\,\textsc{ii} 2796} & narrow & NIRSpec G140M & $0.467\,\substack{+0.061\\-0.052}$ \\
{Mg\,\textsc{ii} 2803} & narrow & NIRSpec G140M & $0.247\,\substack{+0.039\\-0.036}$ \\
{[Ne\,\textsc{v}] 3346} & narrow & NIRSpec G140M & $<0.0592$ \\
{[Ne\,\textsc{v}] 3426} & narrow & NIRSpec G140M & $<0.16$ \\
{[O\,\textsc{ii}] 3727} & narrow & NIRSpec G235M/H & $0.899\,\substack{+0.1\\-0.093}$ \\
{[O\,\textsc{ii}] 3727+3729} & sum & NIRSpec G235M/H & $2.21\,\substack{+0.12\\-0.11}$ \\
{[O\,\textsc{ii}] 3729} & narrow & NIRSpec G235M/H & $1.31\,\substack{+0.11\\-0.1}$ \\
{[Ne\,\textsc{iii}] 3869} & narrow & NIRSpec G235M/H & $2\,\substack{+0.15\\-0.16}$ \\
{H$\delta$ 4102} & narrow & NIRSpec G235M/H & $1.07\,\substack{+0.047\\-0.049}$ \\
{H$\gamma$ 4340} & narrow & NIRSpec G235M/H & $2\,\substack{+0.053\\-0.05}$ \\
{[O\,\textsc{iii}] 4363} & narrow & NIRSpec G235M/H & $0.739\,\substack{+0.051\\-0.048}$ \\
{He\,\textsc{ii} 4686} & narrow & NIRSpec G235M/H & $<0.169$ \\
{H$\beta$ 4861} & broad & NIRSpec G235M/H & $<0.964$ \\
{H$\beta$ 4861} & narrow & NIRSpec G235M/H & $4.55\,\substack{+0.064\\-0.064}$ \\
{[O\,\textsc{iii}] 4959} & narrow & NIRSpec G235M/H & $9.01\,\substack{+0.22\\-0.29}$ \\
{[O\,\textsc{iii}] 4959+5007} & sum & NIRSpec G235M/H & $35.9\,\substack{+0.87\\-1.2}$ \\
{[O\,\textsc{iii}] 5007} & broad & NIRSpec G235M/H & $<1.12$ \\
{[O\,\textsc{iii}] 5007} & intermediate & NIRSpec G235M/H & $2.22\,\substack{+0.94\\-0.79}$ \\
{[O\,\textsc{iii}] 5007} & narrow & NIRSpec G235M/H & $26.9\,\substack{+0.65\\-0.86}$ \\
{[O\,\textsc{i}] 6300} & narrow & NIRSpec G235M & $<0.312$ \\
{[N\,\textsc{ii}] 6548} & narrow & NIRSpec G235M & $<0.0756$ \\
{H$\alpha$ 6563} & broad & NIRSpec G235M & $1.69\,\substack{+0.26\\-0.19}$ \\
{H$\alpha$ 6563} & narrow & NIRSpec G235M & $12.9\,\substack{+0.16\\-0.19}$ \\
{[N\,\textsc{ii}] 6584} & narrow & NIRSpec G235M & $<0.224$ \\
{[S\,\textsc{ii}] 6716} & narrow & NIRSpec G235M & $<0.23$ \\
{[S\,\textsc{ii}] 6716+6731} & sum & NIRSpec G235M & $0.488\,\substack{+0.096\\-0.1}$ \\
{[S\,\textsc{ii}] 6731} & narrow & NIRSpec G235M & $0.325\,\substack{+0.06\\-0.058}$ \\
{[Ar\,\textsc{iii}] 7138} & narrow & NIRSpec G235M & $0.213\,\substack{+0.048\\-0.048}$ \\
\enddata
\tablecomments{Fluxes are in units of $10^{-18}\,\mathrm{erg\,s^{-1}\,cm^{-2}}$ and are quoted before attenuation or morphology-based aperture corrections. For detections, the stacked uncertainties enclose the 16th--84th posterior percentiles; otherwise the adopted one-sided $3\sigma$ upper limit is reported. The model-conditional broad-line limits quoted in the text and Fig.~\ref{fig:hbeta-oiii-fit} instead correspond to the 95th posterior percentiles; the table retains the more conservative $3\sigma$ values for uniformity with the line inventory. NIRSpec narrow components are taken from the joint G140M+G235M+G235H fit, except H$\alpha$ and [N\,\textsc{ii}], which use the dedicated H$\alpha$-complex fit. For the local H$\alpha$ complex, central values are the reference best fit and uncertainties are bootstrap intervals. The intermediate [O\,\textsc{iii}] row comes from the adopted narrow+intermediate fit. Rows labelled `sum' are derived quantities and are not independent of their components. The [O\,\textsc{ii}] component fluxes tabulated here come from the joint grating fit, whereas the density inference uses the better-resolved G235H doublet ratio. The [S\,\textsc{ii}] density likelihood retains the signed $\lambda6716$ measurement even though that component falls below the $3\sigma$ reporting threshold. The apparent MUSE He\,\textsc{ii} $\lambda1640$ amplitude is rejected as a stack-dependent sky residual; its limit includes the stack-difference systematic described in Appendix~\ref{app:muse}. The uncertainty on the MUSE $1.4\arcsec$ Ly$\alpha$ growth-curve flux propagates the empirically rescaled spectral noise and continuum subtraction; it excludes absolute-calibration and aperture-choice systematics. The [Ar\,\textsc{iii}] $\lambda7138$ measurement comes from its dedicated G235M fit and is excluded from the physical-condition and photoionization fits because of the nearby second-order-contaminated region.}
\end{deluxetable*}


\end{document}